\documentclass[final,5p,times,twocolumn]{elsarticle}

\usepackage{graphicx, url, amssymb, booktabs, tabularx}
\usepackage{txfonts, longtable, dcolumn, verbatim, color, lineno, lscape, multirow, pifont, afterpage, microtype}
\usepackage[T1]{fontenc}
\usepackage{natbib}
\bibpunct{(}{)}{;}{a}{}{,}

\usepackage{hyperref}

\journal{Icarus}

\begin{document}

\begin{frontmatter}

   \title{Spin states of asteroids in the Eos collisional family}
   \author[label1]{J.~Hanu{\v s}\corref{cor1}}
   \cortext[cor1]{Corresponding author}
   \ead{hanus.home@gmail.com}

   \author[label2]{M.~Delbo'}

   \author[label3]{V.~Al\'i-Lagoa}

   \author[label2]{B.~Bolin}

   \author[label4]{R.~Jedicke}

   \author[label1]{J.~\v Durech}

   \author[label1]{H.~Cibulkov\'a}

   \author[label5]{P.~Pravec}

   \author[label5]{P.~Ku\v snir\'ak}

   \author[label6]{R.~Behrend}

    \author[label7]{F.~Marchis}

   \author[label8]{P.~Antonini}
          
   \author[label9]{L.~Arnold}

   \author[label10]{M.~Audejean}

   \author[label9]{M.~Bachschmidt}

   \author[label11]{L.~Bernasconi}

   \author[label12]{L.~Brunetto}

   \author[label13]{S.~Casulli}

   \author[label14]{R.~Dymock}

   \author[label15]{N.~Esseiva}

   \author[label16]{M.~Esteban}

   \author[label9]{O.~Gerteis}

   \author[label17]{H.~de~Groot}

   \author[label9]{H.~Gully}

   \author[label18]{H.~Hamanowa}

   \author[label18]{H.~Hamanowa}

   \author[label9]{P.~Krafft}

   \author[label1]{M.~Lehk\'y}

   \author[label19]{F.~Manzini}

   \author[label20]{J.~Michelet}

   \author[label21]{E.~Morelle}
   
   \author[label22]{J.~Oey}

   \author[label23]{F.~Pilcher}

   \author[label24]{F.~Reignier}

   \author[label25]{R.~Roy}

   \author[label16]{P.A.~Salom}

   \author[label26]{B.D.~Warner}

   \address[label1]{Astronomical Institute, Faculty of Mathematics and Physics, Charles University, V~Hole{\v s}ovi{\v c}k{\'a}ch 2, 18000 Prague, Czech Republic}

   \address[label2]{Universit\' e C\^ ote d'Azur, OCA, CNRS, Lagrange, France}

   \address[label3]{Max-Planck-Institut f\"ur extraterrestrische Physik, Giessenbachstra{\ss}e, Postfach 1312, 85741 Garching, Germany}

   \address[label4]{Institute for Astronomy, University of Hawaii at Manoa, Honolulu, HI 96822, USA} 
 
   \address[label5]{Astronomical Institute, Academy of Sciences of the Czech Republic, Fri\v cova 1, CZ-25165 Ond\v rejov, Czech Republic} 
    
   \address[label6]{Geneva Observatory, CH-1290 Sauverny, Switzerland} 

   \address[label7]{SETI Institute, Carl Sagan Center, 189 Bernado Avenue, Mountain View CA 94043, USA}

   \address[label8]{Observatoire des Hauts Patys, F-84410 B\'edoin, France} 

   \address[label9]{Aix Marseille Universit\'e, CNRS, OHP (Observatoire de Haute Provence), Institut Pyth\'eas (UMS 3470) 04870 Saint-Michel-l'Observatoire, France} 

   \address[label10]{Observatoire de Chinon, Mairie de Chinon, 37500 Chinon, France} 

   \address[label11]{Observatoire des Engarouines, 1606 chemin de Rigoy, F-84570 Malemort-du-Comtat, France} 

   \address[label12]{Le Florian, Villa 4, 880 chemin de Ribac-Estagnol, F-06600 Antibes, France} 

   \address[label13]{Vallemare di Bordona, Rieti, Italy} 

   \address[label14]{Waterlooville} 

   \address[label15]{Observatoire St-Martin, 31 grande rue, F-25330 Amathay V\'esigneux, France} 

   \address[label16]{Observatorio CEAM, Caimari, Canary Islands, Spain} 

   \address[label17]{Nijmegen, Netherlands} 

   \address[label18]{Hong Kong Space Museum, Tsimshatsui, Hong Kong, China} 

   \address[label19]{Stazione Astronomica di Sozzago, I-28060 Sozzago, Italy} 

   \address[label20]{Club d'Astronomie Lyon Amp\' ere, 37 rue Paul Cazeneuve, 69008 Lyon, France} 

   \address[label21]{20 parc des Pervenches, F-13012 Marseille, France} 

   \address[label22]{Kingsgrove, NSW, Australia} 

   \address[label23]{4438 Organ Mesa Loop, Las Cruces, NM 88011, USA} 

   \address[label24]{11 rue Fran\c{c}ois-Nouteau, F-49650 Brain-sur-Allonnes, France} 

   \address[label25]{Observatoire de Blauvac, 293 chemin de St Guillaume, F-84570 Blauvac, France} 

   \address[label26]{Center for Solar System Studies, 446 Sycamore Ave, Eaton, CO 80615 USA} 

\begin{abstract} 
Eos family was created during a catastrophic impact about 1.3 Gyr ago. Rotation states of individual family members contain information about the history of the whole population. We aim to increase the number of asteroid shape models and rotation states within the Eos collision family, as well as to revise previously published shape models from the literature. Such results can be used to constrain theoretical collisional and evolution models of the family, or to estimate other physical parameters by a thermophysical modeling of the thermal infrared data. We use all available disk-integrated optical data (i.e., classical dense-in-time photometry obtained from public databases and through a large collaboration network as well as sparse-in-time individual measurements from a few sky surveys) as input for the convex inversion method, and derive 3D shape models of asteroids together with their rotation periods and orientations of rotation axes. We present updated shape models for 15 asteroids and new shape model determinations for 16 asteroids. Together with the already published models from the publicly available DAMIT database, we compiled a sample of 56 Eos family members with known shape models that we used in our analysis of physical properties within the family. Rotation states of asteroids smaller than $\sim$20 km are heavily influenced by the YORP effect, whilst the large objects more or less retained their rotation state properties since the family creation. Moreover, we also present a shape model and bulk density of asteroid (423)~Diotima, an interloper in the Eos family, based on the disk-resolved data obtained by the Near InfraRed Camera (Nirc2) mounted on the W.M. Keck II telescope.
\end{abstract}


\end{frontmatter}


\section{Introduction}\label{introduction}

Asteroid families are groups of asteroids commonly identified in the space of proper orbital elements (proper semi-major axis $a_\mathrm{p}$, eccentricity $e_\mathrm{p}$, and inclination $i_\mathrm{p}$), by the Hierarchical Clustering Method \citep[HCM,][]{Zappala1990,Zappala1994}. This method computes distances between asteroids in ($a_\mathrm{p}$,$e_\mathrm{p}$,$i_\mathrm{p}$) space, and link asteroids to a family when these are separated by less than a selected threshold value of said distance. This threshold, measured in m/s is known as cut-off velocity and is determined by ensuring that the group of the member asteroids has a statistical significance compared to the local background. Some of these families are clearly visible in plots of asteroids in orbital elements space, e.g., ($a_\mathrm{p}$,$e_\mathrm{p}$) and/or ($a_\mathrm{p}$,$i_\mathrm{p}$). 

It is well understood that members of asteroid families are collisional fragments of a parent body that was impacted by another asteroid. The study of asteroid families is thus very important as genetic links can be established between the members of a family and the family parent. In addition, the epochs of the formation of the different families can be estimated, which informs us about the collisional evolution of the Main Belt. Models of the collisional evolution of the solar system predicts that the number of families per unit time, should be roughly constant throughout the age of the solar system, but this is not what is observed \citep[e.g.,][]{Broz2013b, Spoto2015}. The family age can be derived by the orbital dispersion of its members that remove objects from the family center with rate of change of $a_\mathrm{p}$ (d$a$/d$t$) inversely proportional to the asteroid diameter $D$ \citep{Vokrouhlicky2006}, under the effect of a non gravitational force known as the Yarkovsky effect \citep[e.g.,][]{Bottke2006, Vokrouhlicky2015}.  

As a result, the family fragments form along a V-shaped border in ($a_\mathrm{p}$, 1/$D$) space because of the inverse dependence on diameter $D$ of the Yarkovsky drift rate d$a$/d$t$:
\begin{equation}
 \frac{\mathrm{d}a}{\mathrm{d}t} \; = \; \frac{\mathrm{d}a}{\mathrm{d}t_{1\mathrm{km}}} \frac{1}{D} \cos(\epsilon), 
\end{equation}
where $\epsilon$ is the obliquity. The vertex of the V-shape is the center of the family, and the sides of the V-shape are represented by those asteroids moving with the fastest d$a$/d$t$, i.e. with cos($\epsilon$) = 1 and = --1 for the outward and the inward side of the V-shape, respectively (see Fig.~\ref{fig:pv}). Dispersion in $e$ and $i$ is caused by orbital resonances with the planets that the asteroids cross as they drift through regions of the Main Belt.

Not all of the asteroids linked to a family by the HCM method are real members. A fraction of these objects has values of the orbital elements similar to those of the true asteroid family members only by a coincidence: these objects are the so-called interlopers. Interlopers can often be identified from their position with respect to the centre and the sides of the V. 

Another method to identify interlopers \citep{Nesvorny2015} is by comparison of their physical properties with those of the bulk of the family: it is reasonable to assume that family members share similar physical properties, such as homogeneity of the albedo values and the reflectance spectra. For instance, albedo information allowed the decoupling of the Nysa-Polana complex that consists of low- and high-albedo objects indistinguishable by the HCM method \citep{Walsh2013}. \citet{Masiero2013} provide lists of members for a large number of families where both the distance in the proper elements as well as the albedo values are considered. Recently, the WISE satellite \citep{Wright2010} thermal infrared data lead to albedo determinations for more than one hundred thousand asteroids \citep{Masiero2011}. 

Despite the progress made in the last few years, important questions remain to be answered, such as the ages of the families and why they are not uniformly distributed along the history of the solar system. A crucial parameter to determine family ages is the value of d$a$/d$t$, i.e., the strength of the Yarkovsky effect for families of different compositional types. The Yarkovsky effect depends on several physical properties including, size, thermal inertia, rotation period and direction of the spin axis. The last property, is known only for some members of different families, although their number is steadily increasing thanks to specific studies \citep{Slivan2002, Slivan2003, Slivan2009, Torppa2003, Durech2009, Kryszczynska2013a, Hanus2013c, Kim2014a}. 

Here we focus on the Eos asteroid family. Its largest member (221)~Eos is a rare K-type. The family was created by a catastrophic collision about (1.3$\pm$0.2) Gyr ago \citep{Nesvorny2015}, \citet{Broz2013a} provide an age between 1.5 and 1.9 Gyr. The size frequency distribution of the family members suggests that the parent body was fully destroyed and the largest remnant contains only few percent of the mass of the parent body \citep{Broz2013b}. The anhydrous CO, CV, and CK, and hydrated metal-rich CR carbonaceous chondrites meteorites were proposed as meteorites analogs for Eos objects due to the similarity in the visible and near- and mid-infrared spectra \citep{Bell1989, Doressoundiram1998, Clark2009}. Unfortunately, there are only poor observations in the (mid)-infrared that could not shed more light into the mineralogy of the surface and confirm any of the analogs. Also, there are no reliable bulk density estimates, so it is not clear if the Eos K-type objects are consistent with the other K-types in the main belt. The Eos family has a complex structure, with several mean motion and secular resonances transversing it \citep[see, e.g.,][]{Broz2013a}. 

The Eos family provides an opportunity to study the Yarkovsky drift rate of asteroids belonging to a common collisional origin, with the ultimate aim to compare the d$a$/d$t$ from Yarkovsky models to the value determined from the resonance crossing. The list of \citet{Masiero2013} contains 5718 objects, whilst the list of \citet{Nesvorny2015} consists of 9789 asteroids. Although the list of Eos members from \citet{Nesvorny2015} is still based only on the HCM method, the authors also provide a parameter that reflects the distance from the V-shape envelope of the family that corresponds to the extent due to the Yarkovsky drift. A large value of this parameter indicates an interloper. We cross-checked Nesvorn\'y's catalog with a previous version \citet{Nesvorny2012} and added additional asteroids to the list of members of the Eos family. This was done to obtain the largest sample of possible Eos family members for our study. Moreover, we also considered physical properties of selected asteroids to revise their Eos membership.

Here we study shapes and spin states of Eos family members. Lightcurve inversion methods such as the convex inversion of  \citet{Kaasalainen2001a} and \citet{Kaasalainen2001b} are now routinely applied to the optical disk-integrated data, which resulted in the determination of rotation states and shape models for about one thousand asteroids \citep{Kaasalainen2002c, Slivan2003, Torppa2003, Michalowski2005, Michalowski2006, Durech2009, Durech2016, Marciniak2009a, Marciniak2009b, Hanus2011, Hanus2013a, Hanus2016a, Kryszczynska2013a}. Such a number of physical properties revealed unambiguously the influence of the YORP effect on the spin axis directions of small main-belt asteroids \citep[$D < 30$ km,][]{Vokrouhlicky2003, Hanus2011} or that retrograde rotators drift inwards to shorter semi-major axes (left in the 1/$D$ vs. $a$ plot) with respect to the family center and prograde ones outwards (to the right of the plot) \citep{Hanus2013c}. The number of shape models in families is growing. These physical properties should tell us more about the histories of individual families \citep{Vokrouhlicky2003, Hanus2013c, Dykhuis2016}. We aim to increase the number of Eos members with shape models to $\sim$50--60, which will become the largest sample for any family so far. Physical properties can be used to constrain the dynamical models by studying the evolution of the families due to collisions, Yarkovsky and YORP effects \citep[e.g.,][]{Hanus2013c, Hanus2013a}. Moreover, the convex shape models are necessary inputs for the thermophysical modeling \citep[e.g.,][]{Delbo2007a, Muller2011, Muller2014, Rozitis2014, Hanus2015a} or a more detailed shape modeling when stellar occultations or disk-resolved data are available \citep{Durech2011, Viikinkoski2015, Hanus2017a}.

Recently, \citet{Cibulkova2016} estimated ecliptic longitudes and absolute values of ecliptic latitudes |$\beta$| of the spin axes for almost 70\,000 asteroids including more than 2\,500 asteroids from the Eos family. Their model applied to sparse-in-time data from Lowell database by comparing the observed values of mean brightness in each apparition and its dispersion with values computed from the model parameters. The shape is modeled as a triaxial ellipsoid. The estimates for individual asteroids are not fully reliable and the data should be rather treated in a statistical sense.

Asteroid shape models derived by the light curve inversion method and their optical data are usually uploaded to the public Database of Asteroid Models from Inversion Techniques \citep[DAMIT\footnote{\url{http://astro.troja.mff.cuni.cz/projects/asteroids3D}},][]{Durech2010}.

In Sect.~\ref{sec:photometry}, we describe the optical dense- and sparse-in-time disk-integrated data that were used to construct the shape models. We detail the convex inversion method for shape modeling in Sect.~\ref{sec:inversion} and derive updated and new shape models in Sect.~\ref{sec:models}. Rotation state parameters such as the spin axis direction and rotation period are then studied in Sect.~\ref{sec:states}. Finally, we conclude our work in Sect.~\ref{sec:conclusions}.

\section{Optical disk-integrated photometry}\label{sec:photometry}

In this work, we used light curve data from the literature, as well as our own observations. New photometric data were essential, because they allowed us to significantly increase the number of Eos family asteroids with derived shape models.

Besides the dense-in-time disk-integrated photometry, i.e., the classical light curves, we also used the sparse-in-time photometry that consists of a few hundred individual calibrated measurements from several astrometric observatories, typically covering $\sim$15 years. The sparse-in-time data are necessary prerequisites for a successful shape model determination when only few dense light curves are available \citep{Hanus2011,Hanus2016a}. Concerning the sparse data, we used those downloaded from the AstDyS site (Asteroids -- Dynamic Site\footnote{\url{http://hamilton.dm.unipi.it/}}), or from the Lowell Photometric Database \citep{Oszkiewicz2011, Bowell2014a}. All data were processed following the procedure of \citet{Hanus2011, Hanus2013a}.  

The dense-in-time data were acquired from public databases such as the Asteroid Photometric Catalogue \citep[APC\footnote{\url{http://asteroid.astro.helsinki.fi/}}; ][]{Piironen2001} or the Asteroid Lightcurve Data Exchange Format database \citep[ALCDEF\footnote{\url{http://alcdef.org}};][]{Warner2011d}. Many light curves from the European observers are stored in the Courbes de rotation d'ast\' ero\" ides et de com\` etes database (CdR\footnote{\url{http://obswww.unige.ch/\textasciitilde behrend/page\_cou.html}}), maintained by Raoul Behrend at Observatoire de Gen\` eve. He kindly shared the photometry of Eos members with us.

Moreover, we also made use of the thermal infrared data of Eos members acquired by the WISE satellite, in particular the results of the NEOWISE project dealing with the solar system bodies \citep[see, e.g., ][]{Mainzer2011a}. These data were downloaded from the WISE All-Sky Single Exposure L1b Working Database via the IRSA/IPAC archive \footnote{\url{http://irsa.ipac.caltech.edu/Missions/wise.html}}. We employed thermal infrared data from all four filters W1, W2, W3 and W4 (isophotal wavelengths at 3.35, 4.6, 12 and 22 $\mu$m) from the fully cryogenic phase of the mission. The data selection and suitability criteria applied in this work followed those of \citet{AliLagoa2014, Hanus2015a, AliLagoa2016a}. 

For Eos asteroids, the fluxes in filters W1 and W2 usually consist of a significant reflected sunlight component, which allows us to use these fluxes for the shape modeling in combination with the dense and sparse optical photometry. The fluxes in W3 and W4 filters are thermal-emission dominated, however, \citet{Durech2015b} have shown that these data can be treated as reflected if we search for the period as well as for the shape. The reason is that we fit only the relative shape of thermal light curves, not the absolute fluxes. The phase shift between thermal and optical data is usually negligible and thermal light curves are very useful when searching for the sidereal rotation period. The illuminated parts of the surface are the main contributors to the thermal infrared emission as those are the warmest. The contribution from the dark side is negligible, because it is significantly colder and rather small for the typical phase angles of 10--30$^{\circ}$.

In addition, we conducted a one-year observing campaign with the University of Hawaii 2.2-meter telescope (UH88) located near the summit of Maunakea in Hawaii between August 2015 and July 2016. We used the Tektronix 2048x2048 CCD camera, which has a ${7.5}'\mathrm{x}{7.5}'$ field of view corresponding to a pixel scale of ${0.22}''$. The images obtained were Nyquist-sampled corresponding to the typical seeing of $\sim {0.8}'$ during the observations and to reduce the readout time. During each observing night, we acquired biases and evening or morning flats. We reduced the data with our custom-made aperture photometry software Aphot+Redlink developed by Petr Pravec and Miroslav Velen.

Exposure times at UH88 were chosen for each target individually depending on its brightness, so a desired signal-to-noise ratio (usually $\gtrsim100$) could be achieved. We observed in the Sloan ${r}'$ filter. The typical exposure times of 30--180 s allowed us to use the sidereal tracking without significantly corrupting the PSFs of the asteroid. During the observations, we were slewing and looping between 3--5 targets and always taking three exposures during each asteroid visit. As a result, the light curves taken with the UH88 are semi-dense with 15-25 minute gaps between observations. This strategy proved to be efficient, because still, we obtained sufficiently well-sampled light curves due to large enough periods of our targets ($\gtrsim$3~h). Typically, we obtained 5--7 light curves per night. In total, we had 15 clear observing nights, during which we obtained 85 light curves. All data were light-time corrected and treated as relative. Calibrated photometry to absolute scale is not required for the shape modeling. The only non-Eos family asteroid on our UH88 target list was the near-Earth asteroid (3200)~Phaethon, whose data were already used for Phaethon's shape model determination in \citet{Hanus2016b}. In our observing campaign, we prioritized asteroids according to their previous shape modeling attempts: we tried to observe asteroids on the brick of their shape model determination. For example, we observed asteroids, for which more than 2 pole solutions were already derived or whose rotation period was likely to be derived with only few additional light curves. We also targeted asteroids, whose shape models were derived from only Lowell data in \citet{Durech2016}. We aimed to confirm and improve those shape models. Our observing strategy was designed to achieve the highest efficiency in using the allocated telescope time. Similar observing strategy was followed with the BlueEye 600 robotic telescope located in Ond\v rejov, Czech Republic \citep{Durech2017a} with the only exception that the whole asteroid population was targeted.

The summary of the optical data used for the shape model determination, such as the number of dense-in-time light curves and apparitions covered by dense-in-time observations and the total number of sparse-in-time measurements can be found in Table~\ref{tab:models}. Light curve summary and observers are then listed in Table~\ref{tab:references}. We uploaded all previously unpublished light curves to the ALCDEF database.

\section{Convex inversion}\label{sec:inversion}


We apply the lightcurve inversion method of \citet{Kaasalainen2001a} and \citet{Kaasalainen2001b} to derive asteroid rotation states and shape models from their optical disk-integrated data. This technique already led to successful shape model determinations for hundreds of asteroids \citep[see the Asteroids IV chapter of][]{DurechAIV2015}. We employ the convex inversion method, because convex models are usually the only stable or unambiguous inversion result when only disk-integrated data are available \citep{Durech2003}.

We search the parameter space of shape, rotation state, and scattering properties for the best-fitting values in the means of a $\chi^2$-metric

\begin{equation}\label{eq:chi2_rel}
\chi^2=\sum_i \frac{|| L_{\mathrm{OBS}}^{(i)} - L_{\mathrm{MOD}}^{(i)} ||}{\sigma^2_i},
\end{equation}
where the $i$-th brightness measurement $L_{\mathrm{OBS}}^{(i)}$ (with an uncertainty of $\sigma_i$) is compared to the corresponding modeled brightness $L_{\mathrm{MOD}}^{(i)}$.

Ideally, one prominent minimum in the parameter space is found. Several reliability tests are performed on the best-fitting solution prior accepting it -- visual examination of the fit in the rotation period subspace and of the light curves, computation of the principal moments of inertia of the shape model (i.e., we check if the asteroid rotates around its axis with the maximum moment of inertia), and comparison of the derived rotation period with previous independent estimates.

It was already shown and discussed in \citet{Hanus2015a} and \citet{Hanus2016a} that each shape model comes with its uncertainty that is not fully quantified. This uncertainty depends on the amount, variety, and quality of the optical data used for the model determination, and should be considered in further shape model applications. To provide the information about the precision of the derived shape model, we introduce the {\em quality flag ($QF$)} value that is assigned to each shape model solution. These values are now listed for each solution presented in this work, as well as for shape models available in the DAMIT database. Specifically, a shape model based on a large photometric data set and disk-resolved data and/or stellar occultations should well correspond to the real shape of the asteroid, so such a solution has the highest value of the quality flag of $QF=4$. Theoretically, the ground-truth models based on space probe flybys and satellites orbits should have $QF=5$, however, such shape models are not available in the DAMIT database. Next, the value of $QF=3$ is typically assigned to shape models based on large photometric data sets, which usually cover at least three apparitions and consist of tens of single light curves. When only several dense light curves (usually from one or two apparitions) combined with the sparse data provide a shape model, its quality value is set to $QF=2$. Finally, coarse shape models based solely on sparse data have $QF=1$. Intermediate values are used as well. Shape models with $QF<=2$ should be used in further studies with caution. Unfortunately, the variety of available photometry for asteroids is huge, so assigning quality flags cannot be fully generalized and it serves mostly as a tentative indicator of the shape model quality. The quality of the shape model also depends, for example, on the geometry of light curve observations, the light curve amplitude, the aspect angle, or the precision and sampling of the photometry.

We already applied and described in great details the convex inversion method and its application to combined dense and sparse integrated data in our previous studies, so additional information can be found there \citep[e.g.,][]{Hanus2011, Hanus2013c, Hanus2013a, Hanus2016a, Durech2016}. We emphasize the reproducibility of our work -- all shape models as well as the optical light curves and the convex inversion source code are available in the DAMIT database. Moreover, we also uploaded here published light curves to the ALCDEF database.

\section{Results and discussions}\label{sec:results}

\subsection{New and updated shape models}\label{sec:models}

New optical light curves allowed us to revise shape models of 15 asteroids from the Eos family. In most cases, rotation states of updated shape models are consistent within their expected uncertainties with those in the DAMIT database (see Table~\ref{tab:models} for the comparison). Only the second pole solution of asteroid (742)~Edisona is slightly different from the previously published value, however, the revised solution seems to be now more consistent with its mirror solution. Additionally, the revised solution of asteroid (1207)~Ostenia has similar ecliptic latitude of the pole orientation but differs in the ecliptic longitude from the previous solution of \citet{Hanus2011}. The larger photometric data set was sufficient to remove the pole ambiguity of asteroid (573)~Recha. Typically, by revising the shape models, we decreased the uncertainties in rotation states and improved the accuracy of the shape model. All revised shape models replaced their previous versions in the DAMIT database.

We derived new shape models for 11 asteroids. All these shape model determinations are based on combined dense-in-time data and sparse-in-time measurements. Rotation state parameters and information about optical data for asteroids from the Eos family with known shape models are listed in Table~\ref{tab:models}. References to the optical light curves can be found in Table~\ref{tab:references}.

The combined WISE and Lowell optical data sets allowed us to determine 5 additional shape models. We checked that the pole/shape solutions were not sensitive to including thermal data. When repeating the inversion without any WISE data (using the same sidereal rotation period), the optimization converged to pole directions that were not far (within 10 deg) from those that were obtained when using WISE data. Thermal light curves are important for determining the unique rotation period, but for a given period, the models with and without WISE data are similar.

Finally, we compiled a data set of 58 potential Eos family members with known shape models (Table~\ref{tab:models}) and used it for an analysis of physical properties within the family in Sect.~\ref{sec:states}. The uncertainties of rotation state parameters were estimated based on the quality of each photometric dataset.

\subsection{Rotation states in the Eos family}\label{sec:states}

\begin{figure*}
\begin{center}
\resizebox{\hsize}{!}{\includegraphics{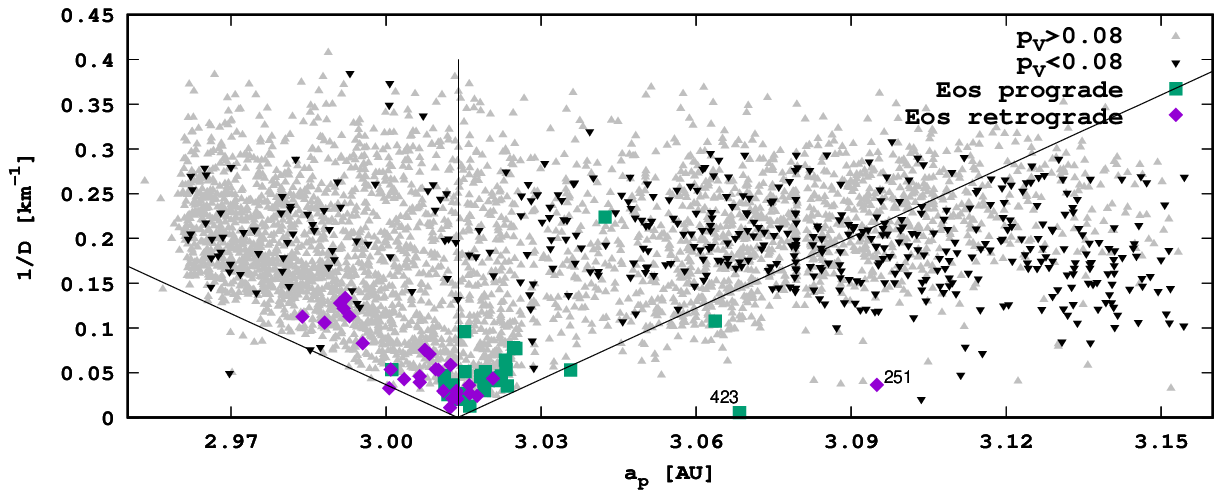}}\\
\resizebox{\hsize}{!}{\includegraphics{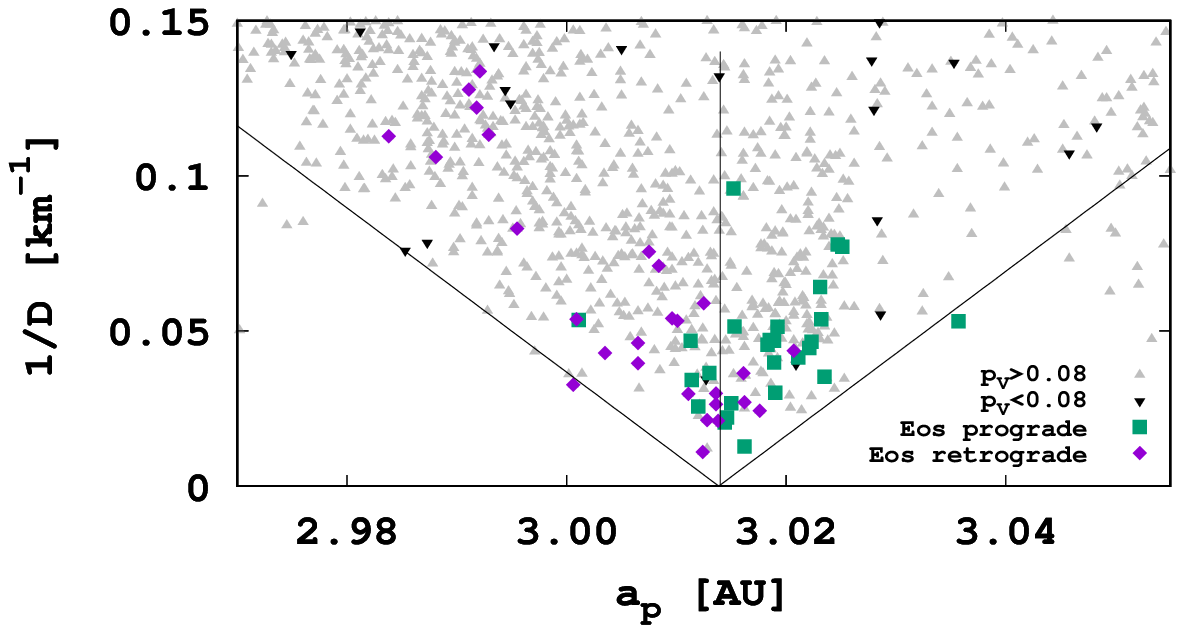}\includegraphics{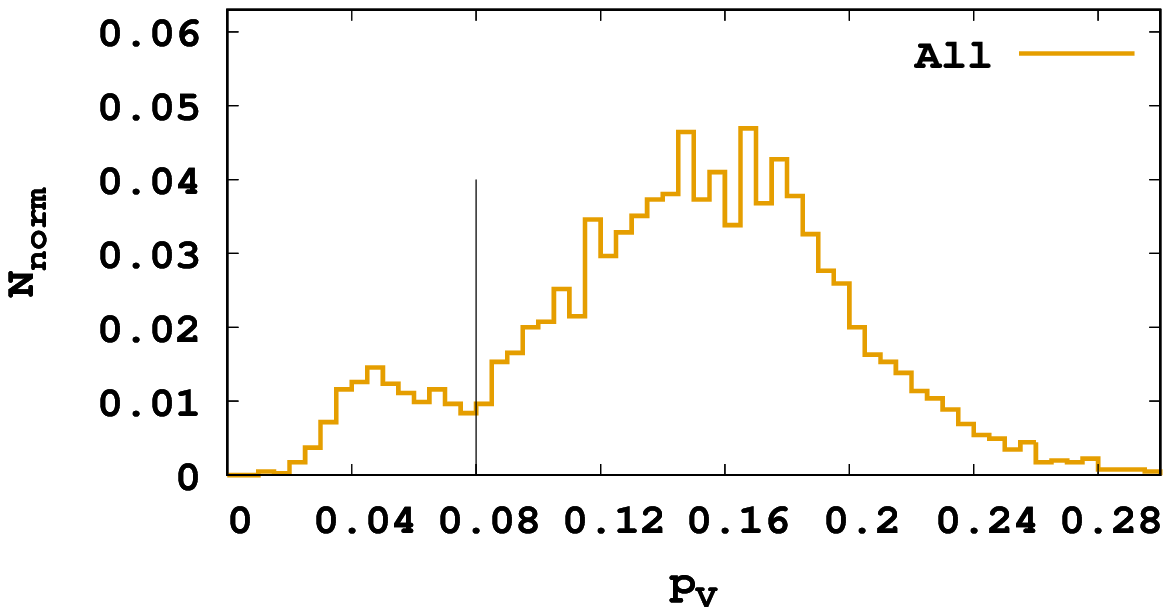}}\\
\end{center}
\caption{\label{fig:pv}Top: Dependence of the inverse size $1/D$ on the proper semi-major axis $a_{\mathrm{p}}$ for the Eos collisional family with the estimated position of the family center (vertical line). The V-shape envelope (two lines) reflects the initial extent of the family and different age and magnitude of the Yarkovsky semi-major axis drift. We also show the positions of the asteroids with known shape models and their sense of rotation along their spin axes. Moreover, we also distinguish objects with geometric visible albedo $p_\mathrm{V}>0.08$ from those with $p_\mathrm{V}<0.08$. Albedo and diameter values are taken from \citet{Masiero2011}. Bottom left: Zoom to the center part of the family, otherwise same as top panel. Bottom right: Distribution of the geometric visible albedo $p_\mathrm{V}$ in the Eos family. The vertical line denotes the border between the bi-modal distribution at $\sim$0.08.}
\end{figure*}

\begin{figure*}
\begin{center}
\resizebox{\hsize}{!}{\includegraphics{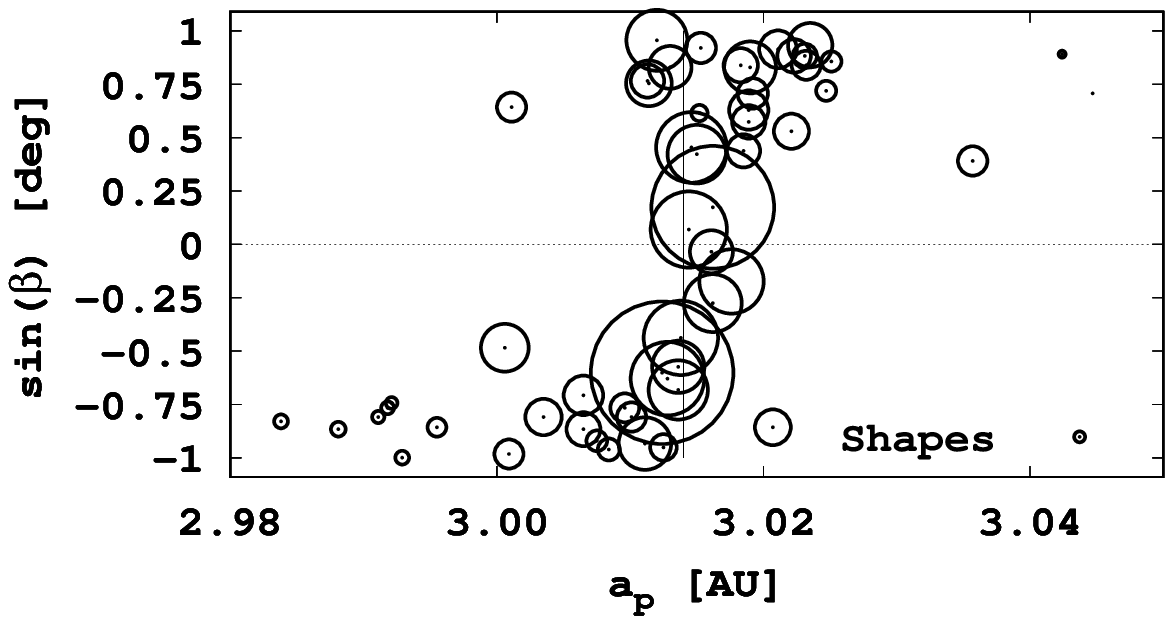}\includegraphics{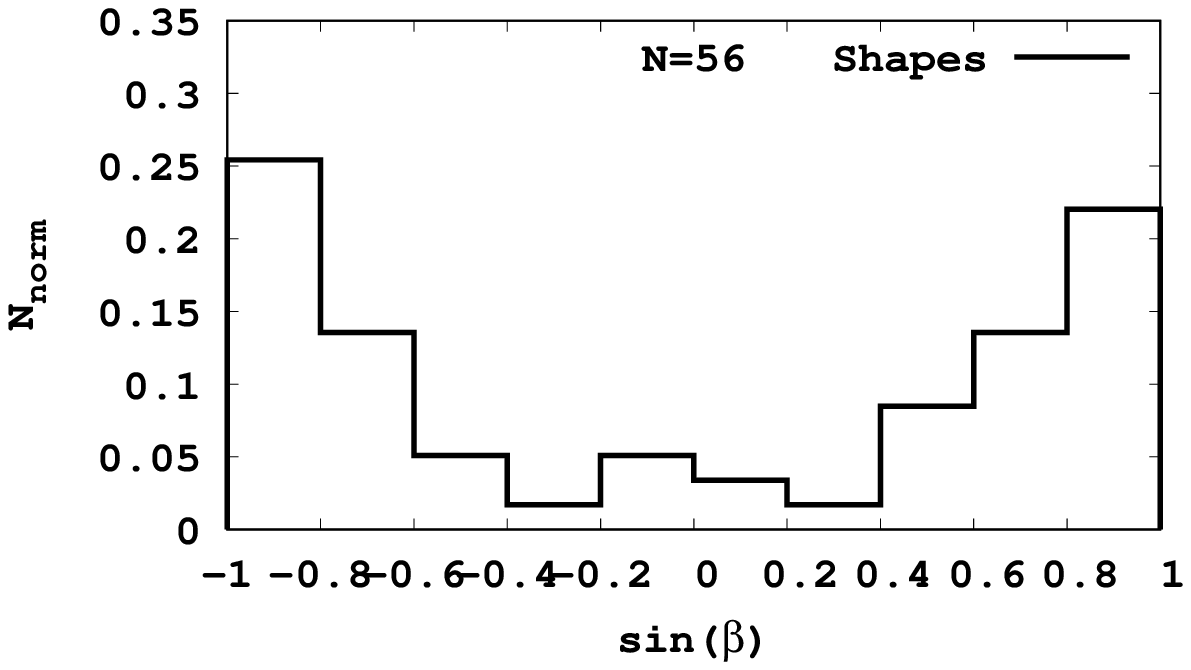}}\\
\resizebox{\hsize}{!}{\includegraphics{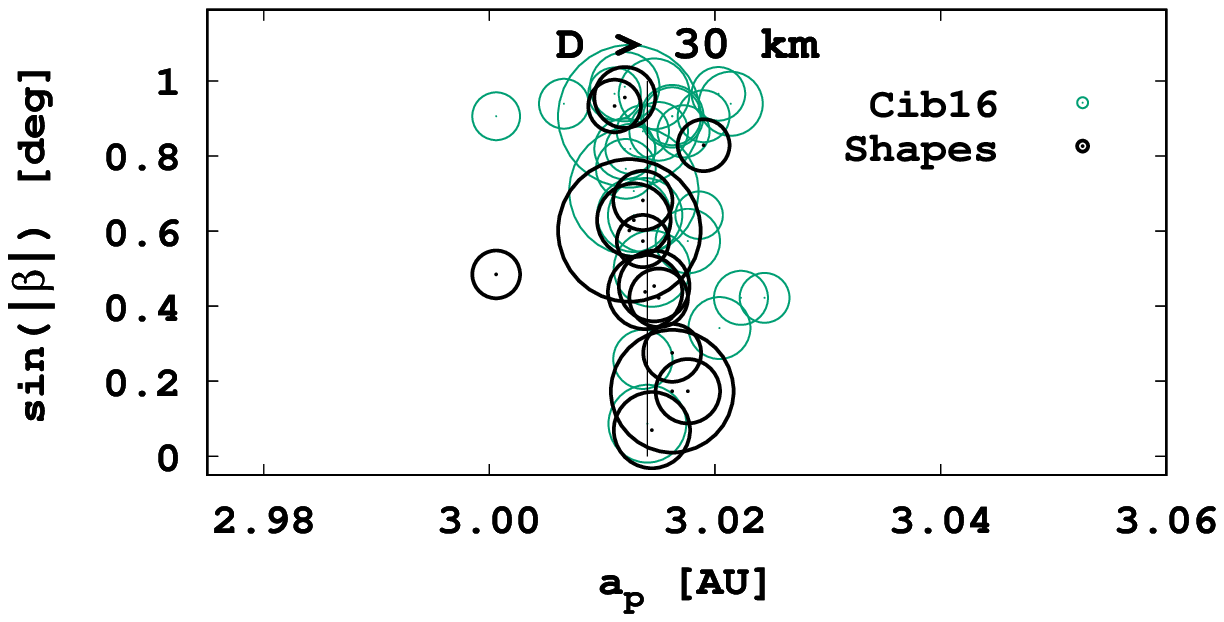}\includegraphics{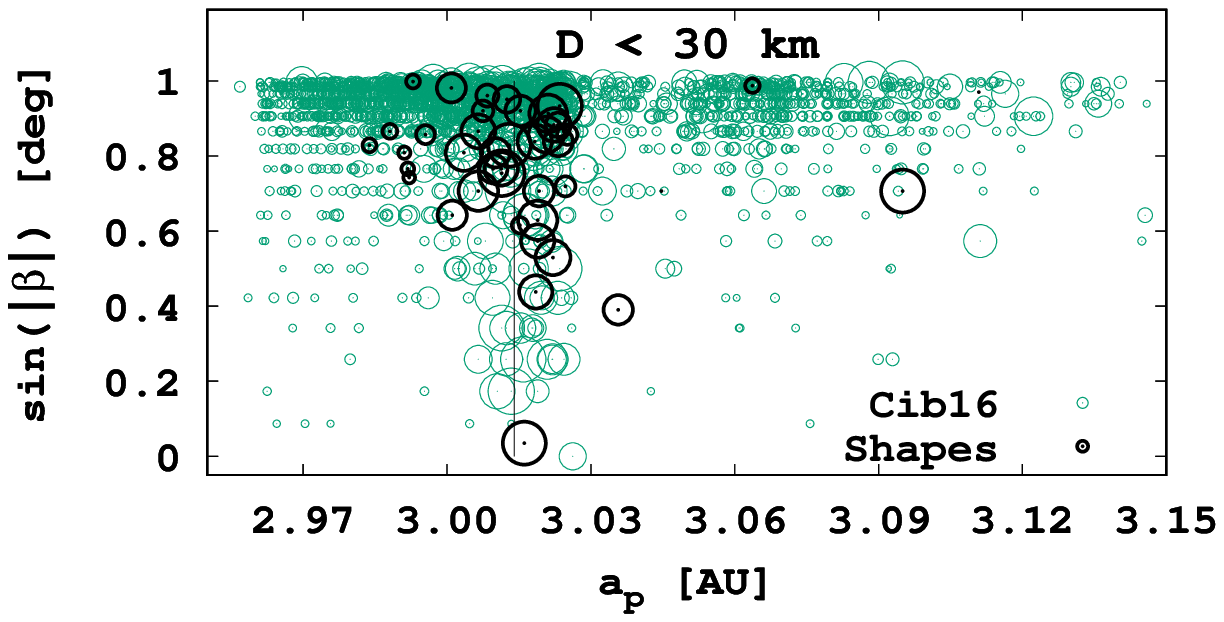}}\\
\resizebox{\hsize}{!}{\includegraphics{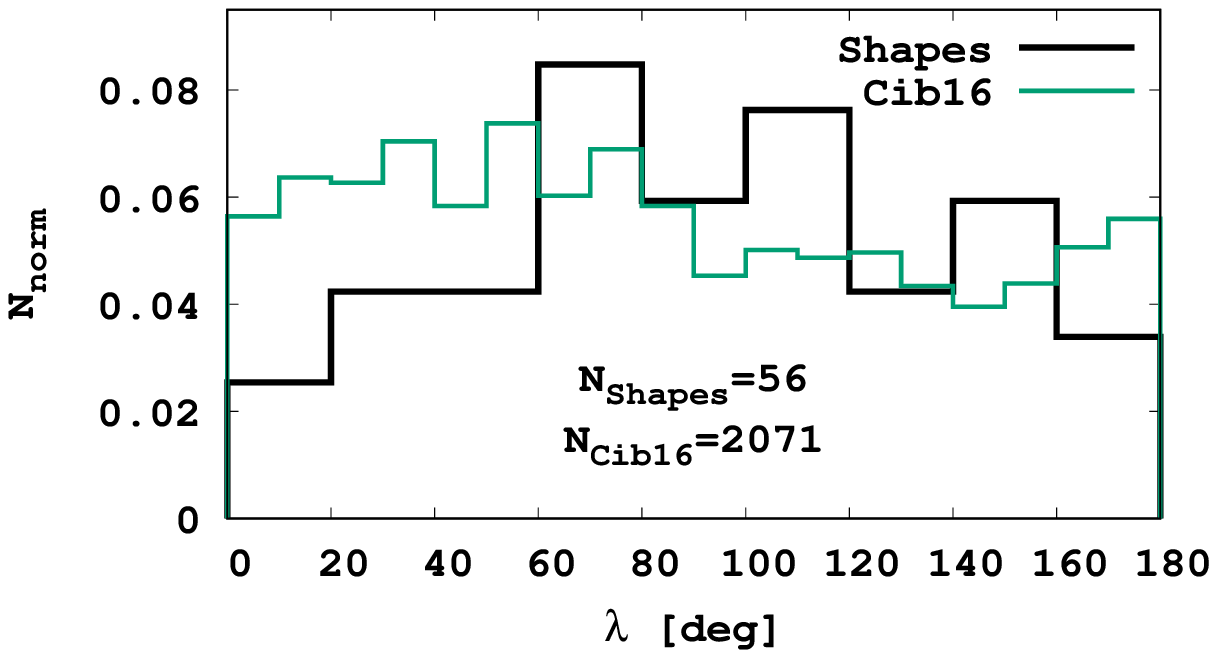}\includegraphics{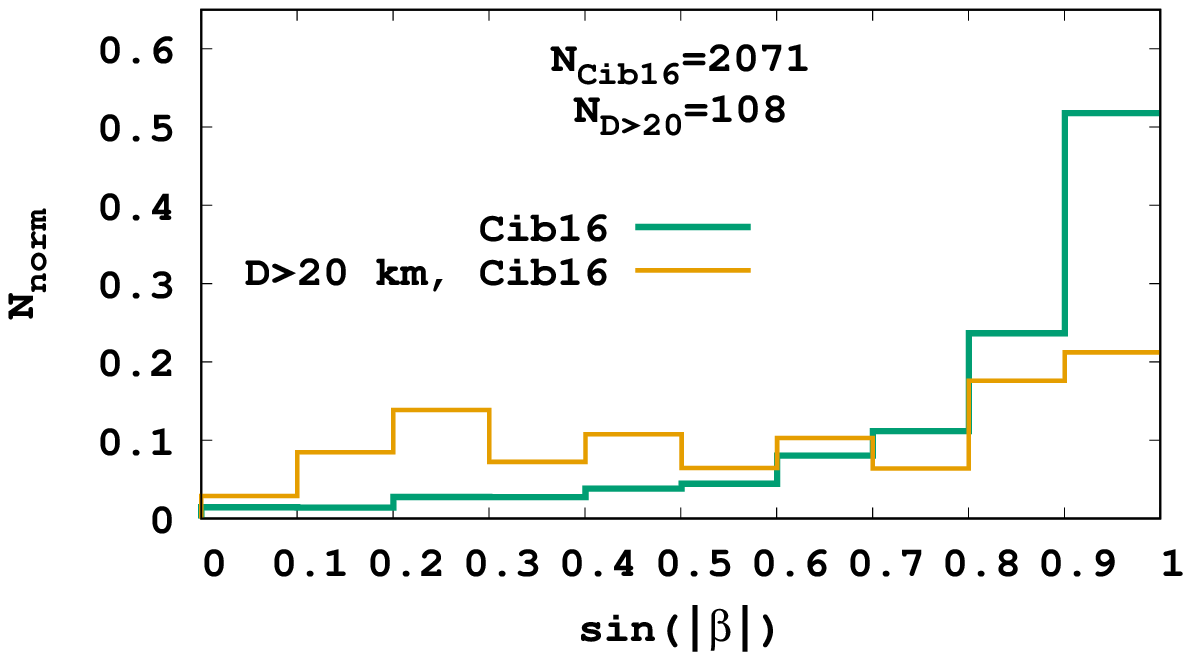}}\\
\end{center}
\caption{\label{fig:states}Top left: Dependence of the pole latitude $\beta$ on the proper semi-major axis $a_{\mathrm{p}}$ for the Eos family asteroids with known shape models. The vertical line corresponds to the estimated center of the asteroid family. The size of the circles is relative to the size of the asteroids. The uncertainties in $\beta$ are usually 5--20$^{\circ}$. Top right: Observed ecliptic pole-latitude distribution in the Eos family (asteroids with shape models). Middle panels: Dependence of |sin($\beta$)| on the proper semi-major axis $a_{\mathrm{p}}$ for the Eos family asteroids with sizes larger and smaller than 30 km, respectively, from \citet{Cibulkova2016} (indicated as Cib16 in the plots). The family center is indicated by a vertical line. Bottom left: Distribution of ecliptic longitudes $\lambda$ for all Eos family objects estimated in \citet{Cibulkova2016} and for asteroids with shape models. Bottom right: De-biased distribution of ecliptic latitudes |$\beta$| for Eos members taken from \citet{Cibulkova2016}. Estimates for latitudes, taken from \citet{Cibulkova2016}, appear as absolute values, so they do not distinguish between prograde and retrograde objects.}
\end{figure*}

\begin{figure*}
\begin{center}
\resizebox{0.5\hsize}{!}{\includegraphics{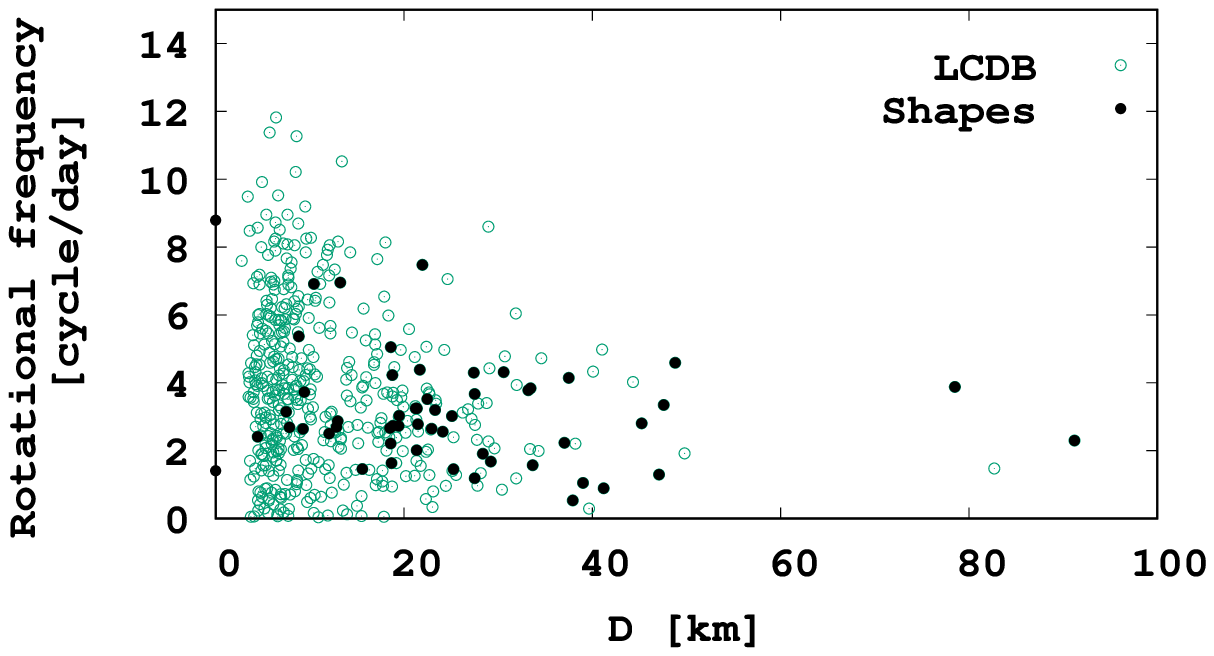}}\\
\resizebox{\hsize}{!}{\includegraphics{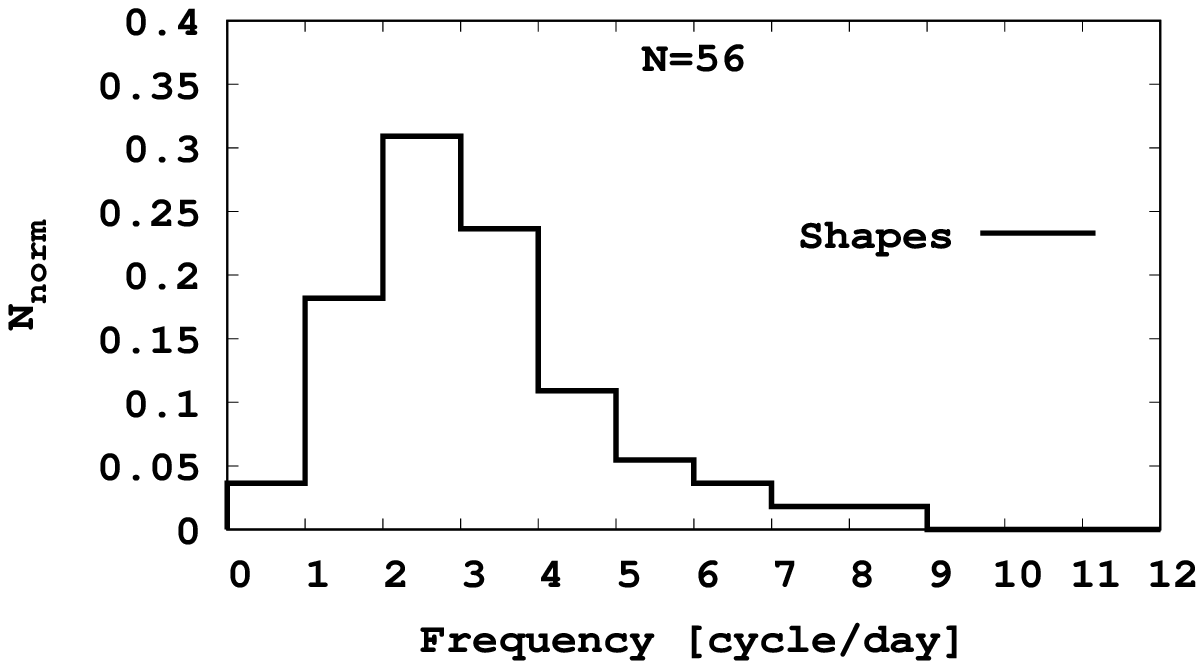}\includegraphics{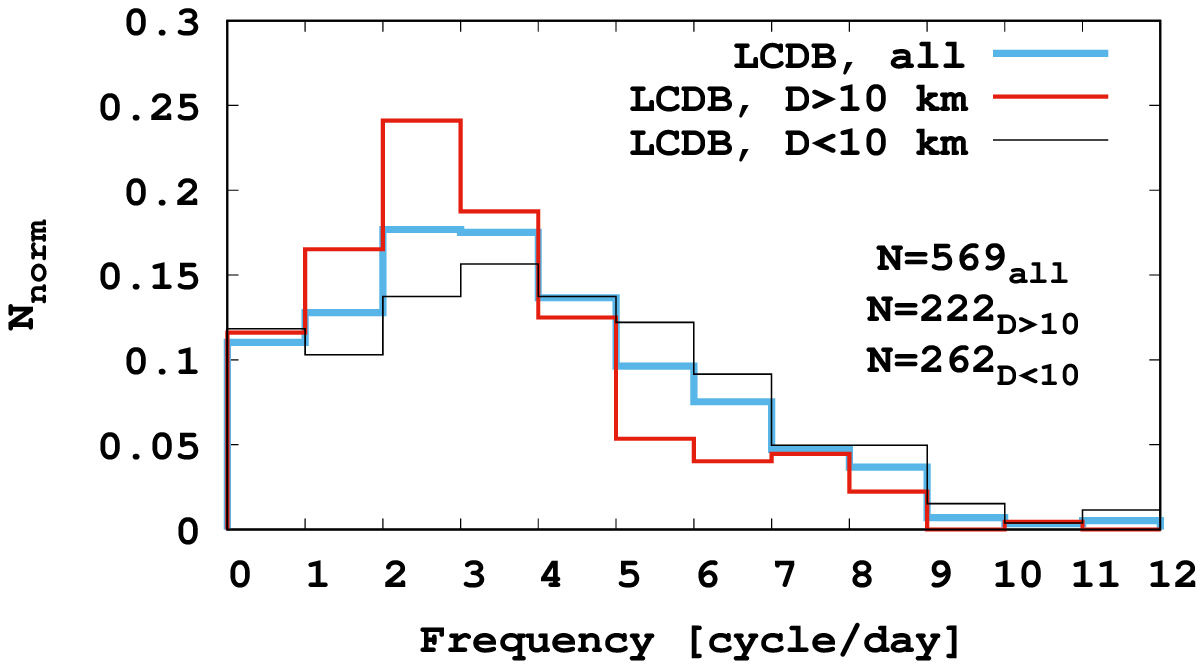}}\\
\end{center}
\caption{\label{fig:frequency}
Top: Dependence of the rotational frequency on the diameter $D$. Asteroids with shape model (filled circles) and for all asteroids with known rotation periods \citep[empty circles; LCDB database,][]{Warner2009}. Bottom left: Rotation period distributions for Eos objects with shape models. Bottom right: Rotation period distributions for the whole Eos family and for Eos asteroids smaller and larger than 10 km.}
\end{figure*}

As already discussed, we consider the Eos collisional family as defined by \citet{Nesvorny2015}. However, we have identified 5 asteroids with shape models that are classified as Eos family members in \citet{Nesvorny2012}, but not in \citet{Nesvorny2015}. We decided to use these five asteroids in our study, but we also checked if their physical parameters are consistent with those of the Eos family. 

We estimated the center of the family from the proper semi-major axis $a_{\mathrm{p}}$ -- size inversion (1/$D$) plot (top panel of Fig.~\ref{fig:pv}). In this representation, the family should fill a V-shape envelope that corresponds to the objects dispersion in the $a_{\mathrm{p}}$ caused by the initial velocity field and the Yarkovsky drift \citep[see, e.g.,][]{Vokrouhlicky2006d, Spoto2015}. Contrary to the commonly used proper semi-major axis -- absolute magnitude plot, the V-shape envelope of the $a_\mathrm{p}-1/D$ plot consists of two straight lines, in general, intersecting at the center of the family. The Eos family is affected by the 9:4 mean-motion resonance with Jupiter at 3.03 au, which breaks the left part of the V-shape. Specifically, most of the objects with $a_\mathrm{p}\gtrsim3.03$ au migrated through the resonance, and so could lie outside the nominal V-shape envelope. We also show in the top panel of Fig.~\ref{fig:pv} the positions of the asteroids with known shape models as well as their sense of rotation around their spin axes. We confirmed the previous findings of \citet{Hanus2013c} suggesting that retrograde spinners drift inwards to shorter semi-major axes with respect to the family center and the prograde ones outwards. We have several objects with known spin state that are situated close to the family center. These objects are rather large to be significantly evolved due to the YORP and by the Yarkovsky drift. The larger prograde and retrograde asteroids are mixed in the proximity of the family center, which could be a result of the initial ejection field after the catastrophic collision. 

Asteroids (251)~Sophia and (423)~Diotima are clearly outside the V-shape envelope (top panel of Fig.~\ref{fig:pv}). There is no doubt that Diotima is an interloper -- it is a C-type object with a low albedo and bulk density. We focus in more details on Diotima in the Appendix \ref{sec:app1}. Sophia is a retrograde rotator situated outward of the family center, which is in the contradiction with the theoretical expectations of the Yarkovsky drift and YORP spin vector evolution applied to a family member. Because the evidence for Sophia and Diotima being interlopers is quite strong, we removed these two asteroids from our following study of physical properties of the Eos family members. Physical properties of the remaining asteroids (60) are rather consistent with the expected ranges within the Eos family.

The top panel of Fig.~\ref{fig:pv} shows that Eos family as defined in \citet{Nesvorny2015} is contaminated by a large number of low-albedo objects. We have chosen the value of $p_\mathrm{V}=0.08$ to divide the low and high albedo objects based on the albedo distribution in the bottom right panel of Fig.~\ref{fig:pv}. The bi-modal albedo distribution is obvious. The concentration of low-albedo objects increases with increasing $a_{\mathrm{p}}$. This could suggest an existence of a low-albedo family in the Eos family region, however, there is no obvious cluster in the full proper element space of $a_{\mathrm{p}}$, $e_{\mathrm{p}}$ and $i_{\mathrm{p}}$. On the contrary, the low-albedo objects are rather spread across the whole Eos family space and could represent the background. On top of that, considering the position of the Eos family in the outer part of the main belt, a background dominated by dark objects is expected. Moreover, the dark objects are more frequent outward of the 9:4 mean-motion resonance with Jupiter. These objects could be dispersed by the resonance and could originate from the few C-type asteroid families in the proximity of the Eos family that are intersected by the resonance. Our sample of shape models represents only the largest asteroids in the family. Their albedo values are higher than 0.08 (with one exception), so are consistent with the Eos taxonomic type (the mean $p_\mathrm{V}$ value within the Eos family is $\sim$0.16). 

The observed ecliptic pole-latitude distribution in the Eos family in the top panels of Fig.~\ref{fig:states} is bi-modal, which is in agreement with previous studies of spin states in the Eos and other asteroid families \citep{Hanus2013c, Kim2014a, Kryszczynska2013a}. A K-S test rejects the null hypothesis that the latitudes are uniformly distributed ($p=0.01$). Our study is based on a significantly larger number of asteroids. The excess of prograde rotators inwards to shorter semi-major axes with respect to the family center and the excess of retrograde rotators outwards is another clear evidence for that (top right panel of Fig.~\ref{fig:states}). The mixture of larger ($\gtrsim20-30$~km) prograde and retrograde asteroids near the center of the family suggests that these objects are not significantly evolved by the YORP and Yarkovsky effects. The number of known ecliptic longitudes of the spin axes (56) is still small to reveal possible trends (bottom left panel of Fig.~\ref{fig:states}). According to the K-S test, the longitude distribution is consistent with the uniform distribution. We do not see any clustering of the spin vectors similar to the so-called Slivan states identified in the Koronis family by \citet{Slivan2002}. Our data sets of latitudes and longitudes are biased towards large asteroids, asteroids with larger light curve amplitudes and asteroids with rotaion periods $\lesssim20$ hours. The data should be de-biased before any interpretation, e.g., before their direct comparison with numerical results. The reliable de-biasing is not straightforward and it is out of the scope of this paper.

The sample of $\lambda$ and $|\beta|$ values for $\sim$2\,500 Eos family members derived by \citet{Cibulkova2016} represents about 25\% of the currently known Eos family. The method used for the estimation of these physical properties is applied to the sparse-in-time optical data from the Lowell database \citep{Bowell2014a}. The most important fact is that the method always produces a solution and that the only selection effect is the amount of the optical data. Consequently, the sample of these $\sim$2\,500 asteroids should be representative in the physical properties of the whole Eos family to some size limit of a few km (the photometric data sets for smaller asteroids are usually poor, so they are excluded from the computations). However, the bias in latitudes comes from the method itself that tends to overestimate the derived |$\beta$| values. Fortunately, \citet{Cibulkova2016} quantified the bias by a synthetic population, so we can correct for it. The method does not bias the longitudes, though. Such data set of longitudes and (de-biased) latitudes should represent the spin-vector properties of the Eos family objects rather reliably. 

In the middle panels of Fig.~\ref{fig:states}, we compare the absolute values of ecliptic latitudes in the Eos family ($a_{\mathrm{p}}$ vs. sin(|$\beta$|)) for large ($D>30$~km) and small ($D<30$~km) objects. As expected, large asteroids are clustered towards the family center and their ecliptic latitudes are not clustered significantly. On the other hand, latitudes of small asteroids are strongly anisotropic with poles almost perpendicular to the ecliptic plane. Moreover, we also show the longitude and debiased latitude distributions in bottom panels of Fig.~\ref{fig:states}. Although there are not many objects larger than 20 km in the sample, the latitude distribution still shows a small excess of larger sin(|$\beta$|) values, however, the most extreme values of sin(|$\beta$|) are less frequent. This probably means that some of the objects are already evolved by the YORP. The longitude distribution does not depend on the size \citep{Cibulkova2016}, so we only show the distribution for all objects. There are no significant differences between latitude distributions for objects smaller than 10 km and objects with sizes between 10 and 30 km. 

Contrary to theoretical expectations, the longitude distribution is not uniform as was already noted by \citet{Bowell2014a} and recently confirmed by \citet{Cibulkova2016}. The K-S test clearly shows that the longitude distribution in the Eos family is not uniform ($p\sim10^{-23}$). Although both authors invested a large effort to find a satisfactory explanation, the reason remains unknown. There is an excess of asteroids with ecliptic longitudes $\sim20-80^{\circ}$ and deficit of asteroids with $\lambda\sim120-160$. The longitude distribution for asteroids with shapes is consistent with the uniform distribution (K-S test: $p=0.80$), but also with the non-uniform distribution based on \citet{Cibulkova2016} Eos data (K-S test: $p=0.18$). This apparent controversy is caused by the statistically poor sample of longitudes for asteroids with shape models. Concerning the latitude distribution, Eos family is dominated by objects with $|\beta|>60^{\circ}$, which is induced by the YORP spin axis and rotation period evolution. 


The Asteroid Lightcurve Database (LCDB\footnote{\url{http://www.minorplanet.info/lightcurvedatabase.html}}) of \citet{Warner2009} contains rotation periods for more than 500 Eos family members, which is larger by an order of magnitude than our sample. We show the rotation frequency in the Eos family in Fig.~\ref{fig:frequency}. Even such extensive sample is largely biased towards periods of several hours. It is more difficult to properly determine a rotation period larger than $\sim$20 hours than of $\sim$5--10 hours by observing asteroid's light curve, because this requires observations from several nights, sometimes even calibrated with respect to a common comparison star. Moreover, fainter fast rotators ($P\sim2-3$ h) require accurate photometry that is difficult to achieve with commonly used telescopes. Specifically, these fast rotators are known to be rather spheroidal \citep[][Fig.~6]{Pravec2000}, and so exhibit light curves with only small amplitudes. These biases has to be properly addressed before any sophisticated interpretation, however, this is a difficult task that we cannot easily do. On the other hand, qualitative estimates can still be made. 

The LCDB period data cover Eos members from various parts of the family in the proper element domain, contrary to the periods of objects with shape models that are mostly close to the center of the family. The distribution of periods for asteroids with shape models is dominated by objects larger than 10 km and can be approximated quite well by a Maxwellian distribution. However, this distribution is strongly biased by the selection effect of the light curve observations similarly to the LCDB data, but also by the light curve inversion method used for the shape model determination. For example, it is not common that there are light curve data for slow rotators from more than one apparition (the aim is usually to obtain a synodic period from one apparition data) or the used lightcurve inversion method cannot handle tumbling asteroids that are common among very slow rotators \citep[$P\sim$hundreds of hours, see, e.g., Fig.~8 of][]{Pravec2014}. As a result, our period distribution underestimates the number of slow rotators. Moreover, the period distributions are different within different size ranges \citep{Warner2009} due to YORP and collisional evolution, so our distribution is rather representative for a size range of $\sim10-20$ km. 

We show the LCDB frequency distribution for the whole sample and for objects larger and smaller than 10 km in the bottom right panel of Fig.~\ref{fig:frequency}. We have chosen only two size ranges because there are not enough objects with sizes larger than 30 km. The distribution for larger objects has a peak at 2--3 revolutions per day similarly to the distribution for the shape models sample, but it shows more slow and fast rotators. These fast and slow rotators are mostly objects with sizes between 10 and 20 km that are already evolved due to YORP and that are not present in the shape models sample. The LCDB period distribution of small asteroids is almost flat for 0--7 revolutions per day with an enhancement near 3--4 revolutions. This period distribution is similar to the period distribution of the whole main-belt population with sizes 7--14 km from \citet{Warner2009} and can be interpreted as YORP dominated with an apparent excess of medium rotators, which may be due to observational biases. The strong excess of slow rotators observed by \citet{Pravec2008} and \citet{Warner2009} is only present for a smaller size range that is not covered in our sample. 

It has been pointed our that an overdensity of asteroids that might be connected to the Eos family, exists inward of the 7:3 mean-motion resonance with Jupiter \citep{Tsirvoulis2017}. Yet, it is being investigated whether Eos members crossed the 7:3 while drifting under the Yarkovsky effect while moving away from the core of the family (which is situated beyond the said resonance) or these objects were directly placed inward the 7:3 by the initial velocity field of the impact that destroyed the parent body of the Eos family. In the case the first hypothesis is correct, limits for d$a$/d$t$ can be derived from dynamical modeling of the 7:3 resonance crossing, which would constitute to one of the first d$a$/d$t$ determination in the Main Belt. Studies of physical properties of asteroids like ours could help to distinguish between the two scenarios, because in the first case, only retrograde rotators should cross the 7:3 resonance, while some prograde rotators should exist inward the 7:3 resonance in the second scenario.

\section{Conclusions}\label{sec:conclusions}

We present updated shape models for 15 asteroids and new shape model determinations for 16 asteroids. These models were determined from optical disk-integrated data by the convex inversion method. Together with the already published models from the publicly available DAMIT database, we compiled a sample of 58 potential Eos family members with known shape models that we used in our analysis of physical properties within the family. Asteroids (251)~Sophia and (423)~Diotima are likely interlopers and were excluded from our further study. Our sample of shape models in the Eos family is the largest sample so far available for any asteroid family. Additionally, we present a shape model of asteroid (423)~Diotima based on the disk-resolved data obtained by the Nirc2 camera mounted on the W.M. Keck II telescope.

Rotation states of asteroids smaller than $\sim$30 km are heavily influenced by the YORP and Yarkovsky effects, whilst the large objects more or less retained their rotation state properties since the family creation. The prograde and retrograde rotators are concentrated outward and inward of the family center, respectively. The pole-ecliptic latitude distribution is bi-modal with most asteroids having their latitudes almost perpendicular to the ecliptic plane. Rotation periods of small asteroids ($D<10$~km) are dominated by the YORP spin up and spin down. Asteroids with $10<D<30$~km are influenced by both the YORP effect and collisions. Unfortunately, the sample of rotation periods for larger asteroids is too small to be interpreted.

Shape models based on large dense-in-time photometric data sets can be also used as inputs for further studies, such as detection of concavities \citep{Devogele2015}, thermophysical modeling with the varied-shape approach by \citet{Hanus2015a,Hanus2016b}, concave shape modeling \citep{Carry2010a, Bartczak2014, Viikinkoski2015, Shepard2017, Hanus2017a}, size optimization by disk-resolved images or occultation data \citep{Hanus2013b, Durech2011}, and density determinations \citep{Carry2010a, Shepard2017, Hanus2017b}.

Asteroids with shape models based only on sparse data or on sparse data and few dense light curves are ideal candidates for follow-up observations. New optical data should lead to improved shape and rotation state solutions that can be then used in further studies. Shape modeling is strongly dependent on a large community of observers that are kind enough to share their data. This tremendous flow of optical data pushes forward our understanding of asteroids.

The plots of sin($\beta$) to $D$ or the distribution of sin($\beta$) of different families should contain information about the degree of the family evolution due to YORP, hence the ages. In general, a young family should contain a larger fraction of objects with lower $|\beta|$ values than an older one, for instance. We note that the surface composition, heliocentric distance or the size frequency distribution should be also kept in mind. However, right now, we cannot draw reliable conclusions based on the sin($\beta$) values. First, the statistical sample based on shape model solutions (i.e., with the complete rotation state solution) is still poor for most families. Second, the data of \citet{Cibulkova2016} provide a large statistical sample, however, this sample is significantly biased: there is dependence on the inclination. This is because we represent the pole orientation in the ecliptic coordinate frame rather than in the orbital one (i.e., with the obliquity). In the YORP theory, the obliquity and the spin rate are evolved. Obliquity is identical to the ecliptic latitude only for the zero inclination. For objects with non-zero inclinations, one should compute the obliquities and compare the plots of sin($\epsilon$) to $D$ of different families. Unfortunately, we do not have the sign of the ecliptic latitude, so we cannot compute the obliquity.

Another link between physical properties of a family members and the YORP theoretical models is offered by comparing the distribution of the $|$sin($\beta$)$|$ and the spin rate. However, this is not straightforward and should be done with caution. The current rotation state of an asteroids is a result of a long past evolution due to YORP effect and collisions which represent a random effect. Also, the initial rotation state is important for the proper interpretation. As a result, only a numerical study with a large statistical sample of individual realizations with different initial conditions can provide reliable results \citep[we used simple YORP and collision models for four families in][]{Hanus2013c}. However, such a study is beyond the scope of this work. It is not possible to estimate the YORP coefficient \citep[e.g., by the equations presented in][]{Jacobson2014b} for an individual object based on its rotation state that is different from the theoretical end state, because we do not know the initial starting point. The YORP coefficient is dependent on the shape and the rotation state, which we do not know accurately enough. Therefore, YORP coefficient can be derived only if we measure the change of the period for the asteroid \citep{Kaasalainen2007b, Taylor2007}. Another complication is that the YORP coefficient can change even due to minor collisions or surface re-shaping if the rotation is fast.

Photometric data from ongoing and future surveys (e.g., Gaia or LSST) has the potential to largely increase our knowledge of asteroids physical properties. Increasing number of asteroids physical properties will also help to better constrain the theoretical models of the family evolution or even confirm prediction of some state-of-the-art theories \citep[e.g.,][]{Carruba2016}.

\section*{Acknowledgments}

The computations have been carried out on the computational cluster of the Astronomical Institute of Charles University in Prague (\url{http://sirrah.troja.mff.cuni.cz/tiger}).

HC and JD were supported by the grant 15-04816S of the Czech Science Foundation. 

VAL: The research leading to these results has received funding from the European Union's Horizon 2020 Research and Innovation Programme, under Grant Agreement no 687378.

We also used data from BlueEye600 observatory. The development of this observatory was supported by the Technology Agency of the Czech Republic (TA \v CR), project no.~TA 03011171, 'Development of Technologies for the Fast Robotic Observatories and Laser Communication Systems'.

For any publications, the user should request the latest data in date to the CdR-CdL team (raoul.behrend@unige.ch). All people involved in the data acquisition and treatment are coauthors of the papers -- exception: PhD thesis where a correct acknowledgement is mandatory.



This research has made use of the Keck Observatory Archive (KOA), which is operated by the W. M. Keck Observatory and the NASA Exoplanet Science Institute (NExScI), under contract with the National Aeronautics and Space Administration.

This publication uses data products from NEOWISE, a project of the Jet Propulsion Laboratory/California Institute of Technology, funded by the Planetary Science Division of the NASA. We made use of the NASA/IPAC Infrared Science Archive, which is operated by the Jet Propulsion Laboratory, California Institute of Technology, under contract with the NASA.

The authors wish to acknowledge and honor the indigenous Hawaiian community because data used in this work were obtained from the summit of Maunakea. We honor and respect the reverence and significance the summit has always had in native Hawaiian culture. We are fortunate and grateful to have had the opportunity to conduct observations from this most sacred mountain.

\bibliography{mybib}

\begin{thebibliography}{91}
\expandafter\ifx\csname natexlab\endcsname\relax\def\natexlab#1{#1}\fi
\expandafter\ifx\csname url\endcsname\relax
  \def\url#1{\texttt{#1}}\fi
\expandafter\ifx\csname urlprefix\endcsname\relax\def\urlprefix{URL }\fi
\providecommand{\eprint}[2][]{\url{#2}}
\providecommand{\bibinfo}[2]{#2}
\ifx\xfnm\relax \def\xfnm[#1]{\unskip,\space#1}\fi
\bibitem[{{Al{\'{\i}}-Lagoa} et~al.(2016){Al{\'{\i}}-Lagoa}, {Licandro},
  {Gil-Hutton}, {Ca{\~n}ada-Assandri}, {Delbo'}, {de Le{\'o}n}, {Campins},
  {Pinilla-Alonso}, {Kelley} and {Hanu{\v s}}}]{AliLagoa2016a}
\bibinfo{author}{{Al{\'{\i}}-Lagoa}, V.}, \bibinfo{author}{{Licandro}, J.},
  \bibinfo{author}{{Gil-Hutton}, R.}, \bibinfo{author}{{Ca{\~n}ada-Assandri},
  M.}, \bibinfo{author}{{Delbo'}, M.}, \bibinfo{author}{{de Le{\'o}n}, J.},
  \bibinfo{author}{{Campins}, H.}, \bibinfo{author}{{Pinilla-Alonso}, N.},
  \bibinfo{author}{{Kelley}, M.S.P.}, \bibinfo{author}{{Hanu{\v s}}, J.},
  \bibinfo{year}{2016}.
\newblock \bibinfo{title}{{Differences between the Pallas collisional family
  and similarly sized B-type asteroids}}.
\newblock \bibinfo{journal}{\aap} \bibinfo{volume}{591}, \bibinfo{pages}{A14}.
\bibitem[{{Al{\'\i}-Lagoa} et~al.(2014){Al{\'\i}-Lagoa}, {Lionni}, {Delbo'},
  {Gundlach}, {Blum} and {Licandro}}]{AliLagoa2014}
\bibinfo{author}{{Al{\'\i}-Lagoa}, V.}, \bibinfo{author}{{Lionni}, L.},
  \bibinfo{author}{{Delbo'}, M.}, \bibinfo{author}{{Gundlach}, B.},
  \bibinfo{author}{{Blum}, J.}, \bibinfo{author}{{Licandro}, J.},
  \bibinfo{year}{2014}.
\newblock \bibinfo{title}{{Thermophysical properties of near-Earth asteroid
  (341843) 2008 EV$\_{5}$ from WISE data}}.
\newblock \bibinfo{journal}{\aap} \bibinfo{volume}{561}, \bibinfo{pages}{A45}.
\newblock \eprint{1310.6715}.
\bibitem[{{Bartczak} et~al.(2014){Bartczak}, {Micha{\l}owski}, {Santana-Ros}
  and {Dudzi{\'n}ski}}]{Bartczak2014}
\bibinfo{author}{{Bartczak}, P.}, \bibinfo{author}{{Micha{\l}owski}, T.},
  \bibinfo{author}{{Santana-Ros}, T.}, \bibinfo{author}{{Dudzi{\'n}ski}, G.},
  \bibinfo{year}{2014}.
\newblock \bibinfo{title}{{A new non-convex model of the binary asteroid 90
  Antiope obtained with the SAGE modelling technique}}.
\newblock \bibinfo{journal}{\mnras} \bibinfo{volume}{443},
  \bibinfo{pages}{1802--1809}.
\newblock \eprint{1406.6555}.
\bibitem[{{Bell} et~al.(1989){Bell}, {Davis}, {Hartmann} and
  {Gaffey}}]{Bell1989}
\bibinfo{author}{{Bell}, J.F.}, \bibinfo{author}{{Davis}, D.R.},
  \bibinfo{author}{{Hartmann}, W.K.}, \bibinfo{author}{{Gaffey}, M.J.},
  \bibinfo{year}{1989}.
\newblock \bibinfo{title}{{Asteroids - The big picture}}, in:
  \bibinfo{editor}{{Binzel}, R.P.}, \bibinfo{editor}{{Gehrels}, T.},
  \bibinfo{editor}{{Matthews}, M.S.} (Eds.), \bibinfo{booktitle}{Asteroids II},
  pp. \bibinfo{pages}{921--945}.
\bibitem[{{Binzel}(1987)}]{Binzel1987a}
\bibinfo{author}{{Binzel}, R.P.}, \bibinfo{year}{1987}.
\newblock \bibinfo{title}{{A photoelectric survey of 130 asteroids}}.
\newblock \bibinfo{journal}{\icarus} \bibinfo{volume}{72},
  \bibinfo{pages}{135--208}.
\bibitem[{{Bottke} et~al.(2006){Bottke}, {Vokrouhlick{\'y}}, {Rubincam} and
  {Nesvorn{\'y}}}]{Bottke2006}
\bibinfo{author}{{Bottke}, J.W.F.}, \bibinfo{author}{{Vokrouhlick{\'y}}, D.},
  \bibinfo{author}{{Rubincam}, D.P.}, \bibinfo{author}{{Nesvorn{\'y}}, D.},
  \bibinfo{year}{2006}.
\newblock \bibinfo{title}{{The Yarkovsky and Yorp Effects: Implications for
  Asteroid Dynamics}}.
\newblock \bibinfo{journal}{Annual Review of Earth and Planetary Sciences}
  \bibinfo{volume}{34}, \bibinfo{pages}{157--191}.
\bibitem[{{Bowell} et~al.(2014){Bowell}, {Oszkiewicz}, {Wasserman}, {Muinonen},
  {Penttil{\"a}} and {Trilling}}]{Bowell2014a}
\bibinfo{author}{{Bowell}, E.}, \bibinfo{author}{{Oszkiewicz}, D.A.},
  \bibinfo{author}{{Wasserman}, L.H.}, \bibinfo{author}{{Muinonen}, K.},
  \bibinfo{author}{{Penttil{\"a}}, A.}, \bibinfo{author}{{Trilling}, D.E.},
  \bibinfo{year}{2014}.
\newblock \bibinfo{title}{{Asteroid spin-axis longitudes from the Lowell
  Observatory database}}.
\newblock \bibinfo{journal}{Meteoritics and Planetary Science}
  \bibinfo{volume}{49}, \bibinfo{pages}{95--102}.
\newblock \eprint{1310.3617}.
\bibitem[{{Brinsfield}(2010)}]{Brinsfield2010a}
\bibinfo{author}{{Brinsfield}, J.W.}, \bibinfo{year}{2010}.
\newblock \bibinfo{title}{{Asteroid Lightcurve Analysis at the Via Capote
  Observatory: 2009 3rd Quarter}}.
\newblock \bibinfo{journal}{Minor Planet Bulletin} \bibinfo{volume}{37},
  \bibinfo{pages}{19--20}.
\bibitem[{{Bro{\v z}} and {Morbidelli}(2013)}]{Broz2013a}
\bibinfo{author}{{Bro{\v z}}, M.}, \bibinfo{author}{{Morbidelli}, A.},
  \bibinfo{year}{2013}.
\newblock \bibinfo{title}{{The Eos family halo}}.
\newblock \bibinfo{journal}{\icarus} \bibinfo{volume}{223},
  \bibinfo{pages}{844--849}.
\newblock \eprint{1302.1447}.
\bibitem[{{Bro\v{z}} et~al.(2013){Bro\v{z}}, {Morbidelli}, {Bottke},
  {Rozehnal}, {Vokrouhlick{\'y}} and {Nesvorn{\'y}}}]{Broz2013b}
\bibinfo{author}{{Bro\v{z}}, M.}, \bibinfo{author}{{Morbidelli}, A.},
  \bibinfo{author}{{Bottke}, W.F.}, \bibinfo{author}{{Rozehnal}, J.},
  \bibinfo{author}{{Vokrouhlick{\'y}}, D.}, \bibinfo{author}{{Nesvorn{\'y}},
  D.}, \bibinfo{year}{2013}.
\newblock \bibinfo{title}{{Constraining the cometary flux through the asteroid
  belt during the late heavy bombardment}}.
\newblock \bibinfo{journal}{\aap} \bibinfo{volume}{551}, \bibinfo{pages}{A117}.
\newblock \eprint{1301.6221}.
\bibitem[{{Carruba} et~al.(2016){Carruba}, {Nesvorn{\'y}} and
  {Vokrouhlick{\'y}}}]{Carruba2016}
\bibinfo{author}{{Carruba}, V.}, \bibinfo{author}{{Nesvorn{\'y}}, D.},
  \bibinfo{author}{{Vokrouhlick{\'y}}, D.}, \bibinfo{year}{2016}.
\newblock \bibinfo{title}{{Detection of the YORP Effect for Small Asteroids in
  the Karin Cluster}}.
\newblock \bibinfo{journal}{\aj} \bibinfo{volume}{151}, \bibinfo{pages}{164}.
\newblock \eprint{1603.09612}.
\bibitem[{{Carry}(2012)}]{Carry2012b}
\bibinfo{author}{{Carry}, B.}, \bibinfo{year}{2012}.
\newblock \bibinfo{title}{{Density of asteroids}}.
\newblock \bibinfo{journal}{\planss} \bibinfo{volume}{73},
  \bibinfo{pages}{98--118}.
\newblock \eprint{1203.4336}.
\bibitem[{{Carry} et~al.(2010){Carry}, {Dumas}, {Kaasalainen}, {Berthier},
  {Merline}, {Erard}, {Conrad}, {Drummond}, {Hestroffer}, {Fulchignoni} and
  {Fusco}}]{Carry2010a}
\bibinfo{author}{{Carry}, B.}, \bibinfo{author}{{Dumas}, C.},
  \bibinfo{author}{{Kaasalainen}, M.}, \bibinfo{author}{{Berthier}, J.},
  \bibinfo{author}{{Merline}, W.J.}, \bibinfo{author}{{Erard}, S.},
  \bibinfo{author}{{Conrad}, A.}, \bibinfo{author}{{Drummond}, J.D.},
  \bibinfo{author}{{Hestroffer}, D.}, \bibinfo{author}{{Fulchignoni}, M.},
  \bibinfo{author}{{Fusco}, T.}, \bibinfo{year}{2010}.
\newblock \bibinfo{title}{{Physical properties of (2) Pallas}}.
\newblock \bibinfo{journal}{\icarus} \bibinfo{volume}{205},
  \bibinfo{pages}{460--472}.
\newblock \eprint{0912.3626}.
\bibitem[{{Cibulkov{\'a}} et~al.(2016){Cibulkov{\'a}}, {{\v D}urech},
  {Vokrouhlick{\'y}}, {Kaasalainen} and {Oszkiewicz}}]{Cibulkova2016}
\bibinfo{author}{{Cibulkov{\'a}}, H.}, \bibinfo{author}{{{\v D}urech}, J.},
  \bibinfo{author}{{Vokrouhlick{\'y}}, D.}, \bibinfo{author}{{Kaasalainen},
  M.}, \bibinfo{author}{{Oszkiewicz}, D.A.}, \bibinfo{year}{2016}.
\newblock \bibinfo{title}{{Distribution of spin-axes longitudes and shape
  elongations of main-belt asteroids}}.
\newblock \bibinfo{journal}{\aap} \bibinfo{volume}{596}, \bibinfo{pages}{A57}.
\newblock \eprint{1610.02790}.
\bibitem[{{Clark} et~al.(2009){Clark}, {Ockert-Bell}, {Cloutis}, {Nesvorny},
  {Moth{\'e}-Diniz} and {Bus}}]{Clark2009}
\bibinfo{author}{{Clark}, B.E.}, \bibinfo{author}{{Ockert-Bell}, M.E.},
  \bibinfo{author}{{Cloutis}, E.A.}, \bibinfo{author}{{Nesvorny}, D.},
  \bibinfo{author}{{Moth{\'e}-Diniz}, T.}, \bibinfo{author}{{Bus}, S.J.},
  \bibinfo{year}{2009}.
\newblock \bibinfo{title}{{Spectroscopy of K-complex asteroids: Parent bodies
  of carbonaceous meteorites?}}
\newblock \bibinfo{journal}{\icarus} \bibinfo{volume}{202},
  \bibinfo{pages}{119--133}.
\bibitem[{{Delbo'} et~al.(2007){Delbo'}, {dell'Oro}, {Harris}, {Mottola} and
  {Mueller}}]{Delbo2007a}
\bibinfo{author}{{Delbo'}, M.}, \bibinfo{author}{{dell'Oro}, A.},
  \bibinfo{author}{{Harris}, A.W.}, \bibinfo{author}{{Mottola}, S.},
  \bibinfo{author}{{Mueller}, M.}, \bibinfo{year}{2007}.
\newblock \bibinfo{title}{{Thermal inertia of near-Earth asteroids and
  implications for the magnitude of the Yarkovsky effect}}.
\newblock \bibinfo{journal}{\icarus} \bibinfo{volume}{190},
  \bibinfo{pages}{236--249}.
\newblock \eprint{0704.1915}.
\bibitem[{{Devog{\`e}le} et~al.(2015){Devog{\`e}le}, {Rivet}, {Tanga},
  {Bendjoya}, {Surdej}, {Bartczak} and {Hanus}}]{Devogele2015}
\bibinfo{author}{{Devog{\`e}le}, M.}, \bibinfo{author}{{Rivet}, J.P.},
  \bibinfo{author}{{Tanga}, P.}, \bibinfo{author}{{Bendjoya}, P.},
  \bibinfo{author}{{Surdej}, J.}, \bibinfo{author}{{Bartczak}, P.},
  \bibinfo{author}{{Hanus}, J.}, \bibinfo{year}{2015}.
\newblock \bibinfo{title}{{A method to search for large-scale concavities in
  asteroid shape models}}.
\newblock \bibinfo{journal}{\mnras} \bibinfo{volume}{453},
  \bibinfo{pages}{2232--2240}.
\bibitem[{{Dimartino}(1986)}]{DiMartino1986}
\bibinfo{author}{{Dimartino}, M.}, \bibinfo{year}{1986}.
\newblock \bibinfo{title}{{A Photoelectric Program for Small and Unusual
  Asteroids}}, in: \bibinfo{editor}{{Lagerkvist}, C.I.},
  \bibinfo{editor}{{Rickman}, H.}, \bibinfo{editor}{{Lindblad}, B.A.},
  \bibinfo{editor}{{Lundstedt}, H.} (Eds.), \bibinfo{booktitle}{Asteroids,
  Comets, Meteors II}.
\bibitem[{{Doressoundiram} et~al.(1998){Doressoundiram}, {Barucci},
  {Fulchignoni} and {Florczak}}]{Doressoundiram1998}
\bibinfo{author}{{Doressoundiram}, A.}, \bibinfo{author}{{Barucci}, M.A.},
  \bibinfo{author}{{Fulchignoni}, M.}, \bibinfo{author}{{Florczak}, M.},
  \bibinfo{year}{1998}.
\newblock \bibinfo{title}{{EOS Family: A Spectroscopic Study}}.
\newblock \bibinfo{journal}{\icarus} \bibinfo{volume}{131},
  \bibinfo{pages}{15--31}.
\bibitem[{{\v Durech} et~al.(2015a){\v Durech}, {Carry}, {Delbo'},
  {Kaasalainen} and {Viikinkoski}}]{DurechAIV2015}
\bibinfo{author}{{\v Durech}, J.}, \bibinfo{author}{{Carry}, B.},
  \bibinfo{author}{{Delbo'}, M.}, \bibinfo{author}{{Kaasalainen}, M.},
  \bibinfo{author}{{Viikinkoski}, M.}, \bibinfo{year}{2015}a.
\newblock \bibinfo{title}{Asteroid models from multiple data sources}, in:
  \bibinfo{editor}{{Michel}, P.}, \bibinfo{editor}{{DeMeo}, F.E.},
  \bibinfo{editor}{{Bottke}, W.F.} (Eds.), \bibinfo{booktitle}{Asteroids IV},
  \bibinfo{publisher}{The University of Arizona Press}. pp.
  \bibinfo{pages}{183--201}.
\bibitem[{{\v Durech} et~al.(2016){\v Durech}, {Hanu\v s}, {Oszkiewicz} and
  {Van\v co}}]{Durech2016}
\bibinfo{author}{{\v Durech}, J.}, \bibinfo{author}{{Hanu\v s}, J.},
  \bibinfo{author}{{Oszkiewicz}, D.}, \bibinfo{author}{{Van\v co}, R.},
  \bibinfo{year}{2016}.
\newblock \bibinfo{title}{{Asteroid models from the Lowell photometric
  database}}.
\newblock \bibinfo{journal}{\aap} \bibinfo{volume}{587}, \bibinfo{pages}{A48}.
\newblock \eprint{1601.02909}.
\bibitem[{{\v Durech} et~al.(2015b){\v Durech}, {Hanu\v s} and {Van\v
  co}}]{Durech2015b}
\bibinfo{author}{{\v Durech}, J.}, \bibinfo{author}{{Hanu\v s}, J.},
  \bibinfo{author}{{Van\v co}, R.}, \bibinfo{year}{2015}b.
\newblock \bibinfo{title}{{Asteroids@home-A BOINC distributed computing project
  for asteroid shape reconstruction}}.
\newblock \bibinfo{journal}{Astronomy and Computing} \bibinfo{volume}{13},
  \bibinfo{pages}{80--84}.
\newblock \eprint{1511.08640}.
\bibitem[{{\v Durech} et~al.(2017){\v Durech}, {Hanu\v{s}}, {Bro\v{z}} and
  {Lehk\'{y}}}]{Durech2017a}
\bibinfo{author}{{\v Durech}, J.}, \bibinfo{author}{{Hanu\v{s}}, J.},
  \bibinfo{author}{{Bro\v{z}}, M.}, \bibinfo{author}{{Lehk\'{y}}, M.},
  \bibinfo{year}{2017}.
\newblock \bibinfo{title}{Shape models of asteroids based on lightcurve
  observations with blueeye600 robotic observatory}.
\newblock \bibinfo{note}{Submitted to \icarus}.
\bibitem[{{\v Durech} and {Kaasalainen}(2003)}]{Durech2003}
\bibinfo{author}{{\v Durech}, J.}, \bibinfo{author}{{Kaasalainen}, M.},
  \bibinfo{year}{2003}.
\newblock \bibinfo{title}{{Photometric signatures of highly nonconvex and
  binary asteroids}}.
\newblock \bibinfo{journal}{\aap} \bibinfo{volume}{404},
  \bibinfo{pages}{709--714}.
\bibitem[{{\v{D}urech} et~al.(2011){\v{D}urech}, {Kaasalainen}, {Herald},
  {Dunham}, {Timerson}, {Hanu\v{s}}, {Frappa}, {Talbot}, {Hayamizu}, {Warner},
  {Pilcher} and {Gal{\'a}d}}]{Durech2011}
\bibinfo{author}{{\v{D}urech}, J.}, \bibinfo{author}{{Kaasalainen}, M.},
  \bibinfo{author}{{Herald}, D.}, \bibinfo{author}{{Dunham}, D.},
  \bibinfo{author}{{Timerson}, B.}, \bibinfo{author}{{Hanu\v{s}}, J.},
  \bibinfo{author}{{Frappa}, E.}, \bibinfo{author}{{Talbot}, J.},
  \bibinfo{author}{{Hayamizu}, T.}, \bibinfo{author}{{Warner}, B.D.},
  \bibinfo{author}{{Pilcher}, F.}, \bibinfo{author}{{Gal{\'a}d}, A.},
  \bibinfo{year}{2011}.
\newblock \bibinfo{title}{{Combining asteroid models derived by lightcurve
  inversion with asteroidal occultation silhouettes}}.
\newblock \bibinfo{journal}{Icarus} \bibinfo{volume}{214},
  \bibinfo{pages}{652--670}.
\newblock \eprint{1104.4227}.
\bibitem[{{\v Durech} et~al.(2009){\v Durech}, {Kaasalainen}, {Warner},
  {Fauerbach}, {Marks}, {Fauvaud}, {Fauvaud}, {Vugnon}, {Pilcher}, {Bernasconi}
  and {Behrend}}]{Durech2009}
\bibinfo{author}{{\v Durech}, J.}, \bibinfo{author}{{Kaasalainen}, M.},
  \bibinfo{author}{{Warner}, B.D.}, \bibinfo{author}{{Fauerbach}, M.},
  \bibinfo{author}{{Marks}, S.A.}, \bibinfo{author}{{Fauvaud}, S.},
  \bibinfo{author}{{Fauvaud}, M.}, \bibinfo{author}{{Vugnon}, J.},
  \bibinfo{author}{{Pilcher}, F.}, \bibinfo{author}{{Bernasconi}, L.},
  \bibinfo{author}{{Behrend}, R.}, \bibinfo{year}{2009}.
\newblock \bibinfo{title}{{Asteroid models from combined sparse and dense
  photometric data}}.
\newblock \bibinfo{journal}{\aap} \bibinfo{volume}{493},
  \bibinfo{pages}{291--297}.
\bibitem[{{\v Durech} et~al.(2010){\v Durech}, {Sidorin} and
  {Kaasalainen}}]{Durech2010}
\bibinfo{author}{{\v Durech}, J.}, \bibinfo{author}{{Sidorin}, V.},
  \bibinfo{author}{{Kaasalainen}, M.}, \bibinfo{year}{2010}.
\newblock \bibinfo{title}{{DAMIT: a database of asteroid models}}.
\newblock \bibinfo{journal}{\aap} \bibinfo{volume}{513}, \bibinfo{pages}{A46}.
\bibitem[{{Dykhuis} et~al.(2016){Dykhuis}, {Molnar}, {Gates}, {Gonzales},
  {Huffman}, {Maat}, {Maat}, {Marks}, {Massey-Plantinga}, {McReynolds},
  {Schut}, {Stoep}, {Stutzman}, {Thomas}, {Vander Tuig}, {Vriesema} and
  {Greenberg}}]{Dykhuis2016}
\bibinfo{author}{{Dykhuis}, M.J.}, \bibinfo{author}{{Molnar}, L.A.},
  \bibinfo{author}{{Gates}, C.J.}, \bibinfo{author}{{Gonzales}, J.A.},
  \bibinfo{author}{{Huffman}, J.J.}, \bibinfo{author}{{Maat}, A.R.},
  \bibinfo{author}{{Maat}, S.L.}, \bibinfo{author}{{Marks}, M.I.},
  \bibinfo{author}{{Massey-Plantinga}, A.R.}, \bibinfo{author}{{McReynolds},
  N.D.}, \bibinfo{author}{{Schut}, J.A.}, \bibinfo{author}{{Stoep}, J.P.},
  \bibinfo{author}{{Stutzman}, A.J.}, \bibinfo{author}{{Thomas}, B.C.},
  \bibinfo{author}{{Vander Tuig}, G.W.}, \bibinfo{author}{{Vriesema}, J.W.},
  \bibinfo{author}{{Greenberg}, R.}, \bibinfo{year}{2016}.
\newblock \bibinfo{title}{{Efficient spin sense determination of Flora-region
  asteroids via the epoch method}}.
\newblock \bibinfo{journal}{\icarus} \bibinfo{volume}{267},
  \bibinfo{pages}{174--203}.
\bibitem[{{Hanu{\v s}} et~al.(2017){Hanu{\v s}}, {Viikinkoski}, {Marchis}, {{\v
  D}urech}, {Kaasalainen}, {Delbo'}, {Herald}, {Frappa}, {Hayamizu}, {Kerr},
  {Preston}, {Timerson}, {Dunham} and {Talbot}}]{Hanus2017b}
\bibinfo{author}{{Hanu{\v s}}, J.}, \bibinfo{author}{{Viikinkoski}, M.},
  \bibinfo{author}{{Marchis}, F.}, \bibinfo{author}{{{\v D}urech}, J.},
  \bibinfo{author}{{Kaasalainen}, M.}, \bibinfo{author}{{Delbo'}, M.},
  \bibinfo{author}{{Herald}, D.}, \bibinfo{author}{{Frappa}, E.},
  \bibinfo{author}{{Hayamizu}, T.}, \bibinfo{author}{{Kerr}, S.},
  \bibinfo{author}{{Preston}, S.}, \bibinfo{author}{{Timerson}, B.},
  \bibinfo{author}{{Dunham}, D.}, \bibinfo{author}{{Talbot}, J.},
  \bibinfo{year}{2017}.
\newblock \bibinfo{title}{{Volumes and bulk densities of forty asteroids from
  ADAM shape modeling}}.
\newblock \bibinfo{journal}{\aap} \bibinfo{volume}{601}, \bibinfo{pages}{A114}.
\newblock \eprint{1702.01996}.
\bibitem[{{Hanu\v{s}} et~al.(2013a){Hanu\v{s}}, {Bro\v{z}}, {\v Durech},
  {Warner}, {Brinsfield}, {Durkee}, {Higgins}, {Koff}, {Oey}, {Pilcher},
  {Stephens}, {Strabla}, {Ulisse} and {Girelli}}]{Hanus2013c}
\bibinfo{author}{{Hanu\v{s}}, J.}, \bibinfo{author}{{Bro\v{z}}, M.},
  \bibinfo{author}{{\v Durech}, J.}, \bibinfo{author}{{Warner}, B.D.},
  \bibinfo{author}{{Brinsfield}, J.}, \bibinfo{author}{{Durkee}, R.},
  \bibinfo{author}{{Higgins}, D.}, \bibinfo{author}{{Koff}, R.A.},
  \bibinfo{author}{{Oey}, J.}, \bibinfo{author}{{Pilcher}, F.},
  \bibinfo{author}{{Stephens}, R.}, \bibinfo{author}{{Strabla}, L.P.},
  \bibinfo{author}{{Ulisse}, Q.}, \bibinfo{author}{{Girelli}, R.},
  \bibinfo{year}{2013}a.
\newblock \bibinfo{title}{{An anisotropic distribution of spin vectors in
  asteroid families}}.
\newblock \bibinfo{journal}{\aap} \bibinfo{volume}{559}, \bibinfo{pages}{A134}.
\newblock \eprint{1309.4296}.
\bibitem[{{Hanu\v{s}} et~al.(2015){Hanu\v{s}}, {Delbo'}, {\v Durech} and
  {Al{\'{\i}}-Lagoa}}]{Hanus2015a}
\bibinfo{author}{{Hanu\v{s}}, J.}, \bibinfo{author}{{Delbo'}, M.},
  \bibinfo{author}{{\v Durech}, J.}, \bibinfo{author}{{Al{\'{\i}}-Lagoa}, V.},
  \bibinfo{year}{2015}.
\newblock \bibinfo{title}{{Thermophysical modeling of asteroids from WISE
  thermal infrared data - Significance of the shape model and the pole
  orientation uncertainties}}.
\newblock \bibinfo{journal}{\icarus} \bibinfo{volume}{256},
  \bibinfo{pages}{101--116}.
\newblock \eprint{1504.04199}.
\bibitem[{{Hanu\v{s}} et~al.(2016a){Hanu\v{s}}, {Delbo'}, {Vokrouhlick{\'y}},
  {Pravec}, {Emery}, {Al{\'{\i}}-Lagoa}, {Bolin}, {Devog{\`e}le}, {Dyvig},
  {Gal{\'a}d}, {Jedicke}, {Korno{\v s}}, {Ku{\v s}nir{\'a}k}, {Licandro},
  {Reddy}, {Rivet}, {Vil{\'a}gi} and {Warner}}]{Hanus2016b}
\bibinfo{author}{{Hanu\v{s}}, J.}, \bibinfo{author}{{Delbo'}, M.},
  \bibinfo{author}{{Vokrouhlick{\'y}}, D.}, \bibinfo{author}{{Pravec}, P.},
  \bibinfo{author}{{Emery}, J.P.}, \bibinfo{author}{{Al{\'{\i}}-Lagoa}, V.},
  \bibinfo{author}{{Bolin}, B.}, \bibinfo{author}{{Devog{\`e}le}, M.},
  \bibinfo{author}{{Dyvig}, R.}, \bibinfo{author}{{Gal{\'a}d}, A.},
  \bibinfo{author}{{Jedicke}, R.}, \bibinfo{author}{{Korno{\v s}}, L.},
  \bibinfo{author}{{Ku{\v s}nir{\'a}k}, P.}, \bibinfo{author}{{Licandro}, J.},
  \bibinfo{author}{{Reddy}, V.}, \bibinfo{author}{{Rivet}, J.P.},
  \bibinfo{author}{{Vil{\'a}gi}, J.}, \bibinfo{author}{{Warner}, B.D.},
  \bibinfo{year}{2016}a.
\newblock \bibinfo{title}{{Near-Earth asteroid (3200) Phaethon:
  Characterization of its orbit, spin state, and thermophysical parameters}}.
\newblock \bibinfo{journal}{\aap} \bibinfo{volume}{592}, \bibinfo{pages}{A34}.
\newblock \eprint{1605.05205}.
\bibitem[{{Hanu\v{s}} et~al.(2013b){Hanu\v{s}}, {\v{D}urech}, {Bro\v{z}},
  {Marciniak}, {Warner}, {Pilcher}, {Stephens}, {Behrend}, {Carry},
  {\v{C}apek}, {Antonini}, {Audejean}, {Augustesen}, {Barbotin}, {Baudouin},
  {Bayol}, {Bernasconi}, {Borczyk}, {Bosch}, {Brochard}, {Brunetto}, {Casulli},
  {Cazenave}, {Charbonnel}, {Christophe}, {Colas}, {Coloma}, {Conjat},
  {Cooney}, {Correira}, {Cotrez}, {Coupier}, {Crippa}, {Cristofanelli},
  {Dalmas}, {Danavaro}, {Demeautis}, {Droege}, {Durkee}, {Esseiva}, {Esteban},
  {Fagas}, {Farroni}, {Fauvaud}, {Fauvaud}, {Del Freo}, {Garcia}, {Geier},
  {Godon}, {Grangeon}, {Hamanowa}, {Hamanowa}, {Heck}, {Hellmich}, {Higgins},
  {Hirsch}, {Husarik}, {Itkonen}, {Jade}, {Kami{\'n}ski}, {Kankiewicz},
  {Klotz}, {Koff}, {Kryszczy{\'n}ska}, {Kwiatkowski}, {Laffont}, {Leroy},
  {Lecacheux}, {Leonie}, {Leyrat}, {Manzini}, {Martin}, {Masi}, {Matter},
  {Micha{\l}owski}, {Micha{\l}owski}, {Micha{\l}owski}, {Michelet},
  {Michelsen}, {Morelle}, {Mottola}, {Naves}, {Nomen}, {Oey}, {Og{\l}oza},
  {Oksanen}, {Oszkiewicz}, {P{\"a}{\"a}kk{\"o}nen}, {Paiella}, {Pallares},
  {Paulo}, {Pavic}, {Payet}, {Poli{\'n}ska}, {Polishook}, {Poncy}, {Revaz},
  {Rinner}, {Rocca}, {Roche}, {Romeuf}, {Roy}, {Saguin}, {Salom}, {Sanchez},
  {Santacana}, {Santana-Ros}, {Sareyan}, {Sobkowiak}, {Sposetti}, {Starkey},
  {Stoss}, {Strajnic}, {Teng}, {Tr{\'e}gon}, {Vagnozzi}, {Velichko},
  {Waelchli}, {Wagrez} and {W{\"u}cher}}]{Hanus2013a}
\bibinfo{author}{{Hanu\v{s}}, J.}, \bibinfo{author}{{\v{D}urech}, J.},
  \bibinfo{author}{{Bro\v{z}}, M.}, \bibinfo{author}{{Marciniak}, A.},
  \bibinfo{author}{{Warner}, B.D.}, \bibinfo{author}{{Pilcher}, F.},
  \bibinfo{author}{{Stephens}, R.}, \bibinfo{author}{{Behrend}, R.},
  \bibinfo{author}{{Carry}, B.}, \bibinfo{author}{{\v{C}apek}, D.},
  \bibinfo{author}{{Antonini}, P.}, \bibinfo{author}{{Audejean}, M.},
  \bibinfo{author}{{Augustesen}, K.}, \bibinfo{author}{{Barbotin}, E.},
  \bibinfo{author}{{Baudouin}, P.}, \bibinfo{author}{{Bayol}, A.},
  \bibinfo{author}{{Bernasconi}, L.}, \bibinfo{author}{{Borczyk}, W.},
  \bibinfo{author}{{Bosch}, J.G.}, \bibinfo{author}{{Brochard}, E.},
  \bibinfo{author}{{Brunetto}, L.}, \bibinfo{author}{{Casulli}, S.},
  \bibinfo{author}{{Cazenave}, A.}, \bibinfo{author}{{Charbonnel}, S.},
  \bibinfo{author}{{Christophe}, B.}, \bibinfo{author}{{Colas}, F.},
  \bibinfo{author}{{Coloma}, J.}, \bibinfo{author}{{Conjat}, M.},
  \bibinfo{author}{{Cooney}, W.}, \bibinfo{author}{{Correira}, H.},
  \bibinfo{author}{{Cotrez}, V.}, \bibinfo{author}{{Coupier}, A.},
  \bibinfo{author}{{Crippa}, R.}, \bibinfo{author}{{Cristofanelli}, M.},
  \bibinfo{author}{{Dalmas}, C.}, \bibinfo{author}{{Danavaro}, C.},
  \bibinfo{author}{{Demeautis}, C.}, \bibinfo{author}{{Droege}, T.},
  \bibinfo{author}{{Durkee}, R.}, \bibinfo{author}{{Esseiva}, N.},
  \bibinfo{author}{{Esteban}, M.}, \bibinfo{author}{{Fagas}, M.},
  \bibinfo{author}{{Farroni}, G.}, \bibinfo{author}{{Fauvaud}, M.},
  \bibinfo{author}{{Fauvaud}, S.}, \bibinfo{author}{{Del Freo}, F.},
  \bibinfo{author}{{Garcia}, L.}, \bibinfo{author}{{Geier}, S.},
  \bibinfo{author}{{Godon}, C.}, \bibinfo{author}{{Grangeon}, K.},
  \bibinfo{author}{{Hamanowa}, H.}, \bibinfo{author}{{Hamanowa}, H.},
  \bibinfo{author}{{Heck}, N.}, \bibinfo{author}{{Hellmich}, S.},
  \bibinfo{author}{{Higgins}, D.}, \bibinfo{author}{{Hirsch}, R.},
  \bibinfo{author}{{Husarik}, M.}, \bibinfo{author}{{Itkonen}, T.},
  \bibinfo{author}{{Jade}, O.}, \bibinfo{author}{{Kami{\'n}ski}, K.},
  \bibinfo{author}{{Kankiewicz}, P.}, \bibinfo{author}{{Klotz}, A.},
  \bibinfo{author}{{Koff}, R.A.}, \bibinfo{author}{{Kryszczy{\'n}ska}, A.},
  \bibinfo{author}{{Kwiatkowski}, T.}, \bibinfo{author}{{Laffont}, A.},
  \bibinfo{author}{{Leroy}, A.}, \bibinfo{author}{{Lecacheux}, J.},
  \bibinfo{author}{{Leonie}, Y.}, \bibinfo{author}{{Leyrat}, C.},
  \bibinfo{author}{{Manzini}, F.}, \bibinfo{author}{{Martin}, A.},
  \bibinfo{author}{{Masi}, G.}, \bibinfo{author}{{Matter}, D.},
  \bibinfo{author}{{Micha{\l}owski}, J.}, \bibinfo{author}{{Micha{\l}owski},
  M.J.}, \bibinfo{author}{{Micha{\l}owski}, T.}, \bibinfo{author}{{Michelet},
  J.}, \bibinfo{author}{{Michelsen}, R.}, \bibinfo{author}{{Morelle}, E.},
  \bibinfo{author}{{Mottola}, S.}, \bibinfo{author}{{Naves}, R.},
  \bibinfo{author}{{Nomen}, J.}, \bibinfo{author}{{Oey}, J.},
  \bibinfo{author}{{Og{\l}oza}, W.}, \bibinfo{author}{{Oksanen}, A.},
  \bibinfo{author}{{Oszkiewicz}, D.}, \bibinfo{author}{{P{\"a}{\"a}kk{\"o}nen},
  P.}, \bibinfo{author}{{Paiella}, M.}, \bibinfo{author}{{Pallares}, H.},
  \bibinfo{author}{{Paulo}, J.}, \bibinfo{author}{{Pavic}, M.},
  \bibinfo{author}{{Payet}, B.}, \bibinfo{author}{{Poli{\'n}ska}, M.},
  \bibinfo{author}{{Polishook}, D.}, \bibinfo{author}{{Poncy}, R.},
  \bibinfo{author}{{Revaz}, Y.}, \bibinfo{author}{{Rinner}, C.},
  \bibinfo{author}{{Rocca}, M.}, \bibinfo{author}{{Roche}, A.},
  \bibinfo{author}{{Romeuf}, D.}, \bibinfo{author}{{Roy}, R.},
  \bibinfo{author}{{Saguin}, H.}, \bibinfo{author}{{Salom}, P.A.},
  \bibinfo{author}{{Sanchez}, S.}, \bibinfo{author}{{Santacana}, G.},
  \bibinfo{author}{{Santana-Ros}, T.}, \bibinfo{author}{{Sareyan}, J.P.},
  \bibinfo{author}{{Sobkowiak}, K.}, \bibinfo{author}{{Sposetti}, S.},
  \bibinfo{author}{{Starkey}, D.}, \bibinfo{author}{{Stoss}, R.},
  \bibinfo{author}{{Strajnic}, J.}, \bibinfo{author}{{Teng}, J.P.},
  \bibinfo{author}{{Tr{\'e}gon}, B.}, \bibinfo{author}{{Vagnozzi}, A.},
  \bibinfo{author}{{Velichko}, F.P.}, \bibinfo{author}{{Waelchli}, N.},
  \bibinfo{author}{{Wagrez}, K.}, \bibinfo{author}{{W{\"u}cher}, H.},
  \bibinfo{year}{2013}b.
\newblock \bibinfo{title}{{Asteroids' physical models from combined dense and
  sparse photometry and scaling of the YORP effect by the observed obliquity
  distribution}}.
\newblock \bibinfo{journal}{\aap} \bibinfo{volume}{551}, \bibinfo{pages}{A67}.
\newblock \eprint{1301.6943}.
\bibitem[{{Hanu\v{s}} et~al.(2011){Hanu\v{s}}, {\v{D}urech}, {Bro\v{z}},
  {Warner}, {Pilcher}, {Stephens}, {Oey}, {Bernasconi}, {Casulli}, {Behrend},
  {Polishook}, {Henych}, {Lehk{\'y}}, {Yoshida} and {Ito}}]{Hanus2011}
\bibinfo{author}{{Hanu\v{s}}, J.}, \bibinfo{author}{{\v{D}urech}, J.},
  \bibinfo{author}{{Bro\v{z}}, M.}, \bibinfo{author}{{Warner}, B.D.},
  \bibinfo{author}{{Pilcher}, F.}, \bibinfo{author}{{Stephens}, R.},
  \bibinfo{author}{{Oey}, J.}, \bibinfo{author}{{Bernasconi}, L.},
  \bibinfo{author}{{Casulli}, S.}, \bibinfo{author}{{Behrend}, R.},
  \bibinfo{author}{{Polishook}, D.}, \bibinfo{author}{{Henych}, T.},
  \bibinfo{author}{{Lehk{\'y}}, M.}, \bibinfo{author}{{Yoshida}, F.},
  \bibinfo{author}{{Ito}, T.}, \bibinfo{year}{2011}.
\newblock \bibinfo{title}{{A study of asteroid pole-latitude distribution based
  on an extended set of shape models derived by the lightcurve inversion
  method}}.
\newblock \bibinfo{journal}{\aap} \bibinfo{volume}{530}, \bibinfo{pages}{A134}.
\newblock \eprint{1104.4114}.
\bibitem[{{Hanu\v{s}} et~al.(2016b){Hanu\v{s}}, {\v{D}urech}, {Oszkiewicz},
  {Behrend}, {Carry}, {Delbo}, {Adam}, {Afonina}, {Anquetin}, {Antonini},
  {Arnold}, {Audejean}, {Aurard}, {Bachschmidt}, {Baduel}, {Barbotin},
  {Barroy}, {Baudouin}, {Berard}, {Berger}, {Bernasconi}, {Bosch}, {Bouley},
  {Bozhinova}, {Brinsfield}, {Brunetto}, {Canaud}, {Caron}, {Carrier},
  {Casalnuovo}, {Casulli}, {Cerda}, {Chalamet}, {Charbonnel}, {Chinaglia},
  {Cikota}, {Colas}, {Coliac}, {Collet}, {Coloma}, {Conjat}, {Conseil},
  {Costa}, {Crippa}, {Cristofanelli}, {Damerdji}, {Deback{\`e}re}, {Decock},
  {D{\'e}hais}, {D{\'e}l{\'e}age}, {Delmelle}, {Demeautis},
  {Dr{\'o}{\.z}d{\.z}}, {Dubos}, {Dulcamara}, {Dumont}, {Durkee}, {Dymock},
  {Escalante del Valle}, {Esseiva}, {Esseiva}, {Esteban}, {Fauchez},
  {Fauerbach}, {Fauvaud}, {Fauvaud}, {Forn{\'e}}, {Fournel}, {Fradet},
  {Garlitz}, {Gerteis}, {Gillier}, {Gillon}, {Giraud}, {Godard}, {Goncalves},
  {Hamanowa}, {Hamanowa}, {Hay}, {Hellmich}, {Heterier}, {Higgins}, {Hirsch},
  {Hodosan}, {Hren}, {Hygate}, {Innocent}, {Jacquinot}, {Jawahar}, {Jehin},
  {Jerosimic}, {Klotz}, {Koff}, {Korlevic}, {Kosturkiewicz}, {Krafft},
  {Krugly}, {Kugel}, {Labrevoir}, {Lecacheux}, {Lehk{\'y}}, {Leroy},
  {Lesquerbault}, {Lopez-Gonzales}, {Lutz}, {Mallecot}, {Manfroid}, {Manzini},
  {Marciniak}, {Martin}, {Modave}, {Montaigut}, {Montier}, {Morelle}, {Morton},
  {Mottola}, {Naves}, {Nomen}, {Oey}, {Og{\l}oza}, {Paiella}, {Pallares},
  {Peyrot}, {Pilcher}, {Pirenne}, {Piron}, {Poli{\'n}ska}, {Polotto}, {Poncy},
  {Previt}, {Reignier}, {Renauld}, {Ricci}, {Richard}, {Rinner}, {Risoldi},
  {Robilliard}, {Romeuf}, {Rousseau}, {Roy}, {Ruthroff}, {Salom}, {Salvador},
  {Sanchez}, {Santana-Ros}, {Scholz}, {S{\'e}n{\'e}}, {Skiff}, {Sobkowiak},
  {Sogorb}, {Sold{\'a}n}, {Spiridakis}, {Splanska}, {Sposetti}, {Starkey},
  {Stephens}, {Stiepen}, {Stoss}, {Strajnic}, {Teng}, {Tumolo}, {Vagnozzi},
  {Vanoutryve}, {Vugnon}, {Warner}, {Waucomont}, {Wertz}, {Winiarski} and
  {Wolf}}]{Hanus2016a}
\bibinfo{author}{{Hanu\v{s}}, J.}, \bibinfo{author}{{\v{D}urech}, J.},
  \bibinfo{author}{{Oszkiewicz}, D.A.}, \bibinfo{author}{{Behrend}, R.},
  \bibinfo{author}{{Carry}, B.}, \bibinfo{author}{{Delbo}, M.},
  \bibinfo{author}{{Adam}, O.}, \bibinfo{author}{{Afonina}, V.},
  \bibinfo{author}{{Anquetin}, R.}, \bibinfo{author}{{Antonini}, P.},
  \bibinfo{author}{{Arnold}, L.}, \bibinfo{author}{{Audejean}, M.},
  \bibinfo{author}{{Aurard}, P.}, \bibinfo{author}{{Bachschmidt}, M.},
  \bibinfo{author}{{Baduel}, B.}, \bibinfo{author}{{Barbotin}, E.},
  \bibinfo{author}{{Barroy}, P.}, \bibinfo{author}{{Baudouin}, P.},
  \bibinfo{author}{{Berard}, L.}, \bibinfo{author}{{Berger}, N.},
  \bibinfo{author}{{Bernasconi}, L.}, \bibinfo{author}{{Bosch}, J.G.},
  \bibinfo{author}{{Bouley}, S.}, \bibinfo{author}{{Bozhinova}, I.},
  \bibinfo{author}{{Brinsfield}, J.}, \bibinfo{author}{{Brunetto}, L.},
  \bibinfo{author}{{Canaud}, G.}, \bibinfo{author}{{Caron}, J.},
  \bibinfo{author}{{Carrier}, F.}, \bibinfo{author}{{Casalnuovo}, G.},
  \bibinfo{author}{{Casulli}, S.}, \bibinfo{author}{{Cerda}, M.},
  \bibinfo{author}{{Chalamet}, L.}, \bibinfo{author}{{Charbonnel}, S.},
  \bibinfo{author}{{Chinaglia}, B.}, \bibinfo{author}{{Cikota}, A.},
  \bibinfo{author}{{Colas}, F.}, \bibinfo{author}{{Coliac}, J.F.},
  \bibinfo{author}{{Collet}, A.}, \bibinfo{author}{{Coloma}, J.},
  \bibinfo{author}{{Conjat}, M.}, \bibinfo{author}{{Conseil}, E.},
  \bibinfo{author}{{Costa}, R.}, \bibinfo{author}{{Crippa}, R.},
  \bibinfo{author}{{Cristofanelli}, M.}, \bibinfo{author}{{Damerdji}, Y.},
  \bibinfo{author}{{Deback{\`e}re}, A.}, \bibinfo{author}{{Decock}, A.},
  \bibinfo{author}{{D{\'e}hais}, Q.}, \bibinfo{author}{{D{\'e}l{\'e}age}, T.},
  \bibinfo{author}{{Delmelle}, S.}, \bibinfo{author}{{Demeautis}, C.},
  \bibinfo{author}{{Dr{\'o}{\.z}d{\.z}}, M.}, \bibinfo{author}{{Dubos}, G.},
  \bibinfo{author}{{Dulcamara}, T.}, \bibinfo{author}{{Dumont}, M.},
  \bibinfo{author}{{Durkee}, R.}, \bibinfo{author}{{Dymock}, R.},
  \bibinfo{author}{{Escalante del Valle}, A.}, \bibinfo{author}{{Esseiva}, N.},
  \bibinfo{author}{{Esseiva}, R.}, \bibinfo{author}{{Esteban}, M.},
  \bibinfo{author}{{Fauchez}, T.}, \bibinfo{author}{{Fauerbach}, M.},
  \bibinfo{author}{{Fauvaud}, M.}, \bibinfo{author}{{Fauvaud}, S.},
  \bibinfo{author}{{Forn{\'e}}, E.}, \bibinfo{author}{{Fournel}, C.},
  \bibinfo{author}{{Fradet}, D.}, \bibinfo{author}{{Garlitz}, J.},
  \bibinfo{author}{{Gerteis}, O.}, \bibinfo{author}{{Gillier}, C.},
  \bibinfo{author}{{Gillon}, M.}, \bibinfo{author}{{Giraud}, R.},
  \bibinfo{author}{{Godard}, J.P.}, \bibinfo{author}{{Goncalves}, R.},
  \bibinfo{author}{{Hamanowa}, H.}, \bibinfo{author}{{Hamanowa}, H.},
  \bibinfo{author}{{Hay}, K.}, \bibinfo{author}{{Hellmich}, S.},
  \bibinfo{author}{{Heterier}, S.}, \bibinfo{author}{{Higgins}, D.},
  \bibinfo{author}{{Hirsch}, R.}, \bibinfo{author}{{Hodosan}, G.},
  \bibinfo{author}{{Hren}, M.}, \bibinfo{author}{{Hygate}, A.},
  \bibinfo{author}{{Innocent}, N.}, \bibinfo{author}{{Jacquinot}, H.},
  \bibinfo{author}{{Jawahar}, S.}, \bibinfo{author}{{Jehin}, E.},
  \bibinfo{author}{{Jerosimic}, L.}, \bibinfo{author}{{Klotz}, A.},
  \bibinfo{author}{{Koff}, W.}, \bibinfo{author}{{Korlevic}, P.},
  \bibinfo{author}{{Kosturkiewicz}, E.}, \bibinfo{author}{{Krafft}, P.},
  \bibinfo{author}{{Krugly}, Y.}, \bibinfo{author}{{Kugel}, F.},
  \bibinfo{author}{{Labrevoir}, O.}, \bibinfo{author}{{Lecacheux}, J.},
  \bibinfo{author}{{Lehk{\'y}}, M.}, \bibinfo{author}{{Leroy}, A.},
  \bibinfo{author}{{Lesquerbault}, B.}, \bibinfo{author}{{Lopez-Gonzales},
  M.J.}, \bibinfo{author}{{Lutz}, M.}, \bibinfo{author}{{Mallecot}, B.},
  \bibinfo{author}{{Manfroid}, J.}, \bibinfo{author}{{Manzini}, F.},
  \bibinfo{author}{{Marciniak}, A.}, \bibinfo{author}{{Martin}, A.},
  \bibinfo{author}{{Modave}, B.}, \bibinfo{author}{{Montaigut}, R.},
  \bibinfo{author}{{Montier}, J.}, \bibinfo{author}{{Morelle}, E.},
  \bibinfo{author}{{Morton}, B.}, \bibinfo{author}{{Mottola}, S.},
  \bibinfo{author}{{Naves}, R.}, \bibinfo{author}{{Nomen}, J.},
  \bibinfo{author}{{Oey}, J.}, \bibinfo{author}{{Og{\l}oza}, W.},
  \bibinfo{author}{{Paiella}, M.}, \bibinfo{author}{{Pallares}, H.},
  \bibinfo{author}{{Peyrot}, A.}, \bibinfo{author}{{Pilcher}, F.},
  \bibinfo{author}{{Pirenne}, J.F.}, \bibinfo{author}{{Piron}, P.},
  \bibinfo{author}{{Poli{\'n}ska}, M.}, \bibinfo{author}{{Polotto}, M.},
  \bibinfo{author}{{Poncy}, R.}, \bibinfo{author}{{Previt}, J.P.},
  \bibinfo{author}{{Reignier}, F.}, \bibinfo{author}{{Renauld}, D.},
  \bibinfo{author}{{Ricci}, D.}, \bibinfo{author}{{Richard}, F.},
  \bibinfo{author}{{Rinner}, C.}, \bibinfo{author}{{Risoldi}, V.},
  \bibinfo{author}{{Robilliard}, D.}, \bibinfo{author}{{Romeuf}, D.},
  \bibinfo{author}{{Rousseau}, G.}, \bibinfo{author}{{Roy}, R.},
  \bibinfo{author}{{Ruthroff}, J.}, \bibinfo{author}{{Salom}, P.A.},
  \bibinfo{author}{{Salvador}, L.}, \bibinfo{author}{{Sanchez}, S.},
  \bibinfo{author}{{Santana-Ros}, T.}, \bibinfo{author}{{Scholz}, A.},
  \bibinfo{author}{{S{\'e}n{\'e}}, G.}, \bibinfo{author}{{Skiff}, B.},
  \bibinfo{author}{{Sobkowiak}, K.}, \bibinfo{author}{{Sogorb}, P.},
  \bibinfo{author}{{Sold{\'a}n}, F.}, \bibinfo{author}{{Spiridakis}, A.},
  \bibinfo{author}{{Splanska}, E.}, \bibinfo{author}{{Sposetti}, S.},
  \bibinfo{author}{{Starkey}, D.}, \bibinfo{author}{{Stephens}, R.},
  \bibinfo{author}{{Stiepen}, A.}, \bibinfo{author}{{Stoss}, R.},
  \bibinfo{author}{{Strajnic}, J.}, \bibinfo{author}{{Teng}, J.P.},
  \bibinfo{author}{{Tumolo}, G.}, \bibinfo{author}{{Vagnozzi}, A.},
  \bibinfo{author}{{Vanoutryve}, B.}, \bibinfo{author}{{Vugnon}, J.M.},
  \bibinfo{author}{{Warner}, B.D.}, \bibinfo{author}{{Waucomont}, M.},
  \bibinfo{author}{{Wertz}, O.}, \bibinfo{author}{{Winiarski}, M.},
  \bibinfo{author}{{Wolf}, M.}, \bibinfo{year}{2016}b.
\newblock \bibinfo{title}{{New and updated convex shape models of asteroids
  based on optical data from a large collaboration network}}.
\newblock \bibinfo{journal}{\aap} \bibinfo{volume}{586}, \bibinfo{pages}{A108}.
\newblock \eprint{1510.07422}.
\bibitem[{{Hanu\v{s}} et~al.(2013c){Hanu\v{s}}, {Marchis} and
  {\v{D}urech}}]{Hanus2013b}
\bibinfo{author}{{Hanu\v{s}}, J.}, \bibinfo{author}{{Marchis}, F.},
  \bibinfo{author}{{\v{D}urech}, J.}, \bibinfo{year}{2013}c.
\newblock \bibinfo{title}{{Sizes of main-belt asteroids by combining shape
  models and Keck adaptive optics observations}}.
\newblock \bibinfo{journal}{\icarus} \bibinfo{volume}{226},
  \bibinfo{pages}{1045--1057}.
\bibitem[{{Hanu\v{s}} et~al.(2017){Hanu\v{s}}, {Marchis}, {Viikinkoski}, {Yang}
  and {Kaasalainen}}]{Hanus2017a}
\bibinfo{author}{{Hanu\v{s}}, J.}, \bibinfo{author}{{Marchis}, F.},
  \bibinfo{author}{{Viikinkoski}, M.}, \bibinfo{author}{{Yang}, B.},
  \bibinfo{author}{{Kaasalainen}, M.}, \bibinfo{year}{2017}.
\newblock \bibinfo{title}{{Shape model of asteroid (130) Elektra from optical
  photometry and disk-resolved images from VLT/SPHERE and Nirc2/Keck}}.
\newblock \bibinfo{journal}{\aap} \bibinfo{volume}{599}, \bibinfo{pages}{A36}.
\newblock \eprint{1611.03632}.
\bibitem[{{Harris} and {Young}(1980)}]{Harris1980}
\bibinfo{author}{{Harris}, A.W.}, \bibinfo{author}{{Young}, J.W.},
  \bibinfo{year}{1980}.
\newblock \bibinfo{title}{{Asteroid rotation. III - 1978 observations}}.
\newblock \bibinfo{journal}{Icarus} \bibinfo{volume}{43},
  \bibinfo{pages}{20--32}.
\bibitem[{{Harris} and {Young}(1983)}]{Harris1983}
\bibinfo{author}{{Harris}, A.W.}, \bibinfo{author}{{Young}, J.W.},
  \bibinfo{year}{1983}.
\newblock \bibinfo{title}{{Asteroid rotation. IV}}.
\newblock \bibinfo{journal}{\icarus} \bibinfo{volume}{54},
  \bibinfo{pages}{59--109}.
\bibitem[{{Jacobson} et~al.(2014){Jacobson}, {Marzari}, {Rossi}, {Scheeres} and
  {Davis}}]{Jacobson2014b}
\bibinfo{author}{{Jacobson}, S.A.}, \bibinfo{author}{{Marzari}, F.},
  \bibinfo{author}{{Rossi}, A.}, \bibinfo{author}{{Scheeres}, D.J.},
  \bibinfo{author}{{Davis}, D.R.}, \bibinfo{year}{2014}.
\newblock \bibinfo{title}{{Effect of rotational disruption on the
  size-frequency distribution of the Main Belt asteroid population}}.
\newblock \bibinfo{journal}{\mnras} \bibinfo{volume}{439},
  \bibinfo{pages}{L95--L99}.
\newblock \eprint{1401.1813}.
\bibitem[{{Kaasalainen} and {Torppa}(2001)}]{Kaasalainen2001a}
\bibinfo{author}{{Kaasalainen}, M.}, \bibinfo{author}{{Torppa}, J.},
  \bibinfo{year}{2001}.
\newblock \bibinfo{title}{{Optimization Methods for Asteroid Lightcurve
  Inversion. I. Shape Determination}}.
\newblock \bibinfo{journal}{Icarus} \bibinfo{volume}{153},
  \bibinfo{pages}{24--36}.
\bibitem[{{Kaasalainen} et~al.(2001){Kaasalainen}, {Torppa} and
  {Muinonen}}]{Kaasalainen2001b}
\bibinfo{author}{{Kaasalainen}, M.}, \bibinfo{author}{{Torppa}, J.},
  \bibinfo{author}{{Muinonen}, K.}, \bibinfo{year}{2001}.
\newblock \bibinfo{title}{{Optimization Methods for Asteroid Lightcurve
  Inversion. II. The Complete Inverse Problem}}.
\newblock \bibinfo{journal}{Icarus} \bibinfo{volume}{153},
  \bibinfo{pages}{37--51}.
\bibitem[{{Kaasalainen} et~al.(2002){Kaasalainen}, {Torppa} and
  {Piironen}}]{Kaasalainen2002c}
\bibinfo{author}{{Kaasalainen}, M.}, \bibinfo{author}{{Torppa}, J.},
  \bibinfo{author}{{Piironen}, J.}, \bibinfo{year}{2002}.
\newblock \bibinfo{title}{{Binary structures among large asteroids}}.
\newblock \bibinfo{journal}{\aap} \bibinfo{volume}{383},
  \bibinfo{pages}{L19--L22}.
\bibitem[{{Kaasalainen} et~al.(2007){Kaasalainen}, {{\v D}urech}, {Warner},
  {Krugly} and {Gaftonyuk}}]{Kaasalainen2007b}
\bibinfo{author}{{Kaasalainen}, M.}, \bibinfo{author}{{{\v D}urech}, J.},
  \bibinfo{author}{{Warner}, B.D.}, \bibinfo{author}{{Krugly}, Y.N.},
  \bibinfo{author}{{Gaftonyuk}, N.M.}, \bibinfo{year}{2007}.
\newblock \bibinfo{title}{{Acceleration of the rotation of asteroid 1862 Apollo
  by radiation torques}}.
\newblock \bibinfo{journal}{\nat} \bibinfo{volume}{446},
  \bibinfo{pages}{420--422}.
\bibitem[{{Kim} et~al.(2014){Kim}, {Choi}, {Moon}, {Byun}, {Brosch}, {Kaplan},
  {Kaynar}, {Uysal}, {G{\"u}zel}, {Behrend}, {Yoon}, {Mottola}, {Hellmich},
  {Hinse}, {Eker} and {Park}}]{Kim2014a}
\bibinfo{author}{{Kim}, M.J.}, \bibinfo{author}{{Choi}, Y.J.},
  \bibinfo{author}{{Moon}, H.K.}, \bibinfo{author}{{Byun}, Y.I.},
  \bibinfo{author}{{Brosch}, N.}, \bibinfo{author}{{Kaplan}, M.},
  \bibinfo{author}{{Kaynar}, S.}, \bibinfo{author}{{Uysal}, {\"O}.},
  \bibinfo{author}{{G{\"u}zel}, E.}, \bibinfo{author}{{Behrend}, R.},
  \bibinfo{author}{{Yoon}, J.N.}, \bibinfo{author}{{Mottola}, S.},
  \bibinfo{author}{{Hellmich}, S.}, \bibinfo{author}{{Hinse}, T.C.},
  \bibinfo{author}{{Eker}, Z.}, \bibinfo{author}{{Park}, J.H.},
  \bibinfo{year}{2014}.
\newblock \bibinfo{title}{{Rotational Properties of the Maria Asteroid
  Family}}.
\newblock \bibinfo{journal}{\aj} \bibinfo{volume}{147}, \bibinfo{pages}{56}.
\newblock \eprint{1311.5318}.
\bibitem[{{Kryszczy{\'n}ska}(2013)}]{Kryszczynska2013a}
\bibinfo{author}{{Kryszczy{\'n}ska}, A.}, \bibinfo{year}{2013}.
\newblock \bibinfo{title}{{Do Slivan states exist in the Flora family? . II.
  Fingerprints of the Yarkovsky and YORP effects}}.
\newblock \bibinfo{journal}{\aap} \bibinfo{volume}{551}, \bibinfo{pages}{A102}.
\bibitem[{{Mainzer} et~al.(2011){Mainzer}, {Bauer}, {Grav}, {Masiero}, {Cutri},
  {Dailey}, {Eisenhardt}, {McMillan}, {Wright}, {Walker}, {Jedicke}, {Spahr},
  {Tholen}, {Alles}, {Beck}, {Brandenburg}, {Conrow}, {Evans}, {Fowler},
  {Jarrett}, {Marsh}, {Masci}, {McCallon}, {Wheelock}, {Wittman}, {Wyatt},
  {DeBaun}, {Elliott}, {Elsbury}, {Gautier}, {Gomillion}, {Leisawitz},
  {Maleszewski}, {Micheli} and {Wilkins}}]{Mainzer2011a}
\bibinfo{author}{{Mainzer}, A.}, \bibinfo{author}{{Bauer}, J.},
  \bibinfo{author}{{Grav}, T.}, \bibinfo{author}{{Masiero}, J.},
  \bibinfo{author}{{Cutri}, R.M.}, \bibinfo{author}{{Dailey}, J.},
  \bibinfo{author}{{Eisenhardt}, P.}, \bibinfo{author}{{McMillan}, R.S.},
  \bibinfo{author}{{Wright}, E.}, \bibinfo{author}{{Walker}, R.},
  \bibinfo{author}{{Jedicke}, R.}, \bibinfo{author}{{Spahr}, T.},
  \bibinfo{author}{{Tholen}, D.}, \bibinfo{author}{{Alles}, R.},
  \bibinfo{author}{{Beck}, R.}, \bibinfo{author}{{Brandenburg}, H.},
  \bibinfo{author}{{Conrow}, T.}, \bibinfo{author}{{Evans}, T.},
  \bibinfo{author}{{Fowler}, J.}, \bibinfo{author}{{Jarrett}, T.},
  \bibinfo{author}{{Marsh}, K.}, \bibinfo{author}{{Masci}, F.},
  \bibinfo{author}{{McCallon}, H.}, \bibinfo{author}{{Wheelock}, S.},
  \bibinfo{author}{{Wittman}, M.}, \bibinfo{author}{{Wyatt}, P.},
  \bibinfo{author}{{DeBaun}, E.}, \bibinfo{author}{{Elliott}, G.},
  \bibinfo{author}{{Elsbury}, D.}, \bibinfo{author}{{Gautier}, I.T.},
  \bibinfo{author}{{Gomillion}, S.}, \bibinfo{author}{{Leisawitz}, D.},
  \bibinfo{author}{{Maleszewski}, C.}, \bibinfo{author}{{Micheli}, M.},
  \bibinfo{author}{{Wilkins}, A.}, \bibinfo{year}{2011}.
\newblock \bibinfo{title}{{Preliminary Results from NEOWISE: An Enhancement to
  the Wide-field Infrared Survey Explorer for Solar System Science}}.
\newblock \bibinfo{journal}{\apj} \bibinfo{volume}{731}, \bibinfo{pages}{53}.
\newblock \eprint{1102.1996}.
\bibitem[{{Marciniak} et~al.(2009a){Marciniak}, {Micha{\l}owski}, {Hirsch},
  {Behrend}, {Bernasconi}, {Descamps}, {Colas}, {Sobkowiak}, {Kami{\'n}ski},
  {Kryszczy{\'n}ska}, {Kwiatkowski}, {Poli{\'n}ska}, {Rudawska}, {Fauvaud},
  {Santacana}, {Bruno}, {Fauvaud}, {Teng-Chuen-Yu} and
  {Peyrot}}]{Marciniak2009b}
\bibinfo{author}{{Marciniak}, A.}, \bibinfo{author}{{Micha{\l}owski}, T.},
  \bibinfo{author}{{Hirsch}, R.}, \bibinfo{author}{{Behrend}, R.},
  \bibinfo{author}{{Bernasconi}, L.}, \bibinfo{author}{{Descamps}, P.},
  \bibinfo{author}{{Colas}, F.}, \bibinfo{author}{{Sobkowiak}, K.},
  \bibinfo{author}{{Kami{\'n}ski}, K.}, \bibinfo{author}{{Kryszczy{\'n}ska},
  A.}, \bibinfo{author}{{Kwiatkowski}, T.}, \bibinfo{author}{{Poli{\'n}ska},
  M.}, \bibinfo{author}{{Rudawska}, R.}, \bibinfo{author}{{Fauvaud}, S.},
  \bibinfo{author}{{Santacana}, G.}, \bibinfo{author}{{Bruno}, A.},
  \bibinfo{author}{{Fauvaud}, M.}, \bibinfo{author}{{Teng-Chuen-Yu}, J.P.},
  \bibinfo{author}{{Peyrot}, A.}, \bibinfo{year}{2009}a.
\newblock \bibinfo{title}{{Photometry and models of selected main belt
  asteroids. VII. 350 Ornamenta, 771 Libera, and 984 Gretia}}.
\newblock \bibinfo{journal}{\aap} \bibinfo{volume}{508},
  \bibinfo{pages}{1503--1507}.
\bibitem[{{Marciniak} et~al.(2009b){Marciniak}, {Micha{\l}owski}, {Hirsch},
  {Poli{\'n}ska}, {Kami{\'n}ski}, {Kwiatkowski}, {Kryszczy{\'n}ska}, {Behrend},
  {Bernasconi}, {Micha{\l}owski}, {Starczewski}, {Fagas} and
  {Sobkowiak}}]{Marciniak2009a}
\bibinfo{author}{{Marciniak}, A.}, \bibinfo{author}{{Micha{\l}owski}, T.},
  \bibinfo{author}{{Hirsch}, R.}, \bibinfo{author}{{Poli{\'n}ska}, M.},
  \bibinfo{author}{{Kami{\'n}ski}, K.}, \bibinfo{author}{{Kwiatkowski}, T.},
  \bibinfo{author}{{Kryszczy{\'n}ska}, A.}, \bibinfo{author}{{Behrend}, R.},
  \bibinfo{author}{{Bernasconi}, L.}, \bibinfo{author}{{Micha{\l}owski}, J.},
  \bibinfo{author}{{Starczewski}, S.}, \bibinfo{author}{{Fagas}, M.},
  \bibinfo{author}{{Sobkowiak}, K.}, \bibinfo{year}{2009}b.
\newblock \bibinfo{title}{{Photometry and models of selected main belt
  asteroids. VI. 160 Una, 747 Winchester, and 849 Ara}}.
\newblock \bibinfo{journal}{\aap} \bibinfo{volume}{498},
  \bibinfo{pages}{313--320}.
\bibitem[{{Marsset} et~al.(2017){Marsset}, {Carry}, {Dumas}, {Hanus},
  {Viikinkoski}, {Vernazza}, {M{\"u}ller}, {Delbo}, {Jehin}, {Gillon}, {Grice},
  {Yang}, {Fusco}, {Berthier}, {Sonnett}, {Kugel}, {Caron} and
  {Behrend}}]{Marsset2017}
\bibinfo{author}{{Marsset}, M.}, \bibinfo{author}{{Carry}, B.},
  \bibinfo{author}{{Dumas}, C.}, \bibinfo{author}{{Hanus}, J.},
  \bibinfo{author}{{Viikinkoski}, M.}, \bibinfo{author}{{Vernazza}, P.},
  \bibinfo{author}{{M{\"u}ller}, T.G.}, \bibinfo{author}{{Delbo}, M.},
  \bibinfo{author}{{Jehin}, E.}, \bibinfo{author}{{Gillon}, M.},
  \bibinfo{author}{{Grice}, J.}, \bibinfo{author}{{Yang}, B.},
  \bibinfo{author}{{Fusco}, T.}, \bibinfo{author}{{Berthier}, J.},
  \bibinfo{author}{{Sonnett}, S.}, \bibinfo{author}{{Kugel}, F.},
  \bibinfo{author}{{Caron}, J.}, \bibinfo{author}{{Behrend}, R.},
  \bibinfo{year}{2017}.
\newblock \bibinfo{title}{{3D shape of asteroid (6)\~{}Hebe from VLT/SPHERE
  imaging: Implications for the origin of ordinary H chondrites}}.
\newblock \bibinfo{journal}{ArXiv e-prints} \eprint{1705.10515}.
\bibitem[{{Masiero} et~al.(2013){Masiero}, {Mainzer}, {Bauer}, {Grav}, {Nugent}
  and {Stevenson}}]{Masiero2013}
\bibinfo{author}{{Masiero}, J.R.}, \bibinfo{author}{{Mainzer}, A.K.},
  \bibinfo{author}{{Bauer}, J.M.}, \bibinfo{author}{{Grav}, T.},
  \bibinfo{author}{{Nugent}, C.R.}, \bibinfo{author}{{Stevenson}, R.},
  \bibinfo{year}{2013}.
\newblock \bibinfo{title}{{Asteroid Family Identification Using the
  Hierarchical Clustering Method and WISE/NEOWISE Physical Properties}}.
\newblock \bibinfo{journal}{\apj} \bibinfo{volume}{770}, \bibinfo{pages}{7}.
\newblock \eprint{1305.1607}.
\bibitem[{{Masiero} et~al.(2011){Masiero}, {Mainzer}, {Grav}, {Bauer}, {Cutri},
  {Dailey}, {Eisenhardt}, {McMillan}, {Spahr}, {Skrutskie}, {Tholen}, {Walker},
  {Wright}, {DeBaun}, {Elsbury}, {Gautier}, {Gomillion} and
  {Wilkins}}]{Masiero2011}
\bibinfo{author}{{Masiero}, J.R.}, \bibinfo{author}{{Mainzer}, A.K.},
  \bibinfo{author}{{Grav}, T.}, \bibinfo{author}{{Bauer}, J.M.},
  \bibinfo{author}{{Cutri}, R.M.}, \bibinfo{author}{{Dailey}, J.},
  \bibinfo{author}{{Eisenhardt}, P.R.M.}, \bibinfo{author}{{McMillan}, R.S.},
  \bibinfo{author}{{Spahr}, T.B.}, \bibinfo{author}{{Skrutskie}, M.F.},
  \bibinfo{author}{{Tholen}, D.}, \bibinfo{author}{{Walker}, R.G.},
  \bibinfo{author}{{Wright}, E.L.}, \bibinfo{author}{{DeBaun}, E.},
  \bibinfo{author}{{Elsbury}, D.}, \bibinfo{author}{{Gautier}, I.T.},
  \bibinfo{author}{{Gomillion}, S.}, \bibinfo{author}{{Wilkins}, A.},
  \bibinfo{year}{2011}.
\newblock \bibinfo{title}{{Main Belt Asteroids with WISE/NEOWISE. I.
  Preliminary Albedos and Diameters}}.
\newblock \bibinfo{journal}{\apj} \bibinfo{volume}{741}, \bibinfo{pages}{68}.
\newblock \eprint{1109.4096}.
\bibitem[{{Micha{\l}owski} et~al.(2005){Micha{\l}owski}, {Kaasalainen},
  {Marciniak}, {Denchev}, {Kwiatkowski}, {Kryszczy{\'n}ska}, {Hirsch},
  {Velichko}, {Erikson}, {Szab{\'o}} and {Kowalski}}]{Michalowski2005}
\bibinfo{author}{{Micha{\l}owski}, T.}, \bibinfo{author}{{Kaasalainen}, M.},
  \bibinfo{author}{{Marciniak}, A.}, \bibinfo{author}{{Denchev}, P.},
  \bibinfo{author}{{Kwiatkowski}, T.}, \bibinfo{author}{{Kryszczy{\'n}ska},
  A.}, \bibinfo{author}{{Hirsch}, R.}, \bibinfo{author}{{Velichko}, F.P.},
  \bibinfo{author}{{Erikson}, A.}, \bibinfo{author}{{Szab{\'o}}, G.M.},
  \bibinfo{author}{{Kowalski}, R.}, \bibinfo{year}{2005}.
\newblock \bibinfo{title}{{Photometry and models of selected main belt
  asteroids. II. 173 Ino, 376 Geometria, and 451 Patientia}}.
\newblock \bibinfo{journal}{\aap} \bibinfo{volume}{443},
  \bibinfo{pages}{329--335}.
\bibitem[{{Micha{\l}owski} et~al.(2006){Micha{\l}owski}, {Kaasalainen},
  {Poli{\'n}ska}, {Marciniak}, {Kwiatkowski}, {Kryszczy{\'n}ska} and
  {Velichko}}]{Michalowski2006}
\bibinfo{author}{{Micha{\l}owski}, T.}, \bibinfo{author}{{Kaasalainen}, M.},
  \bibinfo{author}{{Poli{\'n}ska}, M.}, \bibinfo{author}{{Marciniak}, A.},
  \bibinfo{author}{{Kwiatkowski}, T.}, \bibinfo{author}{{Kryszczy{\'n}ska},
  A.}, \bibinfo{author}{{Velichko}, F.P.}, \bibinfo{year}{2006}.
\newblock \bibinfo{title}{{Photometry and models of selected main belt
  asteroids. III. 283 Emma, 665 Sabine, and 690 Wratislavia}}.
\newblock \bibinfo{journal}{\aap} \bibinfo{volume}{459},
  \bibinfo{pages}{663--668}.
\bibitem[{{M{\"u}ller} et~al.(2011){M{\"u}ller}, {\v{D}urech}, {Hasegawa},
  {Abe}, {Kawakami}, {Kasuga}, {Kinoshita}, {Kuroda}, {Urakawa}, {Okumura},
  {Sarugaku}, {Miyasaka}, {Takagi}, {Weissman}, {Choi}, {Larson}, {Yanagisawa}
  and {Nagayama}}]{Muller2011}
\bibinfo{author}{{M{\"u}ller}, T.G.}, \bibinfo{author}{{\v{D}urech}, J.},
  \bibinfo{author}{{Hasegawa}, S.}, \bibinfo{author}{{Abe}, M.},
  \bibinfo{author}{{Kawakami}, K.}, \bibinfo{author}{{Kasuga}, T.},
  \bibinfo{author}{{Kinoshita}, D.}, \bibinfo{author}{{Kuroda}, D.},
  \bibinfo{author}{{Urakawa}, S.}, \bibinfo{author}{{Okumura}, S.},
  \bibinfo{author}{{Sarugaku}, Y.}, \bibinfo{author}{{Miyasaka}, S.},
  \bibinfo{author}{{Takagi}, Y.}, \bibinfo{author}{{Weissman}, P.R.},
  \bibinfo{author}{{Choi}, Y.J.}, \bibinfo{author}{{Larson}, S.},
  \bibinfo{author}{{Yanagisawa}, K.}, \bibinfo{author}{{Nagayama}, S.},
  \bibinfo{year}{2011}.
\newblock \bibinfo{title}{{Thermo-physical properties of 162173 (1999 JU3), a
  potential flyby and rendezvous target for interplanetary missions}}.
\newblock \bibinfo{journal}{\aap} \bibinfo{volume}{525}, \bibinfo{pages}{A145}.
\newblock \eprint{1011.5029}.
\bibitem[{{M{\"u}ller} et~al.(2014){M{\"u}ller}, {Hasegawa} and
  {Usui}}]{Muller2014}
\bibinfo{author}{{M{\"u}ller}, T.G.}, \bibinfo{author}{{Hasegawa}, S.},
  \bibinfo{author}{{Usui}, F.}, \bibinfo{year}{2014}.
\newblock \bibinfo{title}{{(25143) Itokawa: The power of radiometric techniques
  for the interpretation of remote thermal observations in the light of the
  Hayabusa rendezvous results*}}.
\newblock \bibinfo{journal}{\pasj} \bibinfo{volume}{66}, \bibinfo{pages}{52}.
\newblock \eprint{1404.5842}.
\bibitem[{{Nesvorn{\'y}}(2012)}]{Nesvorny2012}
\bibinfo{author}{{Nesvorn{\'y}}, D.}, \bibinfo{year}{2012}.
\newblock \bibinfo{title}{{Nesvorny HCM Asteroid Families V2.0}}.
\newblock \bibinfo{journal}{NASA Planetary Data System} \bibinfo{volume}{189}.
\bibitem[{{Nesvorn{\'y}} et~al.(2015){Nesvorn{\'y}}, {Bro{\v z}} and
  {Carruba}}]{Nesvorny2015}
\bibinfo{author}{{Nesvorn{\'y}}, D.}, \bibinfo{author}{{Bro{\v z}}, M.},
  \bibinfo{author}{{Carruba}, V.}, \bibinfo{year}{2015}.
\newblock \bibinfo{title}{{Identification and Dynamical Properties of Asteroid
  Families}}.
\newblock pp. \bibinfo{pages}{297--321}.
\bibitem[{{Oey} and {Krajewski}(2008)}]{Oey2008}
\bibinfo{author}{{Oey}, J.}, \bibinfo{author}{{Krajewski}, R.},
  \bibinfo{year}{2008}.
\newblock \bibinfo{title}{{Lightcurve Analysis of Asteroids from Kingsgrove and
  Other Collaborating Observatories in the First Half of 2007}}.
\newblock \bibinfo{journal}{Minor Planet Bulletin} \bibinfo{volume}{35},
  \bibinfo{pages}{47--48}.
\bibitem[{{Oszkiewicz} et~al.(2011){Oszkiewicz}, {Muinonen}, {Bowell},
  {Trilling}, {Penttil{\"a}}, {Pieniluoma}, {Wasserman} and
  {Enga}}]{Oszkiewicz2011}
\bibinfo{author}{{Oszkiewicz}, D.A.}, \bibinfo{author}{{Muinonen}, K.},
  \bibinfo{author}{{Bowell}, E.}, \bibinfo{author}{{Trilling}, D.},
  \bibinfo{author}{{Penttil{\"a}}, A.}, \bibinfo{author}{{Pieniluoma}, T.},
  \bibinfo{author}{{Wasserman}, L.H.}, \bibinfo{author}{{Enga}, M.T.},
  \bibinfo{year}{2011}.
\newblock \bibinfo{title}{{Online multi-parameter phase-curve fitting and
  application to a large corpus of asteroid photometric data}}.
\newblock \bibinfo{journal}{Journal of Quantitative Spectroscopy \& Radiative
  Transfer} \bibinfo{volume}{112}, \bibinfo{pages}{1919--1929}.
\bibitem[{{Owings}(2010)}]{Owings2010}
\bibinfo{author}{{Owings}, L.E.}, \bibinfo{year}{2010}.
\newblock \bibinfo{title}{{Lightcurves for 890 Waltraut, 3162 Nostalgia, and
  6867 Kuwano}}.
\newblock \bibinfo{journal}{Minor Planet Bulletin} \bibinfo{volume}{37},
  \bibinfo{pages}{138}.
\bibitem[{{Piironen} et~al.(2001){Piironen}, {Lagerkvist}, {Torppa},
  {Kaasalainen} and {Warner}}]{Piironen2001}
\bibinfo{author}{{Piironen}, J.}, \bibinfo{author}{{Lagerkvist}, C.},
  \bibinfo{author}{{Torppa}, J.}, \bibinfo{author}{{Kaasalainen}, M.},
  \bibinfo{author}{{Warner}, B.}, \bibinfo{year}{2001}.
\newblock \bibinfo{title}{{Standard Asteroid Photometric Catalogue}}, in:
  \bibinfo{booktitle}{{Bulletin of the American Astronomical Society}}, p.
  \bibinfo{pages}{1562}.
\bibitem[{{Pilcher}(2013)}]{Pilcher2013e}
\bibinfo{author}{{Pilcher}, F.}, \bibinfo{year}{2013}.
\newblock \bibinfo{title}{{Rotation Period Determinations for 102 Miriam, 108
  Hecuba, 221 Eos 225 Oppavia, and 745 Mauritia, and a Note on 871 Amneris}}.
\newblock \bibinfo{journal}{Minor Planet Bulletin} \bibinfo{volume}{40},
  \bibinfo{pages}{158--160}.
\bibitem[{{Pravec} and {Harris}(2000)}]{Pravec2000}
\bibinfo{author}{{Pravec}, P.}, \bibinfo{author}{{Harris}, A.W.},
  \bibinfo{year}{2000}.
\newblock \bibinfo{title}{{Fast and Slow Rotation of Asteroids}}.
\newblock \bibinfo{journal}{\icarus} \bibinfo{volume}{148},
  \bibinfo{pages}{12--20}.
\bibitem[{{Pravec} et~al.(2008){Pravec}, {Harris}, {Vokrouhlick{\'y}},
  {Warner}, {Ku\v{s}nir{\'a}k}, {Hornoch}, {Pray}, {Higgins}, {Oey},
  {Gal{\'a}d}, \v{S}. {Gajdo\v{s}}, {Korno\v{s}}, {Vil{\'a}gi}, {Hus{\'a}rik},
  {Krugly}, {Shevchenko}, {Chiorny}, {Gaftonyuk}, {Cooney}, {Gross}, {Terrell},
  {Stephens}, {Dyvig}, {Reddy}, {Ries}, {Colas}, {Lecacheux}, {Durkee}, {Masi},
  {Koff} and {Goncalves}}]{Pravec2008}
\bibinfo{author}{{Pravec}, P.}, \bibinfo{author}{{Harris}, A.W.},
  \bibinfo{author}{{Vokrouhlick{\'y}}, D.}, \bibinfo{author}{{Warner}, B.D.},
  \bibinfo{author}{{Ku\v{s}nir{\'a}k}, P.}, \bibinfo{author}{{Hornoch}, K.},
  \bibinfo{author}{{Pray}, D.P.}, \bibinfo{author}{{Higgins}, D.},
  \bibinfo{author}{{Oey}, J.}, \bibinfo{author}{{Gal{\'a}d}, A.},
  \bibinfo{author}{\v{S}. {Gajdo\v{s}}}, \bibinfo{author}{{Korno\v{s}}, L.},
  \bibinfo{author}{{Vil{\'a}gi}, J.}, \bibinfo{author}{{Hus{\'a}rik}, M.},
  \bibinfo{author}{{Krugly}, Y.N.}, \bibinfo{author}{{Shevchenko}, V.},
  \bibinfo{author}{{Chiorny}, V.}, \bibinfo{author}{{Gaftonyuk}, N.},
  \bibinfo{author}{{Cooney}, W.R.}, \bibinfo{author}{{Gross}, J.},
  \bibinfo{author}{{Terrell}, D.}, \bibinfo{author}{{Stephens}, R.D.},
  \bibinfo{author}{{Dyvig}, R.}, \bibinfo{author}{{Reddy}, V.},
  \bibinfo{author}{{Ries}, J.G.}, \bibinfo{author}{{Colas}, F.},
  \bibinfo{author}{{Lecacheux}, J.}, \bibinfo{author}{{Durkee}, R.},
  \bibinfo{author}{{Masi}, G.}, \bibinfo{author}{{Koff}, R.A.},
  \bibinfo{author}{{Goncalves}, R.}, \bibinfo{year}{2008}.
\newblock \bibinfo{title}{{Spin rate distribution of small asteroids}}.
\newblock \bibinfo{journal}{\icarus} \bibinfo{volume}{197},
  \bibinfo{pages}{497--504}.
\bibitem[{{Pravec} et~al.(2014){Pravec}, {Scheirich}, {{\v D}urech}, {Pollock},
  {Ku{\v s}nir{\'a}k}, {Hornoch}, {Gal{\'a}d}, {Vokrouhlick{\'y}}, {Harris},
  {Jehin}, {Manfroid}, {Opitom}, {Gillon}, {Colas}, {Oey}, {Vra{\v s}til},
  {Reichart}, {Ivarsen}, {Haislip} and {LaCluyze}}]{Pravec2014}
\bibinfo{author}{{Pravec}, P.}, \bibinfo{author}{{Scheirich}, P.},
  \bibinfo{author}{{{\v D}urech}, J.}, \bibinfo{author}{{Pollock}, J.},
  \bibinfo{author}{{Ku{\v s}nir{\'a}k}, P.}, \bibinfo{author}{{Hornoch}, K.},
  \bibinfo{author}{{Gal{\'a}d}, A.}, \bibinfo{author}{{Vokrouhlick{\'y}}, D.},
  \bibinfo{author}{{Harris}, A.W.}, \bibinfo{author}{{Jehin}, E.},
  \bibinfo{author}{{Manfroid}, J.}, \bibinfo{author}{{Opitom}, C.},
  \bibinfo{author}{{Gillon}, M.}, \bibinfo{author}{{Colas}, F.},
  \bibinfo{author}{{Oey}, J.}, \bibinfo{author}{{Vra{\v s}til}, J.},
  \bibinfo{author}{{Reichart}, D.}, \bibinfo{author}{{Ivarsen}, K.},
  \bibinfo{author}{{Haislip}, J.}, \bibinfo{author}{{LaCluyze}, A.},
  \bibinfo{year}{2014}.
\newblock \bibinfo{title}{{The tumbling spin state of (99942) Apophis}}.
\newblock \bibinfo{journal}{\icarus} \bibinfo{volume}{233},
  \bibinfo{pages}{48--60}.
\bibitem[{{Rozitis} and {Green}(2014)}]{Rozitis2014}
\bibinfo{author}{{Rozitis}, B.}, \bibinfo{author}{{Green}, S.F.},
  \bibinfo{year}{2014}.
\newblock \bibinfo{title}{{Physical characterisation of near-Earth asteroid
  (1620) Geographos. Reconciling radar and thermal-infrared observations}}.
\newblock \bibinfo{journal}{\aap} \bibinfo{volume}{568}, \bibinfo{pages}{A43}.
\newblock \eprint{1407.2127}.
\bibitem[{{Shepard} et~al.(2017){Shepard}, {Richardson}, {Taylor},
  {Rodriguez-Ford}, {Conrad}, {de Pater}, {Adamkovics}, {de Kleer}, {Males},
  {Morzinski}, {Close}, {Kaasalainen}, {Viikinkoski}, {Timerson}, {Reddy},
  {Magri}, {Nolan}, {Howell}, {Benner}, {Giorgini}, {Warner} and
  {Harris}}]{Shepard2017}
\bibinfo{author}{{Shepard}, M.K.}, \bibinfo{author}{{Richardson}, J.},
  \bibinfo{author}{{Taylor}, P.A.}, \bibinfo{author}{{Rodriguez-Ford}, L.A.},
  \bibinfo{author}{{Conrad}, A.}, \bibinfo{author}{{de Pater}, I.},
  \bibinfo{author}{{Adamkovics}, M.}, \bibinfo{author}{{de Kleer}, K.},
  \bibinfo{author}{{Males}, J.R.}, \bibinfo{author}{{Morzinski}, K.M.},
  \bibinfo{author}{{Close}, L.M.}, \bibinfo{author}{{Kaasalainen}, M.},
  \bibinfo{author}{{Viikinkoski}, M.}, \bibinfo{author}{{Timerson}, B.},
  \bibinfo{author}{{Reddy}, V.}, \bibinfo{author}{{Magri}, C.},
  \bibinfo{author}{{Nolan}, M.C.}, \bibinfo{author}{{Howell}, E.S.},
  \bibinfo{author}{{Benner}, L.A.M.}, \bibinfo{author}{{Giorgini}, J.D.},
  \bibinfo{author}{{Warner}, B.D.}, \bibinfo{author}{{Harris}, A.W.},
  \bibinfo{year}{2017}.
\newblock \bibinfo{title}{{Radar observations and shape model of asteroid 16
  Psyche}}.
\newblock \bibinfo{journal}{\icarus} \bibinfo{volume}{281},
  \bibinfo{pages}{388--403}.
\bibitem[{{Slivan}(2002)}]{Slivan2002}
\bibinfo{author}{{Slivan}, S.M.}, \bibinfo{year}{2002}.
\newblock \bibinfo{title}{{Spin vector alignment of Koronis family asteroids}}.
\newblock \bibinfo{journal}{Nature} \bibinfo{volume}{419},
  \bibinfo{pages}{49--51}.
\bibitem[{{Slivan} et~al.(2003){Slivan}, {Binzel}, {Crespo da Silva},
  {Kaasalainen}, {Lyndaker} and {Kr\v{c}o}}]{Slivan2003}
\bibinfo{author}{{Slivan}, S.M.}, \bibinfo{author}{{Binzel}, R.P.},
  \bibinfo{author}{{Crespo da Silva}, L.D.}, \bibinfo{author}{{Kaasalainen},
  M.}, \bibinfo{author}{{Lyndaker}, M.M.}, \bibinfo{author}{{Kr\v{c}o}, M.},
  \bibinfo{year}{2003}.
\newblock \bibinfo{title}{{Spin vectors in the Koronis family: comprehensive
  results from two independent analyses of 213 rotation lightcurves}}.
\newblock \bibinfo{journal}{Icarus} \bibinfo{volume}{162},
  \bibinfo{pages}{285--307}.
\bibitem[{{Slivan} et~al.(2009){Slivan}, {Binzel}, {Kaasalainen}, {Hock},
  {Klesman}, {Eckelman} and {Stephens}}]{Slivan2009}
\bibinfo{author}{{Slivan}, S.M.}, \bibinfo{author}{{Binzel}, R.P.},
  \bibinfo{author}{{Kaasalainen}, M.}, \bibinfo{author}{{Hock}, A.N.},
  \bibinfo{author}{{Klesman}, A.J.}, \bibinfo{author}{{Eckelman}, L.J.},
  \bibinfo{author}{{Stephens}, R.D.}, \bibinfo{year}{2009}.
\newblock \bibinfo{title}{{Spin vectors in the Koronis family. II. Additional
  clustered spins, and one stray}}.
\newblock \bibinfo{journal}{Icarus} \bibinfo{volume}{200},
  \bibinfo{pages}{514--530}.
\bibitem[{{Spoto} et~al.(2015){Spoto}, {Milani} and {Kne{\v
  z}evi{\'c}}}]{Spoto2015}
\bibinfo{author}{{Spoto}, F.}, \bibinfo{author}{{Milani}, A.},
  \bibinfo{author}{{Kne{\v z}evi{\'c}}, Z.}, \bibinfo{year}{2015}.
\newblock \bibinfo{title}{{Asteroid family ages}}.
\newblock \bibinfo{journal}{\icarus} \bibinfo{volume}{257},
  \bibinfo{pages}{275--289}.
\newblock \eprint{1504.05461}.
\bibitem[{{Taylor} et~al.(2007){Taylor}, {Margot}, {Vokrouhlick{\'y}},
  {Scheeres}, {Pravec}, {Lowry}, {Fitzsimmons}, {Nolan}, {Ostro}, {Benner},
  {Giorgini} and {Magri}}]{Taylor2007}
\bibinfo{author}{{Taylor}, P.A.}, \bibinfo{author}{{Margot}, J.L.},
  \bibinfo{author}{{Vokrouhlick{\'y}}, D.}, \bibinfo{author}{{Scheeres}, D.J.},
  \bibinfo{author}{{Pravec}, P.}, \bibinfo{author}{{Lowry}, S.C.},
  \bibinfo{author}{{Fitzsimmons}, A.}, \bibinfo{author}{{Nolan}, M.C.},
  \bibinfo{author}{{Ostro}, S.J.}, \bibinfo{author}{{Benner}, L.A.M.},
  \bibinfo{author}{{Giorgini}, J.D.}, \bibinfo{author}{{Magri}, C.},
  \bibinfo{year}{2007}.
\newblock \bibinfo{title}{{Spin Rate of Asteroid (54509) 2000 PH5 Increasing
  Due to the YORP Effect}}.
\newblock \bibinfo{journal}{Science} \bibinfo{volume}{316},
  \bibinfo{pages}{274}.
\bibitem[{{Torppa} et~al.(2003){Torppa}, {Kaasalainen}, {Micha{\l}owski},
  {Kwiatkowski}, {Kryszczy{\'n}ska}, {Denchev} and {Kowalski}}]{Torppa2003}
\bibinfo{author}{{Torppa}, J.}, \bibinfo{author}{{Kaasalainen}, M.},
  \bibinfo{author}{{Micha{\l}owski}, T.}, \bibinfo{author}{{Kwiatkowski}, T.},
  \bibinfo{author}{{Kryszczy{\'n}ska}, A.}, \bibinfo{author}{{Denchev}, P.},
  \bibinfo{author}{{Kowalski}, R.}, \bibinfo{year}{2003}.
\newblock \bibinfo{title}{{Shapes and rotational properties of thirty asteroids
  from photometric data}}.
\newblock \bibinfo{journal}{Icarus} \bibinfo{volume}{164},
  \bibinfo{pages}{346--383}.
\bibitem[{{Tsirvoulis} et~al.(2017){Tsirvoulis}, {Morbidelli}, {Delbo} and
  {Tsiganis}}]{Tsirvoulis2017}
\bibinfo{author}{{Tsirvoulis}, G.}, \bibinfo{author}{{Morbidelli}, A.},
  \bibinfo{author}{{Delbo}, M.}, \bibinfo{author}{{Tsiganis}, K.},
  \bibinfo{year}{2017}.
\newblock \bibinfo{title}{{Reconstructing the size distribution of the
  primordial Main Belt}}.
\newblock \bibinfo{journal}{ArXiv e-prints} \eprint{1706.02091}.
\bibitem[{{Viikinkoski}(2016)}]{Viikinkoski2016}
\bibinfo{author}{{Viikinkoski}, M.}, \bibinfo{year}{2016}.
\newblock \bibinfo{title}{{Shape Reconstruction from Generalized Projections}}.
\newblock Ph.D. thesis. Tampere University of Technology.
\bibitem[{{Viikinkoski} et~al.(2015a){Viikinkoski}, {Kaasalainen} and
  {\v{D}urech}}]{Viikinkoski2015}
\bibinfo{author}{{Viikinkoski}, M.}, \bibinfo{author}{{Kaasalainen}, M.},
  \bibinfo{author}{{\v{D}urech}, J.}, \bibinfo{year}{2015}a.
\newblock \bibinfo{title}{{ADAM: a general method for using various data types
  in asteroid reconstruction}}.
\newblock \bibinfo{journal}{\aap} \bibinfo{volume}{576}, \bibinfo{pages}{A8}.
\newblock \eprint{1501.05958}.
\bibitem[{{Viikinkoski} et~al.(2015b){Viikinkoski}, {Kaasalainen}, {{\v
  D}urech}, {Carry}, {Marsset}, {Fusco}, {Dumas}, {Merline}, {Yang},
  {Berthier}, {Kervella} and {Vernazza}}]{Viikinkoski2015b}
\bibinfo{author}{{Viikinkoski}, M.}, \bibinfo{author}{{Kaasalainen}, M.},
  \bibinfo{author}{{{\v D}urech}, J.}, \bibinfo{author}{{Carry}, B.},
  \bibinfo{author}{{Marsset}, M.}, \bibinfo{author}{{Fusco}, T.},
  \bibinfo{author}{{Dumas}, C.}, \bibinfo{author}{{Merline}, W.J.},
  \bibinfo{author}{{Yang}, B.}, \bibinfo{author}{{Berthier}, J.},
  \bibinfo{author}{{Kervella}, P.}, \bibinfo{author}{{Vernazza}, P.},
  \bibinfo{year}{2015}b.
\newblock \bibinfo{title}{{VLT/SPHERE- and ALMA-based shape reconstruction of
  asteroid (3) Juno}}.
\newblock \bibinfo{journal}{\aap} \bibinfo{volume}{581}, \bibinfo{pages}{L3}.
\bibitem[{{Vokrouhlick{\'y}} et~al.(2015){Vokrouhlick{\'y}}, {Bottke},
  {Chesley}, {Scheeres} and {Statler}}]{Vokrouhlicky2015}
\bibinfo{author}{{Vokrouhlick{\'y}}, D.}, \bibinfo{author}{{Bottke}, W.F.},
  \bibinfo{author}{{Chesley}, S.R.}, \bibinfo{author}{{Scheeres}, D.J.},
  \bibinfo{author}{{Statler}, T.S.}, \bibinfo{year}{2015}.
\newblock \bibinfo{title}{{The Yarkovsky and YORP Effects}}.
\newblock pp. \bibinfo{pages}{509--531}.
\bibitem[{{Vokrouhlick{\'y}} et~al.(2006a){Vokrouhlick{\'y}}, {Bro{\v z}},
  {Bottke}, {Nesvorn{\'y}} and {Morbidelli}}]{Vokrouhlicky2006d}
\bibinfo{author}{{Vokrouhlick{\'y}}, D.}, \bibinfo{author}{{Bro{\v z}}, M.},
  \bibinfo{author}{{Bottke}, W.F.}, \bibinfo{author}{{Nesvorn{\'y}}, D.},
  \bibinfo{author}{{Morbidelli}, A.}, \bibinfo{year}{2006}a.
\newblock \bibinfo{title}{{Yarkovsky/YORP chronology of asteroid families}}.
\newblock \bibinfo{journal}{\icarus} \bibinfo{volume}{182},
  \bibinfo{pages}{118--142}.
\bibitem[{{Vokrouhlick{\'y}} et~al.(2003){Vokrouhlick{\'y}}, {Nesvorn{\'y}} and
  {Bottke}}]{Vokrouhlicky2003}
\bibinfo{author}{{Vokrouhlick{\'y}}, D.}, \bibinfo{author}{{Nesvorn{\'y}}, D.},
  \bibinfo{author}{{Bottke}, W.F.}, \bibinfo{year}{2003}.
\newblock \bibinfo{title}{{The vector alignments of asteroid spins by thermal
  torques}}.
\newblock \bibinfo{journal}{Nature} \bibinfo{volume}{425},
  \bibinfo{pages}{147--151}.
\bibitem[{{Vokrouhlick{\'y}} et~al.(2006b){Vokrouhlick{\'y}}, {Nesvorn{\'y}}
  and {Bottke}}]{Vokrouhlicky2006}
\bibinfo{author}{{Vokrouhlick{\'y}}, D.}, \bibinfo{author}{{Nesvorn{\'y}}, D.},
  \bibinfo{author}{{Bottke}, W.F.}, \bibinfo{year}{2006}b.
\newblock \bibinfo{title}{{Secular spin dynamics of inner main-belt
  asteroids}}.
\newblock \bibinfo{journal}{\icarus} \bibinfo{volume}{184},
  \bibinfo{pages}{1--28}.
\bibitem[{{Walsh} et~al.(2013){Walsh}, {Delb{\'o}}, {Bottke},
  {Vokrouhlick{\'y}} and {Lauretta}}]{Walsh2013}
\bibinfo{author}{{Walsh}, K.J.}, \bibinfo{author}{{Delb{\'o}}, M.},
  \bibinfo{author}{{Bottke}, W.F.}, \bibinfo{author}{{Vokrouhlick{\'y}}, D.},
  \bibinfo{author}{{Lauretta}, D.S.}, \bibinfo{year}{2013}.
\newblock \bibinfo{title}{{Introducing the Eulalia and new Polana asteroid
  families: Re-assessing primitive asteroid families in the inner Main Belt}}.
\newblock \bibinfo{journal}{\icarus} \bibinfo{volume}{225},
  \bibinfo{pages}{283--297}.
\newblock \eprint{1305.2821}.
\bibitem[{{Warner}(2006a)}]{Warner2006}
\bibinfo{author}{{Warner}, B.D.}, \bibinfo{year}{2006}a.
\newblock \bibinfo{title}{{Asteroid lightcurve analysis at the Palmer Divide
  Observatory - late 2005 and early 2006}}.
\newblock \bibinfo{journal}{Minor Planet Bulletin} \bibinfo{volume}{33},
  \bibinfo{pages}{58--62}.
\bibitem[{{Warner}(2006b)}]{Warner2006c}
\bibinfo{author}{{Warner}, B.D.}, \bibinfo{year}{2006}b.
\newblock \bibinfo{title}{{Asteroid lightcurve analysis at the Palmer Divide
  Observatory - March - June 2006}}.
\newblock \bibinfo{journal}{Minor Planet Bulletin} \bibinfo{volume}{33},
  \bibinfo{pages}{85--88}.
\bibitem[{{Warner}(2010)}]{Warner2010a}
\bibinfo{author}{{Warner}, B.D.}, \bibinfo{year}{2010}.
\newblock \bibinfo{title}{{Asteroid Lightcurve Analysis at the Palmer Divide
  Observatory: 2009 December - 2010 March}}.
\newblock \bibinfo{journal}{Minor Planet Bulletin} \bibinfo{volume}{37},
  \bibinfo{pages}{112--118}.
\bibitem[{{Warner} et~al.(2009){Warner}, {Harris} and {Pravec}}]{Warner2009}
\bibinfo{author}{{Warner}, B.D.}, \bibinfo{author}{{Harris}, A.W.},
  \bibinfo{author}{{Pravec}, P.}, \bibinfo{year}{2009}.
\newblock \bibinfo{title}{{The asteroid lightcurve database}}.
\newblock \bibinfo{journal}{Icarus} \bibinfo{volume}{202},
  \bibinfo{pages}{134--146}.
\bibitem[{{Warner} et~al.(2011){Warner}, {Stephens} and {Harris}}]{Warner2011d}
\bibinfo{author}{{Warner}, B.D.}, \bibinfo{author}{{Stephens}, R.D.},
  \bibinfo{author}{{Harris}, A.W.}, \bibinfo{year}{2011}.
\newblock \bibinfo{title}{{Save the Lightcurves}}.
\newblock \bibinfo{journal}{Minor Planet Bulletin} \bibinfo{volume}{38},
  \bibinfo{pages}{172--174}.
\bibitem[{{Wright} et~al.(2010){Wright}, {Eisenhardt}, {Mainzer}, {Ressler},
  {Cutri}, {Jarrett}, {Kirkpatrick}, {Padgett}, {McMillan}, {Skrutskie},
  {Stanford}, {Cohen}, {Walker}, {Mather}, {Leisawitz}, {Gautier}, {McLean},
  {Benford}, {Lonsdale}, {Blain}, {Mendez}, {Irace}, {Duval}, {Liu}, {Royer},
  {Heinrichsen}, {Howard}, {Shannon}, {Kendall}, {Walsh}, {Larsen}, {Cardon},
  {Schick}, {Schwalm}, {Abid}, {Fabinsky}, {Naes} and {Tsai}}]{Wright2010}
\bibinfo{author}{{Wright}, E.L.}, \bibinfo{author}{{Eisenhardt}, P.R.M.},
  \bibinfo{author}{{Mainzer}, A.K.}, \bibinfo{author}{{Ressler}, M.E.},
  \bibinfo{author}{{Cutri}, R.M.}, \bibinfo{author}{{Jarrett}, T.},
  \bibinfo{author}{{Kirkpatrick}, J.D.}, \bibinfo{author}{{Padgett}, D.},
  \bibinfo{author}{{McMillan}, R.S.}, \bibinfo{author}{{Skrutskie}, M.},
  \bibinfo{author}{{Stanford}, S.A.}, \bibinfo{author}{{Cohen}, M.},
  \bibinfo{author}{{Walker}, R.G.}, \bibinfo{author}{{Mather}, J.C.},
  \bibinfo{author}{{Leisawitz}, D.}, \bibinfo{author}{{Gautier}, I.T.N.},
  \bibinfo{author}{{McLean}, I.}, \bibinfo{author}{{Benford}, D.},
  \bibinfo{author}{{Lonsdale}, C.J.}, \bibinfo{author}{{Blain}, A.},
  \bibinfo{author}{{Mendez}, B.}, \bibinfo{author}{{Irace}, W.R.},
  \bibinfo{author}{{Duval}, V.}, \bibinfo{author}{{Liu}, F.},
  \bibinfo{author}{{Royer}, D.}, \bibinfo{author}{{Heinrichsen}, I.},
  \bibinfo{author}{{Howard}, J.}, \bibinfo{author}{{Shannon}, M.},
  \bibinfo{author}{{Kendall}, M.}, \bibinfo{author}{{Walsh}, A.L.},
  \bibinfo{author}{{Larsen}, M.}, \bibinfo{author}{{Cardon}, J.G.},
  \bibinfo{author}{{Schick}, S.}, \bibinfo{author}{{Schwalm}, M.},
  \bibinfo{author}{{Abid}, M.}, \bibinfo{author}{{Fabinsky}, B.},
  \bibinfo{author}{{Naes}, L.}, \bibinfo{author}{{Tsai}, C.W.},
  \bibinfo{year}{2010}.
\newblock \bibinfo{title}{{The Wide-field Infrared Survey Explorer (WISE):
  Mission Description and Initial On-orbit Performance}}.
\newblock \bibinfo{journal}{\aj} \bibinfo{volume}{140},
  \bibinfo{pages}{1868--1881}.
\newblock \eprint{1008.0031}.
\bibitem[{{Zappal{\`a}} et~al.(1990){Zappal{\`a}}, {Cellino}, {Farinella} and
  {Kne\v{z}evi{\'c}}}]{Zappala1990}
\bibinfo{author}{{Zappal{\`a}}, V.}, \bibinfo{author}{{Cellino}, A.},
  \bibinfo{author}{{Farinella}, P.}, \bibinfo{author}{{Kne\v{z}evi{\'c}}, Z.},
  \bibinfo{year}{1990}.
\newblock \bibinfo{title}{{Asteroid families. I - Identification by
  hierarchical clustering and reliability assessment}}.
\newblock \bibinfo{journal}{\aj} \bibinfo{volume}{100},
  \bibinfo{pages}{2030--2046}.
\bibitem[{{Zappal{\`a}} et~al.(1994){Zappal{\`a}}, {Cellino}, {Farinella} and
  {Milani}}]{Zappala1994}
\bibinfo{author}{{Zappal{\`a}}, V.}, \bibinfo{author}{{Cellino}, A.},
  \bibinfo{author}{{Farinella}, P.}, \bibinfo{author}{{Milani}, A.},
  \bibinfo{year}{1994}.
\newblock \bibinfo{title}{{Asteroid families. 2: Extension to unnumbered
  multiopposition asteroids}}.
\newblock \bibinfo{journal}{\aj} \bibinfo{volume}{107},
  \bibinfo{pages}{772--801}.

\end{thebibliography}
\bibliographystyle{model2-names}

\appendix

\section{(423)~Diotima -- an interloper in the Eos family}
\label{sec:app1}

\begin{figure}
    \centering
        \resizebox{0.24\hsize}{!}{\includegraphics{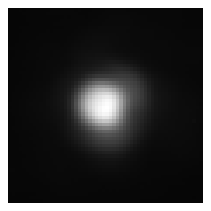}}\resizebox{0.24\hsize}{!}{\includegraphics{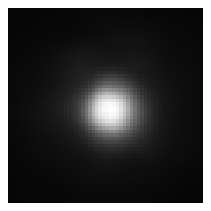}}\resizebox{0.24\hsize}{!}{\includegraphics{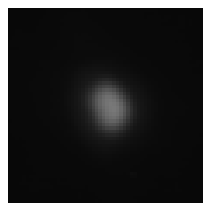}}\resizebox{0.24\hsize}{!}{\includegraphics{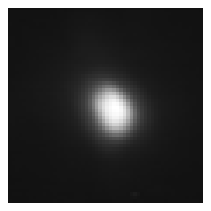}}\\
        \resizebox{0.24\hsize}{!}{\includegraphics{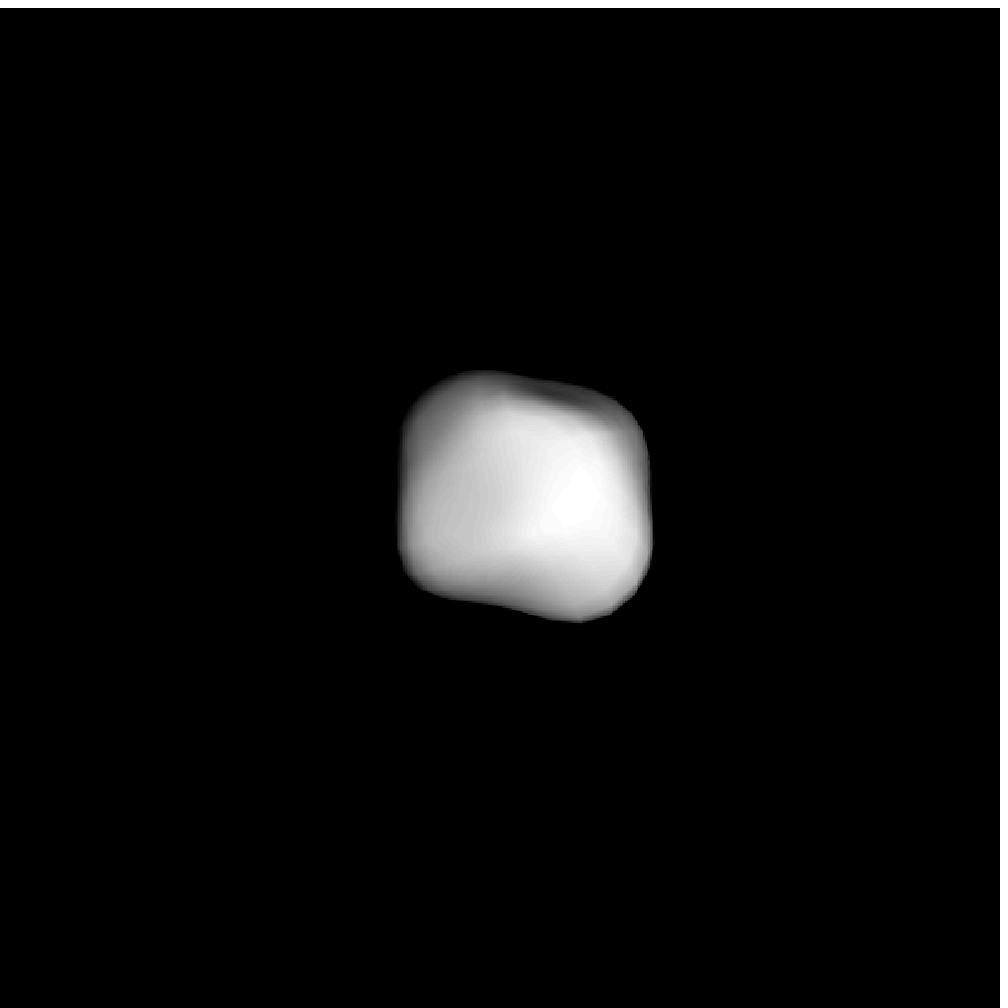}}\resizebox{0.24\hsize}{!}{\includegraphics{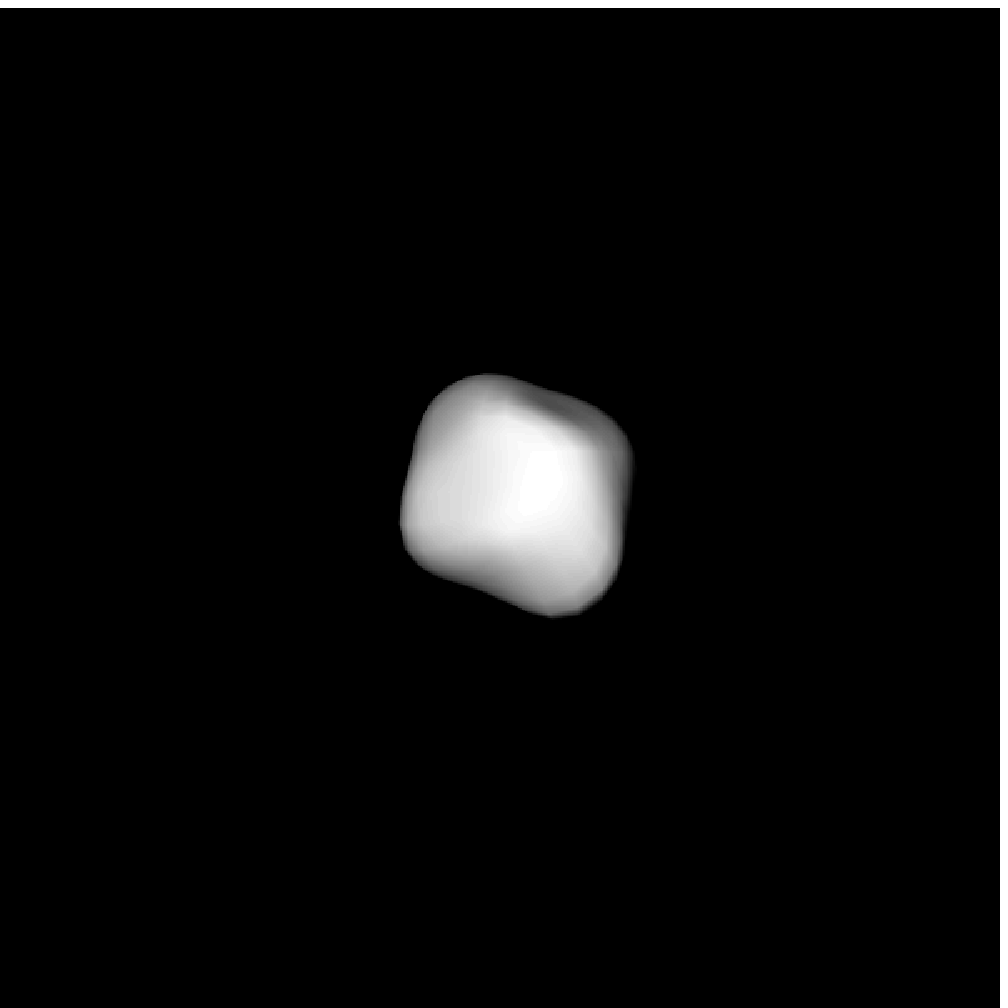}}\resizebox{0.24\hsize}{!}{\includegraphics{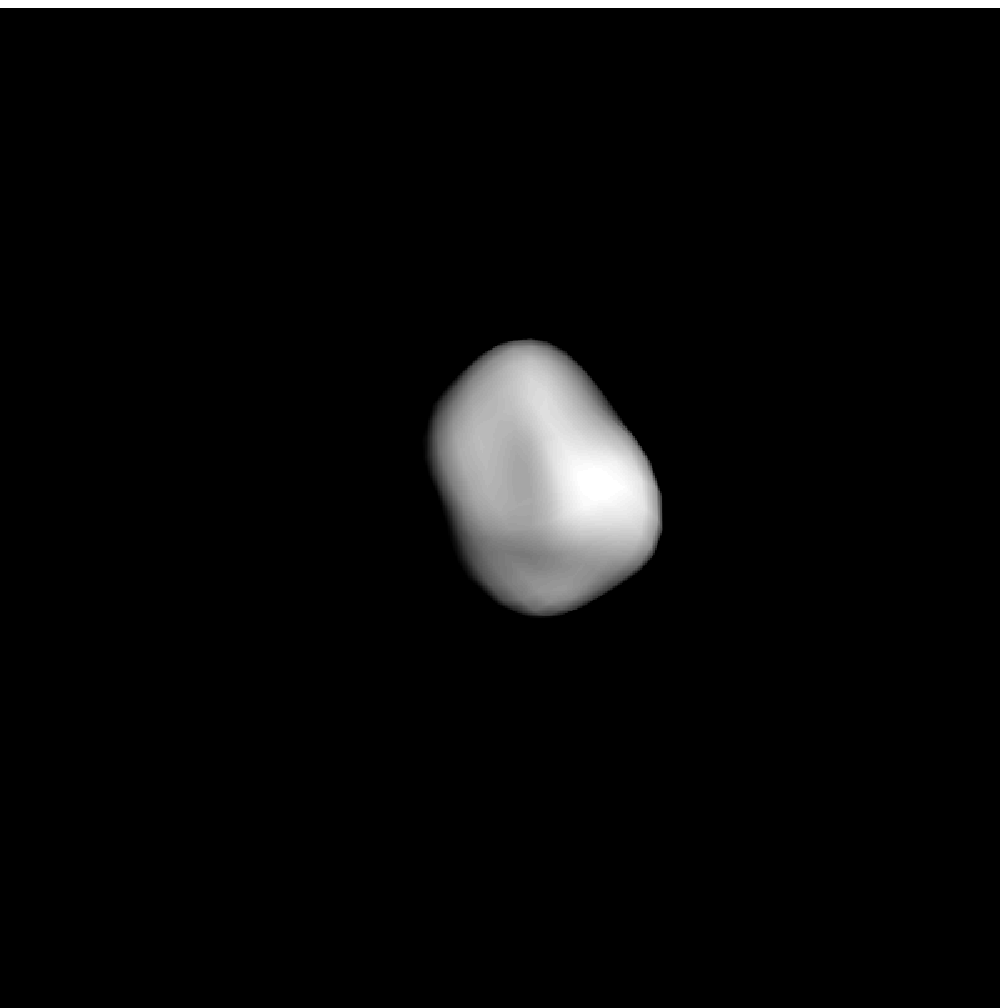}}\resizebox{0.24\hsize}{!}{\includegraphics{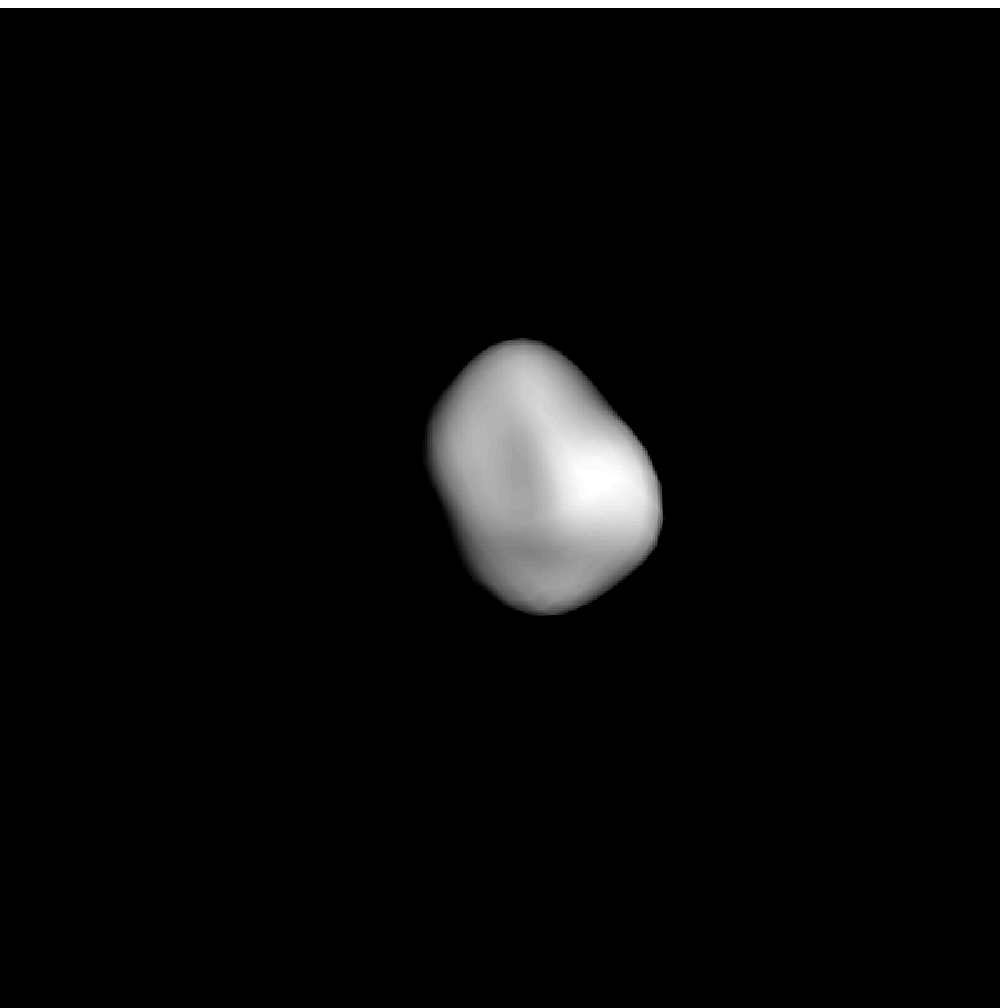}}\\
    \caption{\label{fig:423}Comparison between model projections (bottom) and corresponding AO images (top) for asteroid (423) Diotima.}
\end{figure}

A particular interesting case is offered by the asteroid (423)~Diotima, which is identified as an Eos member by previous published list of the Eos family based on the HCM \citep{Nesvorny2012}, but not in the \citet{Nesvorny2015}. However, Diotima is a C-type object, so very likely an interloper, which was already suggested. Diotima's bulk density of (1.39$\pm$0.50) g\,cm$^{−3}$ \citep{Carry2012b} is much lower than the typical density of K-type asteroids \citep[3.54$\pm$0.21 g\,cm$^{−3}$,][]{Carry2012b}. Moreover, the low geometric visible albedo of Diotima also suggests that this asteroid does not belong to the Eos family. Despite that, we would like to take advantage of the existence of the disk-resoled data of Diotima we gathered in the framework of \citet{Hanus2017b} from the Keck observatory archive to derive a 3D shape model and revise the density estimate of \citet{Carry2012b}.

In our recent studies \citep{Hanus2017a,Hanus2017b}, we processed all asteroid disk-resolved images obtained by the Near InfraRed Camera 2 (Nirc2) mounted on the W.M. Keck II telescope on Maunakea in Hawaii that are publicly available in the Keck Observatory Archive. We identified more than 500 disk-resolved images of almost 80 asteroids, out of them four belong to asteroid (423)~Diotima. We applied the All-Data Asteroid Modeling inversion algorithm \citep[ADAM,][]{Viikinkoski2015} to the combined optical light curves and disk-resolved images and derived a scaled shape model of Diotima. We used the spin state of the convex shape model from \citet{Hanus2016a} as a necessary initial input for the shape modeling with ADAM. We confirmed the results of \citet{Hanus2013b} that the mirror pole solution does not fit well the disk-resolved images, so a single pole solution is presented. Our volume-equivalent diameter of (205$\pm$7)~km is consistent with the estimates of \citet{Hanus2013b} (194$\pm$18 km) and \citet{Carry2012b} (211$\pm$16~km). The former is based on scaling the size to fit a single asteroid contour extracted from the disk-resolved image, while the latter one is based on a compilation of various literature values. We show the comparison between model projections and corresponding AO images in Fig.~\ref{fig:423}. The combination of the mass of (6.91$\pm$1.93)$\cdot$10$^{18}$~kg \citep{Carry2012b} and our size estimate leads to a bulk density of (1.5$\pm$0.4)~g$\cdot$cm$^{-3}$. Such density is typical for primitive asteroids from the C-complex, and in an agreement with the C-type classification of Diotima. While K-type asteroids should have larger bulk densities \citep{Carry2012b}, the low density of Diotima is another confirmation that this asteroid is not a true member of the Eos family. To derive the shape model by the ADAM algorithm, we followed the procedure already applied in \citet{Viikinkoski2015b, Hanus2017a, Hanus2017b,Marsset2017}. Additional details about this method can be found in these studies or in the general introduction of the method in \citet{Viikinkoski2015, Viikinkoski2016}. We uploaded the scaled shape model of Diotima to the DAMIT database.

\section{On-line tables}
\label{sec:app2}

\onecolumn
\scriptsize{
\begin{longtable}{r@{\,\,\,}l rrrr D{.}{.}{6} rrrc rrr}
\caption{\label{tab:models}Rotational states of members of Eos collision family and summary of used photometry for their shape model determinations. The table gives ecliptic coordinates $\lambda_1$ and $\beta_1$ of the best-fitting pole solution, ecliptic coordinates $\lambda_2$ and $\beta_2$ for the possible second (mirror) pole solution, sidereal rotational period $P$, the number of dense light curves $N_{\mathrm{lc}}$ spanning $N_{\mathrm{app}}$ apparitions, the number of sparse-in-time measurements $N_{\mathrm{sp}}$, the reference to the model, the size $D$ and geometric visible albedo $p_\mathrm{V}$ based on WISE thermal infrared measurements \citep{Masiero2011}, and the shape model quality flag $QF$.}\\
\hline
\multicolumn{2}{c} {Asteroid} & \multicolumn{1}{c} {$\lambda_1$} & \multicolumn{1}{c} {$\beta_1$} & \multicolumn{1}{c} {$\lambda_2$} & \multicolumn{1}{c} {$\beta_2$} & \multicolumn{1}{c} {$P$} & $N_{\mathrm{lc}}$ & $N_{\mathrm{app}}$  & $N_{\mathrm{sp}}$ & \multicolumn{1}{c}{Reference} & \multicolumn{1}{c}{$D$} & \multicolumn{1}{c}{$p_\mathrm{V}$} & \multicolumn{1}{c}{$QF$} \\
\multicolumn{2}{l} { } & [deg] & [deg] & [deg] & [deg] & \multicolumn{1}{c} {[hours]} &  &  &  & \multicolumn{1}{c}{} & [km] &  & \\
\hline\hline

\endfirsthead
\caption{continued.}\\

\hline
\multicolumn{2}{c} {Asteroid} & \multicolumn{1}{c} {$\lambda_1$} & \multicolumn{1}{c} {$\beta_1$} & \multicolumn{1}{c} {$\lambda_2$} & \multicolumn{1}{c} {$\beta_2$} & \multicolumn{1}{c} {$P$} & $N_{\mathrm{lc}}$ & $N_{\mathrm{app}}$  & $N_{\mathrm{sp}}$ & \multicolumn{1}{c}{Reference} & \multicolumn{1}{c}{$D$} & \multicolumn{1}{c}{$p_\mathrm{V}$} & \multicolumn{1}{c}{$QF$} \\
\multicolumn{2}{l} { } & [deg] & [deg] & [deg] & [deg] & \multicolumn{1}{c} {[hours]} &  &  &  & \multicolumn{1}{c}{} & [km] &  & \\
\hline\hline
\endhead
\hline
\endfoot
\hline
221     &  Eos            &  119$\pm$5   & --37$\pm$5   &  289$\pm$5   & --23$\pm$5   &  10.42213(2)   &  27  &  6   &  304  &  New                 & 91.2$\pm$2.2 & 0.18$\pm$0.02 & 3 \\
251     &  Sophia\footnotemark[1]    &  235$\pm$10  & --52$\pm$10  &  47$\pm$10   & --84$\pm$10  &  20.2222(1)    &  7   &  3   &  438  &  New      & 27.5$\pm$0.2 & 0.23$\pm$0.04 & 2 \\
423     &  Diotima\footnotemark[1]   &  351$\pm$5   &  4$\pm$5     &              &   &  4.775377(4)   &  58  &  12  &  540  &  \citet{Hanus2016a}  & 175.9$\pm$3.9 & 0.07$\pm$0.01 & 3 \\
        &                 &  354$\pm$2   &  0$\pm$2     &              &              &  4.775377(2)   &  58  &  12  &       &  Revised, ADAM       &  &  & 4 \\
450     &  Brigitta       &  32$\pm$10   & --16$\pm$10  &  203$\pm$10  & --26$\pm$10  &  10.76437(4)   &  12  &  2   &  253  &  New                 & 37.0$\pm$0.1 & 0.09$\pm$0.02 & 2 \\
513     &  Centesima      &  149$\pm$20  &  4$\pm$20    &  332$\pm$20  &  15$\pm$20   &  5.32399(3)    &      &      &  460  &  \citet{Durech2016}  & 48.8$\pm$0.7 & 0.09$\pm$0.02 & 2 \\
520     &  Franziska      &  114$\pm$10  & --45$\pm$10  &  282$\pm$10  & --79$\pm$10  &  16.50449(5)   &  9   &  2   &  434  &  \citet{Hanus2016a}  & 25.3$\pm$0.2 & 0.15$\pm$0.03 & 2 \\
529     &  Preziosa       &  178$\pm$15  & --10$\pm$15  &  2$\pm$15    & --25$\pm$15  &  25.9391(5)    &  1   &  1   &  399  &  New, WISE           & 41.2$\pm$0.5 & 0.10$\pm$0.00 & 1.5 \\
562     &  Salome         &  78$\pm$10   &  40$\pm$10   &  275$\pm$10  &  28$\pm$10   &  6.35031(2)    &  9   &  3   &  475  &  \citet{Hanus2016a} & 33.2$\pm$0.5 & 0.15$\pm$0.03 & 2 \\
        &                 &  54$\pm$10   &  56$\pm$10   &  268$\pm$10  &  44$\pm$10   &  6.35030(1)    &  25  &  5   &  425  &  Revised            &  &  & 3 \\
573     &  Recha          &  252$\pm$15  & --48$\pm$15  &  74$\pm$15   & --24$\pm$15  &  7.16586(3)    &  3   &  1   &  246  &  \citet{Hanus2011}  & 47.6$\pm$0.5 & 0.11$\pm$0.02 & 1.5 \\
        &                 &              &              &  76$\pm$10   & --26$\pm$10  &  7.16585(2)    &  9   &  3   &  356  &  Revised            &  &  & 2 \\
590     &  Tomyris        &  120$\pm$15  & --46$\pm$15  &  273$\pm$15  & --47$\pm$15  &  5.55248(2)    &  3   &  1   &  123  &  \citet{Hanus2011}  & 30.6$\pm$0.2 & 0.19$\pm$0.04 & 1.5 \\
        &                 &  113$\pm$10  & --35$\pm$10  &  274$\pm$10  & --29$\pm$10  &  5.55248(1)    &  5   &  2   &  257  &  Revised            &  &  & 2 \\
639     &  Latona         &  204$\pm$10  &  10$\pm$10   &  25$\pm$10   &  12$\pm$10   &  6.19127(1)    &  5   &  2   &  202  &  \citet{Hanus2016a} & 78.5$\pm$1.2 & 0.12$\pm$0.02 & 2 \\
669     &  Kypria         &  189$\pm$15  &  49$\pm$15   &  31$\pm$15   &  40$\pm$15   &  14.2789(1)    &  5   &  1   &  268  &  \citet{Hanus2013a} & 29.2$\pm$0.4 & 0.17$\pm$0.03 & 1.5 \\
742     &  Edisona        &  175$\pm$10  & --43$\pm$10  &  46$\pm$10   & --54$\pm$10  &  18.5833(1)    &  15  &  3   &  356  &  \citet{Hanus2016a} & 47.1$\pm$0.4 & 0.12$\pm$0.02 & 2 \\
        &                 &  170$\pm$10  & --39$\pm$10  &  9$\pm$10    & --65$\pm$10  &  18.5833(1)    &  21  &  4   &  256  &  Revised            &  &  & 3 \\
798     &  Ruth           &  84$\pm$10   &  27$\pm$10   &              &              &  8.55068(2)    &  18  &  5   &  426  &  \citet{Hanus2016a} & 45.2$\pm$0.6 & 0.15$\pm$0.01 & 3 \\
807     &  Ceraskia       &  132$\pm$15  &  26$\pm$15   &  325$\pm$15  &  23$\pm$15   &  7.37391(3)    &  2   &  1   &  243  &  \citet{Hanus2013a} & 21.2$\pm$0.3 & 0.21$\pm$0.02 & 1.5 \\
890     &  Waltraut       &  30$\pm$10   &  69$\pm$10   &  123$\pm$10  &  72$\pm$10   &  12.5831(1)    &  26  &  3   &  262  &  New                & 28.4$\pm$0.2 & 0.12$\pm$0.02 & 3 \\
1075    &  Helina         &  123$\pm$20  & --33$\pm$20  &  284$\pm$20  & --34$\pm$20  &  44.678(1)     &      &      &  421  &  \citet{Durech2016} & 37.9$\pm$0.8 & 0.11$\pm$0.01 & 1 \\
        &                 &  127$\pm$15  & --43$\pm$15  &  280$\pm$15  & --44$\pm$15  &  44.677(1)     &  5   &  1   &  312  &  Revised            &  &  & 2 \\
1087    &  Arabis         &  155$\pm$15  &  12$\pm$15   &  334$\pm$15  & --7$\pm$15   &  5.79500(1)    &  3   &  1   &  248  &  \citet{Hanus2011}  & 37.5$\pm$0.5 & 0.15$\pm$0.03 & 1.5 \\
        &                 &  155$\pm$15  &  25$\pm$15   &  331$\pm$15  &  5$\pm$15    &  5.79500(1)    &  4   &  2   &  302  &  Revised            &  &  & 2 \\
1148    &  Rarahu         &  148$\pm$20  & --9$\pm$20   &  322$\pm$20  & --9$\pm$20   &  6.54449(3)    &      &      &  159  &  \citet{Hanus2011}  & 27.5$\pm$0.4 & 0.20$\pm$0.03 & 1 \\
        &                 &  146$\pm$10  & --2$\pm$10   &  326$\pm$10  & --2$\pm$10   &  6.54448(2)    &  5   &  3   &  280  &  Revised            &  &  & 2 \\
1207    &  Ostenia        &  124$\pm$15  & --51$\pm$15  &  310$\pm$15  & --77$\pm$15  &  9.07129(1)    &  2   &  1   &  158  &  \citet{Hanus2011}  & 22.9$\pm$1.3 & 0.13$\pm$0.02 & 1.5 \\
        &                 &  183$\pm$10  & --63$\pm$10  &  51$\pm$10   & --59$\pm$10  &  9.07130(1)    &  7   &  3   &  403  &  Revised            &  &  & 2 \\
1210    &  Morosovia      &  62$\pm$10   & --69$\pm$10  &  245$\pm$10  & --77$\pm$10  &  15.26088(4)   &  12  &  4   &  295  &  New                & 33.7$\pm$0.3 & 0.21$\pm$0.03 & 2 \\
1286    &  Banachiewicza  &  214$\pm$20  &  62$\pm$20   &  64$\pm$20   &  60$\pm$20   &  8.63043(2)    &      &      &  132  &  \citet{Hanus2013c} & 21.5$\pm$0.2 & 0.17$\pm$0.01 & 1 \\
1291    &  Phryne         &  106$\pm$15  &  35$\pm$15   &  277$\pm$15  &  59$\pm$15   &  5.584137(1)   &  2   &  1   &  201  &  \citet{Hanus2011}  & 27.4$\pm$0.1 & 0.13$\pm$0.02 & 1.5 \\
        &                 &  109$\pm$15  &  33$\pm$15   &  281$\pm$15  &  56$\pm$15   &  5.584139(1)   &  5   &  2   &  267  &  Revised            &  &  & 2 \\
1339    &  Desagneauxa    &  225$\pm$20  &  42$\pm$20   &  63$\pm$20   &  53$\pm$20   &  9.37514(5)    &      &      &  465  &  \citet{Durech2016} & 24.1$\pm$0.3 & 0.14$\pm$0.02 & 1 \\
        &                 &  230$\pm$15  &  66$\pm$15   &  106$\pm$15  &  72$\pm$15   &  9.37510(4)    &  4   &  1   &  248  &  Revised            &  &  & 1.5 \\
1353    &  Maartje        &  266$\pm$20  &  73$\pm$20   &  92$\pm$20   &  57$\pm$20   &  22.9926(4)    &      &      &  293  &  \citet{Hanus2013c} & 39.0$\pm$0.5 & 0.07$\pm$0.04 & 1 \\
        &                 &  285$\pm$10  &  73$\pm$10   &  119$\pm$10  &  41$\pm$10   &  22.9924(2)    &  7   &  3   &  343  &  Revised            &  &  & 2 \\
1364    &  Safara         &  197$\pm$10  &  32$\pm$10   &  10$\pm$10   &  12$\pm$10   &  7.14908(4)    &  6   &  2   &  251  &  New                & 21.2$\pm$0.2 & 0.23$\pm$0.03 & 2 \\
1388    &  Aphrodite      &  325$\pm$10  &  35$\pm$10   &  137$\pm$10  &  66$\pm$10   &  11.94389(5)   &  16  &  3   &  220  &  New                & 21.4$\pm$0.3 & 0.18$\pm$0.05 & 2 \\
1464    &  Armisticia     &  194$\pm$15  & --54$\pm$15  &  35$\pm$15   & --69$\pm$15  &  7.46699(4)    &  2   &  1   &  298  &  \citet{Hanus2013c} & 23.3$\pm$5.1 & 0.12$\pm$0.36 & 1.5 \\
1552    &  Bessel         &  61$\pm$20   & --50$\pm$20  &  221$\pm$20  & --57$\pm$20  &  8.96318(5)    &  2   &  1   &  338  &  New, WISE          & 18.5$\pm$0.1 & 0.16$\pm$0.02 & 1.5 \\
1557    &  Roehla         &  124$\pm$20  & --38$\pm$20  &  329$\pm$20  & --57$\pm$20  &  5.67899(2)    &      &      &  334  &  \citet{Durech2016} & 18.8$\pm$0.3 & 0.15$\pm$0.04 & 1 \\
        &                 &  124$\pm$15  & --41$\pm$15  &  326$\pm$15  & --54$\pm$15  &  5.67900(1)    &  2   &  1   &  384  &  Revised            &  &  & 1.5 \\
1641    &  Tana           &  102$\pm$15  &  39$\pm$15   &  276$\pm$15  &  18$\pm$15   &  8.01971(1)    &  4   &  2   &  217  &  New                & 25.1$\pm$1.0 & 0.20$\pm$0.02 & 2 \\
1723    &  Klemola        &  239$\pm$20  & --56$\pm$20  &  52$\pm$20   & --54$\pm$20  &  6.25609(2)    &      &      &  484  &  \citet{Durech2016} & 33.5$\pm$0.2 & 0.16$\pm$0.03 & 1 \\
        &                 &  252$\pm$10  & --35$\pm$10  &  80$\pm$10   & --58$\pm$10  &  6.25610(1)    &  14  &  5   &  348  &  Revised            &  &  & 3 \\
1753    &  Mieke          &  121$\pm$20  &  67$\pm$20   &  321$\pm$20  &  35$\pm$20   &  10.1994(2)    &      &      &  413  &  \citet{Durech2016} & 19.4$\pm$0.2 & 0.17$\pm$0.03 & 1 \\
1758    &  Naantali       &  150$\pm$20  & --60$\pm$20  &              &              &  5.47369(3)    &      &      &  446  &  \citet{Durech2016} & 21.7$\pm$1.3 & 0.17$\pm$0.02 & 1 \\
1801    &  Titicaca       &  260$\pm$20  &  57$\pm$20   &  48$\pm$20   &  51$\pm$20   &  3.21123(1)    &      &      &  379  &  \citet{Durech2016} & 22.0$\pm$0.3 & 0.15$\pm$0.03 & 1 \\
2180    &  Marjaleena     &  85$\pm$15   &  50$\pm$15   &  238$\pm$15  &  54$\pm$15   &  8.34626(4)    &  2   &  1   &  122  &  New                & 21.4$\pm$0.2 & 0.15$\pm$0.02 & 1.5 \\
2358    &  Bahner         &  0$\pm$10    &  57$\pm$10   &  193$\pm$10  &  52$\pm$10   &  10.8528(1)    &  13  &  1   &  69   &  \citet{Hanus2016a} & 18.6$\pm$0.2 & 0.20$\pm$0.03 & 2 \\
2425    &  Shenzhen       &  265$\pm$20  &  40$\pm$20   &  50$\pm$20   &  58$\pm$20   &  9.83818(2)    &      &      &  551  &  \citet{Durech2016} & 18.7$\pm$0.4 & 0.18$\pm$0.02 & 1 \\
2836    &  Sobolev        &  270$\pm$20  & --79$\pm$20  &  82$\pm$20   & --51$\pm$20  &  4.75488(1)    &      &      &  560  &  \citet{Durech2016} & 18.6$\pm$0.2 & 0.14$\pm$0.01 & 1 \\
2957    &  Tatsuo         &  248$\pm$10  &  32$\pm$10   &  81$\pm$10   &  45$\pm$10   &  6.82038(1)    &  13  &  1   &  135  &  \citet{Hanus2013a} & 22.5$\pm$0.1 & 0.26$\pm$0.04 & 2 \\
4077    &  Asuka          &  57$\pm$10   &  45$\pm$10   &  266$\pm$10  &  44$\pm$10   &  7.92309(2)    &  9   &  2   &  527  &  New                & 19.5$\pm$0.2 & 0.19$\pm$0.03 & 2 \\
4529    &  Webern         &  26$\pm$20   &  38$\pm$20   &  196$\pm$20  &  59$\pm$20   &  3.46903(1)    &  1   &  1   &  438  &  New, WISE          & 10.4$\pm$0.1 & 0.21$\pm$0.03 & 1.5 \\
4800    &  Veveri         &  274$\pm$15  & --40$\pm$15  &  95$\pm$15   & --65$\pm$15  &  6.21570(2)    &  4   &  1   &  477  &  New                & 14.1$\pm$0.1 & 0.17$\pm$0.04 & 1.5 \\
5281    &  Lindstrom      &  238$\pm$15  & --72$\pm$15  &  84$\pm$15   & --81$\pm$15  &  9.2511(1)     &  2   &  1   &  76   &  \citet{Hanus2013a} & 17.0$\pm$0.3 & 0.15$\pm$0.03 & 1.5 \\
5488    &  Kiyosato       &  19$\pm$20   &  23$\pm$20   &  242$\pm$20  &  62$\pm$20   &  8.76307(3)    &      &      &  449  &  \citet{Durech2016} & 18.8$\pm$0.3 & 0.15$\pm$0.03 & 1 \\
6136    &  Gryphon        &  134$\pm$20  &  57$\pm$20   &  337$\pm$20  &  63$\pm$20   &  16.4683(1)    &      &      &  602  &  \citet{Durech2016} & 15.6$\pm$0.3 & 0.22$\pm$0.03 & 1 \\
        &                 &  87$\pm$15   &  52$\pm$15   &  310$\pm$15  &  62$\pm$15   &  16.4684(1)    &  6   &  2   &  652  &  Revised            &  &  & 2 \\
6166    &  Univsima       &  242$\pm$20  & --59$\pm$20  &  80$\pm$20   & --42$\pm$20  &  11.3766(1)    &      &      &  425  &  \citet{Durech2016} & 12.1$\pm$0.4 & 0.18$\pm$0.03 & 1 \\
6590    &  Barolo         &  171$\pm$20  &  59$\pm$20   &  313$\pm$20  &  63$\pm$20   &  8.35926(3)    &      &      &  460  &  \citet{Durech2016} & 13.0$\pm$1.1 & 0.24$\pm$0.04 & 1 \\
7905    &  Juzoitami      &  105$\pm$20  & --76$\pm$20  &  226$\pm$20  & --55$\pm$20  &  2.72744(1)    &      &      &  118  &  \citet{Hanus2013a} & 0.0$\pm$0.0 & 0.00$\pm$0.00 & 1 \\
12384   &  Luigimartella  &  160$\pm$20  & --60$\pm$20  &  4$\pm$20    & --50$\pm$20  &  6.44220(2)    &      &      &  517  &  \citet{Durech2016} & 9.4$\pm$1.4 & 0.15$\pm$0.07 & 1 \\
14203   &  Hocking        &  106$\pm$20  &  81$\pm$20   &  240$\pm$20  &  52$\pm$20   &  9.08564(3)    &      &      &  393  &  \citet{Durech2016} & 9.3$\pm$0.3 & 0.16$\pm$0.01 & 1 \\
17111   & 1999 JH$_{52}$  &  157$\pm$20  &  46$\pm$20   &  328$\pm$20  &  62$\pm$20   &  8.89847(3)    &      &      &  629  &  \citet{Durech2016} & 12.8$\pm$0.3 & 0.10$\pm$0.02 & 1 \\
19848   &  Yeungchuchiu   &  190$\pm$20  & --67$\pm$20  &  66$\pm$20   & --70$\pm$20  &  3.451038(3)   &      &      &  107  &  \citet{Hanus2013a} & 13.2$\pm$0.3 & 0.21$\pm$0.03 & 1 \\
22018   & 1999 XK$_{105}$ &  61$\pm$20   &  45$\pm$20   &  228$\pm$20  &  31$\pm$20   &  17.0575(1)    &      &      &  373  &  \citet{Durech2016} & 0.0$\pm$0.0 & 0.00$\pm$0.00 & 1 \\
40104   & 1998 QL$_{4}$   &  324$\pm$20  & --87$\pm$20  &              &              &  4.475241(2)   &      &      &  443  &  \citet{Durech2016} & 8.8$\pm$0.1 & 0.13$\pm$0.03 & 1 \\
44417   & 1998 SS$_{146}$ &  353$\pm$15  & --48$\pm$15  &              &              &  7.63346(2)    &  2   &  1   &  371  &  New, WISE          & 7.5$\pm$0.2 & 0.20$\pm$0.02 & 1.5 \\
63166   & 2000 YW$_{17}$  &  68$\pm$20   &  63$\pm$20   &  304$\pm$20  &  50$\pm$20   &  9.94934(3)    &  1   &  1   &  192  &  New, WISE          & 4.5$\pm$0.5 & 0.20$\pm$0.05 & 1.5 \\
66076   & 1998 RD$_{53}$  &  100$\pm$20  & --54$\pm$20  &  233$\pm$20  & --66$\pm$20  &  8.93853(2)    &      &      &  485  &  \citet{Durech2016} & 7.8$\pm$0.2 & 0.10$\pm$0.01 & 1 \\
\end{longtable}
}
\footnotetext[1]{Likely interlopers that are not dynamically associated with the Eos family}

\onecolumn

\begin{longtable}{r@{\,\,\,}l lll}
\caption{\label{tab:references}New observations used for updating the shape models and observations that are not included in the UAPC used for new shape model determinations. Observations with the UH88 telescope were obtained as a joint effort of Victor Al\'i-Lagoa, Bryce Bolin, Josef Hanu\v s, Robert Jedicke and Marco Delbo'.}\\
\hline
 \multicolumn{2}{c} {Asteroid} & Date & $N_{\mathrm{LC}}$ & Observer \\ \hline\hline

\endfirsthead
\caption{continued.}\\

\hline
 \multicolumn{2}{c} {Asteroid} & Date & $N_{\mathrm{LC}}$ & Observer \\ \hline\hline
\endhead
\hline
\endfoot
  221 &             Eos & 1978 07 --  1978 08 &  6 & \citet{Harris1980}\\
      &                 & 1979 11 --  1979 12 &  6 & \citet{Harris1983} \\
      &                 & 2003 02 --  2003 02 &  3 & Martin Lehk\' y \\
      &                 & 2007 02 --  2007 02 &  2 & Ren\'e Roy \\
      &                 & 2013 01 --  2013 03 &  6 & \citet{Pilcher2013e} \\
      &                 & 2014 05 --  2014 06 &  4 & Nicolas Esseiva, Raoul Behrend \\
  251 &          Sophia & 2005 10 --  2005 10 &  1 & Laurent Bernasconi\\
      &                 & 2008 05 --  2008 05 &  5 & Etienne Morelle \\
      &                 & 2014 05 --  2014 06 &  1 & Ren\'e Roy \\
  450 &        Brigitta & 2004 12 --  2004 12 &  4 & Laurent Bernasconi\\
      &                 & 2016 01 --  2016 02 &  5 & Henk de Groot, Raoul Behrend \\
  562 &          Salome & 2005 05 --  2005 06 &  3 & Ren\'e Roy \\
      &                 & 2007 11 --  2007 12 & 13 & Hiromi Hamanowa, Hiroko Hamanowa \\
  573 &           Recha & 2005 12 --  2005 12 &  2 & \citet{Warner2006}\\
      &                 & 2010 10 --  2010 12 &  4 & Maurice Audejean \\
  590 &         Tomyris & 2005 05 --  2005 05 &  2 & Federico Manzini \\
  742 &         Edisona & 2003 04 --  2003 04 &  2 & Ren\'e Roy \\
      &                 & 2007 01 --  2007 02 &  4 & Pierre Antonini \\
  890 &        Waltraut & 2009 07 --  2009 07 &  4 & \citet{Brinsfield2010a}\\
      &                 & 2009 07 --  2009 08 &  3 & Pierre Antonini \\
      &                 & 2009 07 --  2009 09 & 17 & \citet{Owings2010} \\
      &                 & 2015 08 --  2015 09 &  2 & UH88 \\
 1075 &          Helina & 2013 04 --  2013 04 &  4 & Paul Krafft, Olivier Gerteis, Hubert Gully, \\
      &                 &                     &    & Luc Arnold, Matthieu Bachschmidt\\
      &                 & 2013 05 --  2013 05 &  1 & Matthieu Bachschmidt \\
 1087 &          Arabis & 2003 02 --  2003 02 &  1 & Martin Lehk\' y \\
 1148 &          Rarahu & 2002 07 --  2002 07 &  1 & Laurent Brunetto\\
      &                 & 2007 05 --  2007 05 &  2 & Pierre Antonini \\
      &                 & 2011 03 --  2011 03 &  2 & Ren\'e Roy \\
 1207 &         Ostenia & 2006 01 --  2006 03 &  5 & \citet{Warner2006} \\
 1210 &       Morosovia & 1984 08 --  1984 08 &  2 & \citet{DiMartino1986}\\
      &                 & 2003 02 --  2003 02 &  2 & Martin Lehk\' y \\
      &                 & 2009 05 --  2009 06 &  7 & Pierre Antonini \\
      &                 & 2014 04 --  2014 04 &  1 & Ren\'e Roy \\
 1291 &          Phryne & 2006 07 --  2006 09 &  3 & Pierre Antonini \\
 1339 &     Desagneauxa & 2006 07 --  2006 09 &  4 & Ren\'e Roy\\
 1353 &         Maartje & 2005 08 --  2005 08 &  1 & Laurent Bernasconi\\
      &                 & 2005 08 --  2005 09 &  4 & Pierre Antonini \\
      &                 & 2010 07 --  2010 07 &  1 & Fabien Reignier \\
      &                 & 2015 07 --  2015 07 &  1 & Nicolas Esseiva, Raoul Behrend \\
 1364 &          Safara & 2010 02 --  2010 02 &  1 & \citet{Warner2010a}\\
      &                 & 2016 06 --  2016 06 &  1 & Lehk\'y \\
      &                 & 2016 05 --  2016 05 &  1 & UH88 \\
 1388 &       Aphrodite & 2006 02 --  2006 02 &  2 & Ren\'e Roy\\
      &                 & 2007 04 --  2007 07 & 11 & \citet{Oey2008} \\
      &                 & 2016 01 --  2016 03 &  3 & UH88 \\
 1557 &          Roehla & 2016 03 --  2016 03 &  2 & UH88 \\
 1641 &            Tana & 1983 09 --  1983 09 &  1 & \citet{Binzel1987a}\\
      &                 & 2009 10 --  2009 10 &  1 & Jacques Michelet \\
 1723 &         Klemola & 1984 04 --  1984 05 &  2 & \citet{Binzel1987a}\\
      &                 & 2002 12 --  2002 12 &  4 & Martin Lehk\' y \\
      &                 & 2005 04 --  2005 05 &  2 & Ondrejov \\
      &                 & 2005 05 --  2005 05 &  1 & Laurent Bernasconi \\
      &                 & 2011 07 --  2011 07 &  4 & Ren\'e Roy \\
      &                 & 2016 05 --  2016 05 &  1 & UH88 \\
 2180 &      Marjaleena & 2013 09 --  2013 09 &  2 & Silvano Casulli\\
 4077 &           Asuka & 2006 04 --  2006 04 &  3 & \citet{Warner2006c}\\
      &                 & 2016 01 --  2016 04 &  6 & UH88 \\
 4800 &          Veveri & 2015 08 --  2015 08 &  1 & UH88 \\
 6136 &         Gryphon & 2008 04 --  2008 04 &  1 & Julian Oey\\
      &                 & 2015 09 --  2015 10 &  5 & UH88 \\ \hline
\end{longtable}





\end{document}